\documentclass[]{openjournal}

\usepackage{orcidlink}
\usepackage{url}

\usepackage{siunitx}
\usepackage{physics}
\usepackage{hyperref}
\hypersetup{
    colorlinks = true,
    linkcolor = blue,
    citecolor = blue,
}
\usepackage{array}
\usepackage{booktabs}
\usepackage{graphicx}
\usepackage[export]{adjustbox}
\usepackage{makecell}
\usepackage[rightcaption]{sidecap}
\sidecaptionvpos{figure}{b}
\usepackage{wrapfig}
\usepackage{capt-of}

\newcommand{\Msun}{\mathrm{M_\odot}}
\newcommand{\Zsun}{\mathrm{Z_\odot}}

\begin{document}

\title{\textit{How can we finally see the first light?}\\
Status and perspectives in the search for Population III stars}

\author{Alessandra Venditti \orcidlink{0000-0003-2237-0777}$^{1,2\star}$
} 
\thanks{
$^\star$ Cosmic Frontier Center Prize Fellow\\
Email: \href{mailto:alessandra.venditti@utexas.edu}{alessandra.venditti@utexas.edu}
}
    \author{Daniel Schaerer
        \orcidlink{0000-0001-7144-7182}$^3$
    } 
    \author{Erik Zackrisson
        \orcidlink{0000-0003-1096-2636}$^4$
    } 
    \author{Yoshihisa Asada
        \orcidlink{0000-0003-3983-5438}$^5$
    } 
    \author{Harley Katz\orcidlink{0000-0003-1561-3814}$^6$} 
    \author{Stefania Salvadori\orcidlink{0000-0001-7298-2478}$^{7,8}$} 
    \author{Eros Vanzella
        \orcidlink{0000-0001-5228-9326}$^{9}$
    } 
    \author{Julian B. Muñoz
        \orcidlink{0000-0002-8984-0465}$^1$
    }
    \author{Anatole Storck
        \orcidlink{0009-0000-3245-7951}$^{10}$
    } 
    \author{Andrew J. Bunker\orcidlink{0000-0002-8651-9879}$^{11}$}
    \author{Alessandro Trinca\orcidlink{0000-0002-1899-4360}$^{12}$}
    \author{Dirk Scholte\orcidlink{0000-0002-6867-1244}$^{12}$}
    \author{Fabio Pacucci\orcidlink{0000-0001-9879-7780}$^{13}$}
    \author{Pablo G. Pérez-González\orcidlink{0000-0003-4528-5639}$^{14}$}
    \author{Seiji Fujimoto\orcidlink{0000-0001-7201-5066}$^5$}
    \author{Corinne Charbonnel\orcidlink{0000-0002-6449-6194}$^3$}
    \author{Roberto Maiolino\orcidlink{0000-0002-4985-3819}$^{15,16,17}$}
    \author{Andrea Ferrara\orcidlink{0000-0002-9400-7312}$^{18}$}
    \author{Mauro Giavalisco\orcidlink{0000-0002-7831-8751}$^{19}$}
    \author{Raffaella Schneider\orcidlink{0000-0001-9317-2888}$^{20,21,22}$}
    \author{Josephine Baggen\orcidlink{0009-0005-2295-7246}$^{23}$}
    \author{Hakim Atek\orcidlink{0000-0002-7570-0824}$^{24}$}
    \author{Volker Bromm\orcidlink{0000-0003-0212-2979}$^{1,25}$}
    \author{Karina Caputi\orcidlink{0000-0001-8183-1460}$^{26}$}
    \author{Laure Ciesla\orcidlink{0000-0003-0541-2891}$^{27}$}
    \author{Pratika Dayal\orcidlink{0000-0001-8460-1564}$^{28,5,29}$}
    \author{Chiaki Kobayashi\orcidlink{0000-0002-4343-0487}$^{30}$}
\author{Marco Castellano\orcidlink{0000-0001-9875-8263}$^{22}$}
\author{Paola Santini\orcidlink{0000-0002-9334-8705}$^{22}$}

\affiliation{
$^1$ Dpt. of Astronomy, University of Texas at Austin, 2515 Speedway, Stop C1400, Austin, TX 78712, USA \\
$^2$ Cosmic Frontier Center, The University of Texas at Austin, Austin, TX 78712 \\
$^3$ Dpt. of Astronomy, University of Geneva, Chemin Pegasi 51, 1290 Versoix, Switzerland \\
$^4$ Observational Astrophysics, Dpt. of Physics and Astronomy, Uppsala University, Box 516, SE-751 20 Uppsala, Sweden \\
$^5$ Dunlap Institute for Astronomy and Astrophysics, 50 St. George Street, Toronto, ON M5S 3H4, Canada \\
$^6$ Dpt. of Astronomy \& Astrophysics, University of Chicago, Chicago, IL 60637, USA \\
$^7$ Dipartimento di Fisica e Astrofisica, Universitá degli Studi di Firenze, Via G. Sansone 1, 50019 Sesto Fiorentino, Italy \\
$^8$ INAF/Osservatorio Astrofisico di Arcetri, Largo E. Fermi 5, 50125 Firenze, Italy \\
$^9$ INAF-OAS, Osservatorio di Astrofisica e Scienza dello Spazio di Bologna, via Gobetti 93/3, I-40129, Bologna, Italy \\
$^{10}$ Sub-dpt. of Astrophysics, University of Oxford, DWB, Keble Road, Oxford OX1 3RH, UK \\
$^{11}$ Dpt. of Physics, University of Oxford, Denys Wilkinson Building, Keble Road, Oxford OX1 3RH, UK \\
$^{12}$ Institute for Astronomy, University of Edinburgh, Royal Observatory, Edinburgh EH9 3HJ, UK \\
$^{13}$ Center for Astrophysics, Harvard \& Smithsonian, Cambridge, MA 02138, USA \\
$^{14}$ Centro de Astrobiolog\'{\i}a (CAB), CSIC-INTA, Ctra. de Ajalvir km 4, Torrej\'on de Ardoz, E-28850, Madrid, Spain \\
$^{15}$ Kavli Institute for Cosmology, University of Cambridge, Madingley Road, Cambridge CB3 0HA, UK \\
$^{16}$ Cavendish Laboratory, University of Cambridge, 19 JJ Thomson
Avenue, Cambridge CB3 0HE, UK \\
$^{17}$ Dpt. of Physics and Astronomy, University College London,
Gower Street, London WC1E 6BT, UK \\
$^{18}$ Scuola Normale Superiore, Piazza dei Cavalieri 7, 50126 Pisa, Italy \\
$^{19}$ University of Massachusetts Amherst, 710 North Pleasant Street, Amherst, MA 01003-9305, USA \\
$^{20}$ Dipartimento di Fisica, Sapienza, Università di Roma, Piazzale Aldo Moro 5, 00185, Roma, Italy \\
$^{21}$ INFN, Sezione di Roma I, Piazzale Aldo Moro 2, 00185, Roma, Italy \\
$^{22}$ INAF-OAR, Osservatorio Astronomico di Roma, Via di Frascati 33, 00078, Monte Porzio Catone, Italy \\
$^{23}$ Dpt. of Astronomy, Yale University, New Haven, CT 06511, USA \\
$^{24}$ Institut d'Astrophysique de Paris, CNRS, Sorbonne Université, 98bis Boulevard Arago, 75014, Paris, France \\
$^{25}$ Weinberg Institute for Theoretical Physics, University of Texas at Austin, Austin, TX 78712, USA \\
$^{26}$ Kapteyn Astronomical Institute, University of Groningen, Landleven 12, 9747 AD Groningen, the Netherlands \\
$^{27}$ Aix Marseille University, CNRS, CNES, LAM, Marseille, France \\
$^{28}$ Canadian Institute for Theoretical Astrophysics, 60 St George St, University of Toronto, Toronto, ON M5S 3H8, Canada \\
$^{29}$ Dpt. of Physics, 60 St George St, University of Toronto, Toronto, ON M5S 3H8, Canada \\
$^{30}$ Centre for Astrophysics Research, Dpt. of Physics, Astronomy and Mathematics, University of Hertfordshire, Hatfield AL10 9AB, UK \\
}



\begin{abstract}
Finding the first (Population III or Pop~III) stars is one of the fundamental quests of astronomy, aiming to deliver the missing link in how stars form at early cosmic times. Yet their initial mass function, formation sites and feedback remain highly uncertain, as well as the timing and topology of the transition to metal-enriched star formation. The observability of their peculiar spectral features is also debated, due to their short lifetime and faintness.
This review summarizes current theoretical expectations for Pop~III star formation, and the main observational strategies that have been adopted to constrain their properties across cosmic time, including near-field cosmology studies, direct searches for extremely metal-poor star-forming complexes and/or hard-ionizing spectral signatures at high and intermediate redshifts, and prospects for identifying Pop~III activity up to Cosmic Dawn.
The combination of JWST spectroscopy, time-domain searches, lensing surveys, stellar archaeology, absorption-line studies, as well as improved simulations, is yielding a growing number of observational candidates and narrowing the allowed parameter space for the first stars, setting the stage for a ``golden era'' of Pop~III searches.
\end{abstract}


\section{Introduction}

The first stars mark the transition from a chemically pristine Universe to the complex, metal-enriched cosmos observed today. Forming out of primordial gas, Pop~III stars set the conditions for the emergence of the first galaxies, the onset of cosmic reionization, and the transition to later generations of metal-enriched stars through their radiative, mechanical, and chemical feedback. Despite decades of theoretical and observational efforts \citep[e.g.][]{BrommLarson04, Bromm13, KlessenGlover23}, their basic properties remain highly uncertain.

However, the search for Pop~III stars is now entering a particularly timely phase. JWST opened a direct observational window into the first billion years of cosmic history, revealing galaxies up to $z \approx 14.4$ \citep{Naidu+26}, and providing unprecedented constraints on the stellar populations, ionizing spectra, and chemical enrichment of early galaxies. At the same time, these observations shed further light on the challenges involved in the search for genuinely pristine systems: metal enrichment appears to be widespread even at the highest probed redshifts \citep[e.g.][]{Bunker+23, Castellano+24, DEugenio+24, Carniani+25, Naidu+26}, and several non-Pop~III sources can mimic the hard ionizing spectra expected from metal-free stars. A key problem is therefore no longer only whether Pop~III stars can be detected, but how their signatures can be robustly separated from those of other high-$z$ contaminants and coeval metal-enriched star formation.

The present paper summarizes the research status and the directions for future progress as emerged in the relevant session ``The search for Pop~III stars'' at the \textit{CSI: Sesto workshop}, held in Sesto, January $26 - 30$, 2026, with the goal of providing useful guidance throughout this transformative era. We first outline the theoretical framework for Pop~III star formation from minihalos and the expected initial mass function (Section~\ref{sec:PopIII_IMF_theory}). We then review indirect constraints from cosmic archaeology, including (carbon-enhanced) metal-poor stars, possible pair-instability supernova signatures, and a discussion of the proposed connection with nitrogen-loud systems (Section~\ref{sec:cosmic_archaeology}). We subsequently present recent searches at high and intermediate redshifts, focusing on extremely metal-poor star-forming complexes, hard-ionizing spectral diagnostics -- including potential confusing sources --, hybrid Pop~II/III systems, and the role of gravitational lensing (Section~\ref{sec:high-intermediate_zs}). Finally, we discuss some prospects for identifying Pop~III activity at ultra-high redshifts, including the interpretation of UV-bright galaxy candidates, bright Pop~III supernovae, and complementary constraints from future facilities and 21-cm observations, as well as from observations of the cosmic infrared background (Section~\ref{sec:ultra-high_zs}).

\vspace{.75em}
\section{Classical theoretical framework of Pop~III star formation and the Pop~III initial mass function}
\label{sec:PopIII_IMF_theory}

\begin{wrapfigure}{r}{.475\textwidth}
    \centering
    \includegraphics[width=\linewidth]{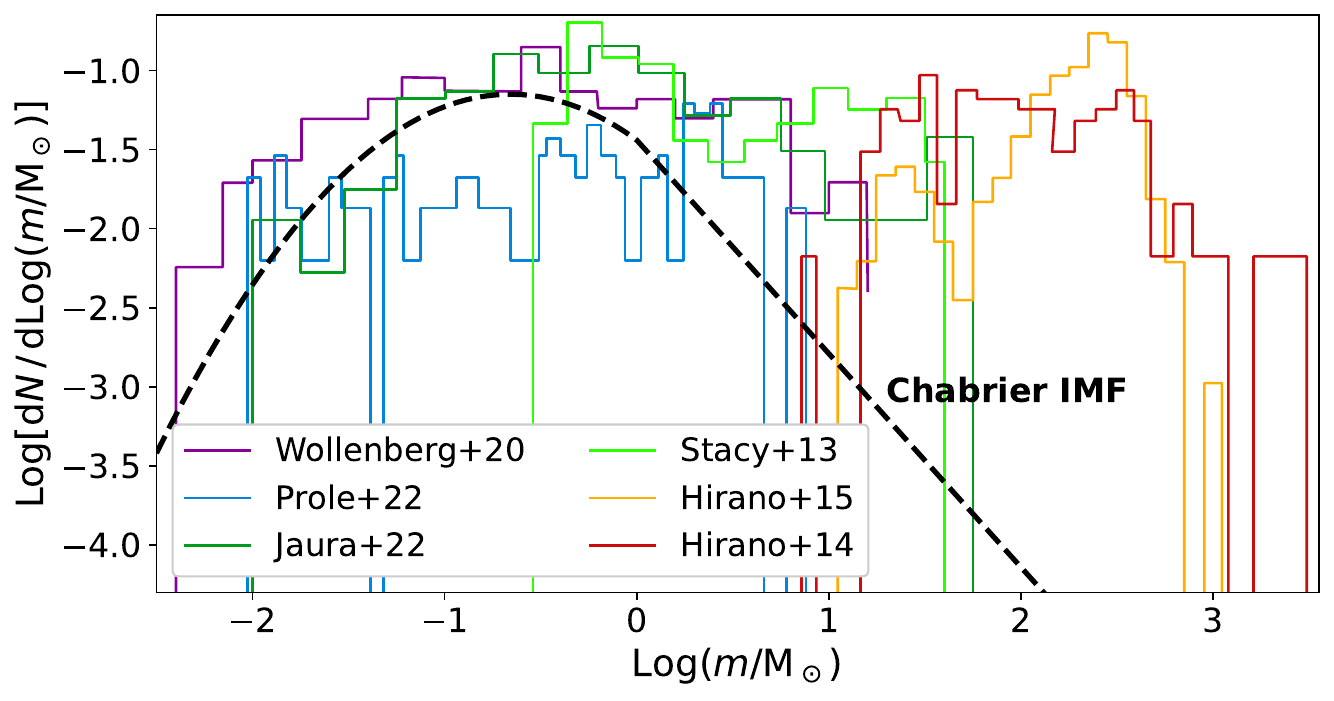}
    \caption{Compilation of predictions on the Pop~III IMF from simulations 
    (\textcolor{violet}{[1]} \citealt{Wollenberg+20},
    \textcolor{cyan}{[2]} \citealt{Prole+22}, 
    \textcolor{teal}{[3]} \citealt{Jaura+22}, 
    \textcolor{green}{[4]} \citealt{StacyBromm13}, 
    \textcolor{orange}{[5]} \citealt{Hirano+15}, 
    \textcolor{red}{[6]} \citealt{Hirano+14},  
    \textit{colored, solid lines}), from \citet{KlessenGlover23}. Models that include stellar feedback from the central proto-star (e.g. \textcolor{teal}{[3]}, \textcolor{orange}{[5]}, \textcolor{red}{[6]}) predict larger stellar masses on average. However results also sensitively depend on the spatial resolution/time span achieved, and on details of the numerical implementation: 
    lower-resolution (e.g. \textcolor{green}{[4]}, \textcolor{orange}{[5]}, \textcolor{red}{[6]}) and 2D simulations (e.g. \textcolor{orange}{[5]}, \textcolor{red}{[6]}) exhibit less fragmentation -- possibly due to a less detailed modelling of the inner accretion disk --, 
    while high-resolution, 3D simulations (e.g. \textcolor{violet}{[1]}, \textcolor{cyan}{[2]}, \textcolor{teal}{[3]}) only cover a small fraction of the full accretion history, leaving the question open on the final fate of the remaining gas reservoir and of existing proto-stars. 
    See \citet{KlessenGlover23} for more details on the simulations and their comparison.
    Despite the large variation in mass range, however, all these predicted IMFs are approximately logarithmically flat, i.e. much more ``top-heavy'' compared to a standard ``bottom-heavy'' Chabrier IMF, typical of Pop~II/I stars (\citealt{Chabrier03}, \textit{black, dashed line}).}
    \label{fig:PopIII_IMF}
\end{wrapfigure}

According to the standard theoretical framework, the bulk of Pop~III star formation occurs in minihalos with a virial mass of $\sim 10^5 - 10^7 ~\Msun$, starting at Cosmic Dawn ($z \sim 20 - 30$) (see e.g. the reviews of \citealt{BarkanaLoeb01}, \citealt{Bromm13}, and \citealt{KlessenGlover23}). As primordial gas lacks metals and dust that would otherwise provide more efficient cooling channels, the collapse of gas clouds in these environments is predominantly fueled by inefficient H$_2$ cooling \citep{Haiman+96, Tegmark+97, Yoshida+03}, resulting in significantly lower star-formation efficiencies (SFEs) and top-heavier initial mass functions (IMFs) with respect to present-day star-forming clouds. Once formed, these massive Pop~III stars synthesize metals and release them into the surrounding medium after a short lifetime, eventually allowing Pop~II stars to form from the enriched gas.

The exact shape and mass range of the Pop~III IMF are however extremely uncertain. The scatter among available predictions reflects the sensitivity of the problem to both physical and numerical assumptions, including the thermochemical evolution of metal-free gas, the treatment of turbulence, rotation, radiative and magnetic feedback, as well as numerical prescriptions and resolution adopted to follow fragmentation, and the duration over which accretion and dynamical interactions are evolved. A compilation of Pop~III IMF predictions from \citet{KlessenGlover23} is shown for illustrative purposes in Figure~\ref{fig:PopIII_IMF}, including a comparison with a classical Chabrier IMF \citep{Chabrier03}.

Early studies from \citet{Abel02}, \citet{Abel+02}, \citet{Bromm+02}, and \citet{Yoshida+08}, simulating primordial star-formation from zero-metallicity clouds, found that the star-formation process in these environments may result in the formation of a single star with mass $\gg 100 ~\Msun$.
On the contrary, later simulations employing sink particles\footnote{Sink particles are a numerical technique used to represent unresolved, high-density gas regions undergoing gravitational collapse, which can accrete surrounding gas and interact gravitationally with the rest of the system \citep[e.g.][]{Bate+95, Federrath+10}.} for the study of unresolved collapsing protostellar fragments in minihalos \citep[e.g.][]{Clark+08, Clark+11a, Clark+11b, Greif+11}, as well as radiation-hydrodynamics (RHD) simulations including radiative feedback from the central proto-star \citep[e.g.][]{Hosokawa+11}, showed that fragmentation may actually be important, and lower-mass stars can form, with \citet{Susa+19} providing forecasts on the number of fragments that should form as a function of time. 
Radiation-magneto-hydrodynamics (RMHD) simulations from \citet{Sharda+20}, \citet{Sharda+21}, \citet{Sharda+25}, and \citet{ShardaMenon25} further demonstrated the importance of magnetic fields in limiting the maximum mass that can be achieved, although independent studies found magnetic fields to have instead a negligible impact in shaping the Pop~III IMF \citep{Prole+22}, or even suppress fragmentation \citep{Sadanari+21, Sadanari+24}; these apparently different outcomes may reflect the sensitivity of the problem to the assumed initial strength and structure of the magnetic field, and to how efficiently it is amplified during collapse.

Nonetheless, recent works consistently predict a top-heavy IMF emerging from pristine (or very-metal-poor) clouds, with typical masses of $\sim 10s - 100s ~\Msun$ \citep[e.g.][]{StacyBromm13, Susa+14, Safranek-Shrader+14, Stacy+16, HiranoBromm17, Sugimura+20, Wollenberg+20, Chon+21, Chon+22, Chon+24, Latif+22, Jaura+22, TangChen24, ItoOmukai+24}, and possibly even extended up to $\sim 10^3 ~\Msun$ \citep[e.g.][]{Hirano+14, Hirano+15, Hosokawa+16, Chon+18, Regan+20}. In Section~\ref{sec:low-metallicity}, we will lay out the theoretical framework for the transition from the top-heavy IMF predicted for the first stars in minihalos to a present-day IMF, and the impact of environmental factors such as the intensity of the radiation field on this transition. The implications of similar environmental considerations on the SFE in metal-free star-forming clouds will be further discussed in Section~\ref{sec:massive_halos}.

\vspace{.75em}
\section{Constraints from cosmic archaeology}
\label{sec:cosmic_archaeology}

Indirect constraints on both the Pop~III IMF and the properties of Pop~III supernovae (SNe) \citep[e.g.][]{UmedaNomoto02, HegerWoosley02, MeynetMaeder02, Meynet+06, Nomoto+06, Ekstrom+08, ChatzopoulosWheeler12, Yoon+12, Murphy+21, Jeena+23, Martinet+23, Roberti+24} can be achieved through near-field cosmology studies, i.e. by analyzing the fossil records of these early formation sites in the local Universe; see e.g. the reviews of \citet{BeersChristlieb05}, \citet{Ricotti+10}, \citet{FrebelNorris15}, and \citet{Bonifacio25}. A more detailed and pedagogical discussion of data-driven constraints on the Pop~III IMF is presented in \citet{Salvadori+26}, while here we summarize available archaeological constraints at the low-/intermediate- (Section~\ref{sec:CEMPs}) and high-mass end (Section~\ref{sec:PISNe}), as well as chemical signatures recently proposed to be associated with a very-massive or even super-massive component (Section~\ref{sec:N-emitters})\footnote{We note that, while this topic has not been discussed in the present review, an additional channel to indirectly constrain the properties of Pop~III stars comes from the study of their remnants at both low and high redshifts. We refer the reader to the review of \citet{Hammerle+20} for a thorough discussion of black hole seeds from Pop~III stars.}.

\vspace{.6em}
\subsection{Low/intermediate-mass end: search for Pop~III survivors and carbon-enhanced extremely-metal-poor stars}
\label{sec:CEMPs}

When looking for extremely metal-poor (EMP, $\lesssim 0.1\% ~\Zsun$) stars in the Milky Way (MW) halo \citep[e.g.][]{Christlieb+02, Christlieb03, Caffau+11, Caffau+13, Yong+13a, Bonifacio+21, Starkenburg+18, Aguado+18b, Aguado+19, Francois+18}
and in nearby ultra-faint dwarfs (UFDs, e.g. \citealt{Kirby+08, KirbyCohen12, Norris+10a, Norris+10b, Norris+10c, Simon+11, Yoon+19}), 
we find ubiquitous evidence of carbon enhancement in their atmosphere (CEMP stars, e.g. \citealt{Frebel+05, Frebel+15, Norris+07, KirbyCohen12, Yong+13b, Lee+13, Placco+13, Keller+14, Spite+14, Hansen+14, Bonifacio+15, Aguado+18a, Nordlander+19}). Some of this carbon enhancement can be explained by binary evolution, traced by $s$-process elements (CEMP-$s$ stars, \citealt{Lucatello+05})\footnote{The slow neutron-capture process ($s$-process)
mainly occurs in low- and intermediate-mass asymptotic giant branch (AGB) stars, on longer evolutionary timescales than massive-star core-collapse SNe \citep{Busso+99}. In CEMP-$s$ stars, the carbon and $s$-process enhancement is therefore commonly interpreted as the result of mass transfer from an AGB companion in a binary system.}. However, many of these stars do not show evidence of $s$-process elements (CEMP-no stars, e.g. \citealt{Norris+10b, Starkenburg+14, Placco+14a, Placco+14b, Hansen+15, Yoon+19, Aguado+23a}), and their abundance pattern is instead consistent with enrichment from various types of core-collapse SNe (CCSNe) from the first stars.

\begin{figure}[htbp]
    \centering

    \begin{minipage}{\textwidth}
        \centering
        \begin{minipage}[c]{0.73\textwidth}
            \centering
            \includegraphics[width=\linewidth]{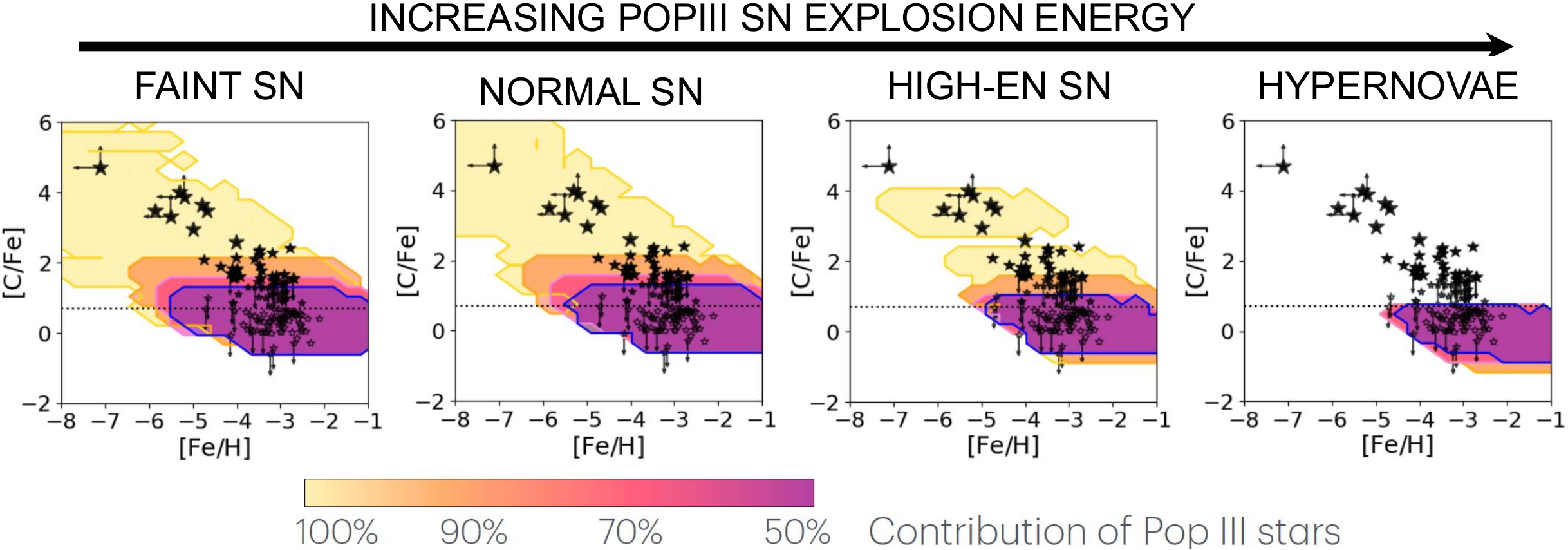}
        \end{minipage}
        \hfill
        \begin{minipage}[c]{0.25\textwidth}
            \caption{Carbon enhancement ([C/Fe]) vs metallicity ([Fe/H]) resulting from Pop~III SN enrichment episodes with an increasing explosion energy (from \textbf{left} to \textbf{right}), when considering an incremental level of contamination from subsequent generations of Pop~II stars (from \textit{yellow} to \textit{purple}), adapted from \citet{Vanni+23a}.}
            \label{fig:CEMPs_signature}
        \end{minipage}
    \end{minipage}

\end{figure}

\citet{Ishigaki+18} provided constraints on the Pop~III IMF through abundance fitting of $\sim 200$ CEMP stars in the solar neighborhood to theoretical SN yields. They found the primary enrichment sources for these systems to be CCSNe (consistent with progenitor masses $< 40 ~\Msun$), with more than half the stars best fitted by a $25 ~\Msun$ hypernova (HN) model with an explosion energy of $10^{52} ~\si{erg}$, and some contribution from faint SNe/HNe.
The parametric study of \citet{Vanni+23b} further investigated the chemical imprints of Pop~III SNe and the contamination from subsequent generations of Pop~II stars (Figure~\ref{fig:CEMPs_signature}), finding that the region at high [C/Fe] and low [Fe/H] is progressively depopulated, and that stars with $\mathrm{[C/Fe]} > 2.5$ are consistent with being enriched solely by Pop~III SNe of faint-to-normal energy. However, the distinction appears less clear when higher-energy Pop~III SNe dominate the enrichment, as their nucleosynthetic signature shows a significant overlap with highly contaminated signatures (contributed more than 50\% by Pop~II SNe) on the [C/Fe] vs [Fe/H] plane.
This result was confirmed by studies with the semi-analytical model \texttt{NEFERTITI} \citep{Koutsouridou+23}, including Pop~III yields from \citet{HegerWoosley+10} and Pop~II yields from \citet{LimongiChieffi18}, as well as stochastic sampling of the stellar IMF (also see e.g. \citealt{Salvadori+15, Graziani+15, Graziani+17, deBennassuti+17, Hartwig+18a, Hartwig+19, Hartwig+23, Welsh+19, Welsh+21, Sarmento+19, Rossi+23, Koutsouridou+25}). By matching the observed fraction of CEMP stars as a function of metallicity to model predictions, \citet{Koutsouridou+23} infer a dominant contribution from low-energy, faint Pop~III SNe, deriving a characteristic mass of the Pop~III IMF\footnote{\citet{Tumlinson07b}, \citet{Salvadori+07}, and \citet{deBennassuti+14} further investigated the role of a metallicity-driven vs dust-driven transition from a top-heavy Pop~III IMF to a present-day, bottom-heavy IMF, including a potential impact of Cosmic Microwave Background (CMB) radiation \citep{Tumlinson07a}. See Section~\ref{sec:low-metallicity} for a broader discussion on the Pop~III-II IMF transition and on the impact of dust cooling and CMB radiation.} $m_\mathrm{ch} \sim 1 - 20 ~\Msun$.

These studies imply that Pop~III masses should extend down to $\sim 10 ~\Msun$ or lower. On the other hand, the non-detection of metal-free stars in the MW halo and in local UFDs is consistent with a picture of predominantly massive Pop~III stars that do not survive until present day, allowing a lower limit of $\sim 0.8 ~\Msun$ to be placed on the mass of the first stars \citep{Hartwig+15, Magg+18, Magg+19, Rossi+21}.

\vspace{.6em}
\subsection{High-mass end: pair-instability supernova signatures}
\label{sec:PISNe}

If Pop~III stars reach the $140 - 260 ~\Msun$ regime, they should produce pair-instability SNe (PISNe), releasing up to about 50\% of their initial mass in metals, with energies of $\sim 10^{52} - 10^{53}$~erg \citep{HegerWoosley+10}.

Candidate transients consistent with the typical light curve expected from Ni decay after a PISN event have been proposed by \citet{GalYam+09}, \citet{Quimby+11}, \citet{Cooke+12}, and \citet{Schulze+24}; however, the observed light curve could also be produced by some types of \ion{H}{2} region-circumstellar medium interactions or magnetars (see e.g. \citealt{Moriya+10, Dessart+12} for possible alternative interpretations of the PISN candidate SN 2007bi). A $z \approx 15$ PISN interpretation has been initially discussed for the ``Capotauro'' source \citep{Gandolfi+26a, Ferrara+26}, although the recent detection of non-zero proper motion for this source has ruled out the possibility that it is of extragalactic origin (\citealt{Liu+26}, see the discussion in Section~\ref{sec:ultra-high_zs}). The detectability of these events at Cosmic Dawn and during the Epoch of Reionization (EoR, $6 \lesssim z \lesssim 10$) with the James Webb Space Telescope (JWST) and with the Nancy Grace Roman Space Telescope has been discussed in \citet{WeinmannLilly+05}, \citet{Whalen+13b}, \citet{Wang+17}, \citet{Hartwig+18b}, \citet{Regos+20}, \citet{Moriya+22a}, \citet{LazarBromm22}, \citet{Venditti+24a}, and \citet{Jeon+26a}, while the possibility of detecting PISNe in the Euclid Deep Survey (EDS, \citealt{Laureijs+11, EuclidCollaboration+22}) at $z \lesssim 3.5$ is studied in \citet{Moriya+22b}; we refer the reader to Section~\ref{sec:ultra-high_zs} for a more in-depth discussion of Pop III transients at high redshifts.

\begin{figure}
    \centering
    \begin{minipage}{0.49\textwidth}
        \centering
        \includegraphics[width=\linewidth]{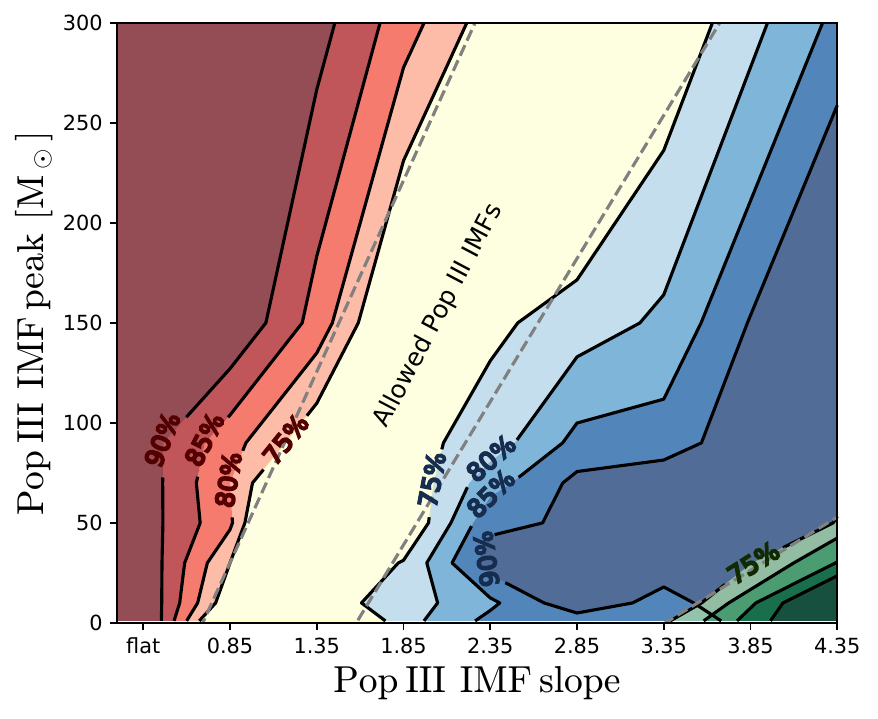}
        \caption{Confidence levels at which a Pop~III IMF with given peak mass and slope can be excluded based on (i) the non-detection of mono-enriched PISNe descendants in the SAGA catalog \citep{Suda+08, Suda+17} at $\mathrm{[Fe/H]} < - 2.5$ (\textit{red contours}), and (ii) the assumption of a PISN origin interpretation for the J1010+2358 star, discovered among 15000 very-metal-poor LAMOST stars (\textit{blue/green contours}), derived from a statistical comparison with predictions of the cosmological galaxy formation model \texttt{NEFERTITI} (adapted from \citealt{Koutsouridou+24}). The \textit{blue contours} demonstrate how detecting even a single PISN descendant in the Galactic halo will enable strong constraints on the Pop~III IMF in the regime corresponding to lower characteristic masses and/or steeper slopes.}
        \label{fig:PISN_non-detection_constraints}
    \end{minipage}
    \hfill
    \begin{minipage}{0.49\textwidth}
        \centering
        \includegraphics[width=\linewidth]{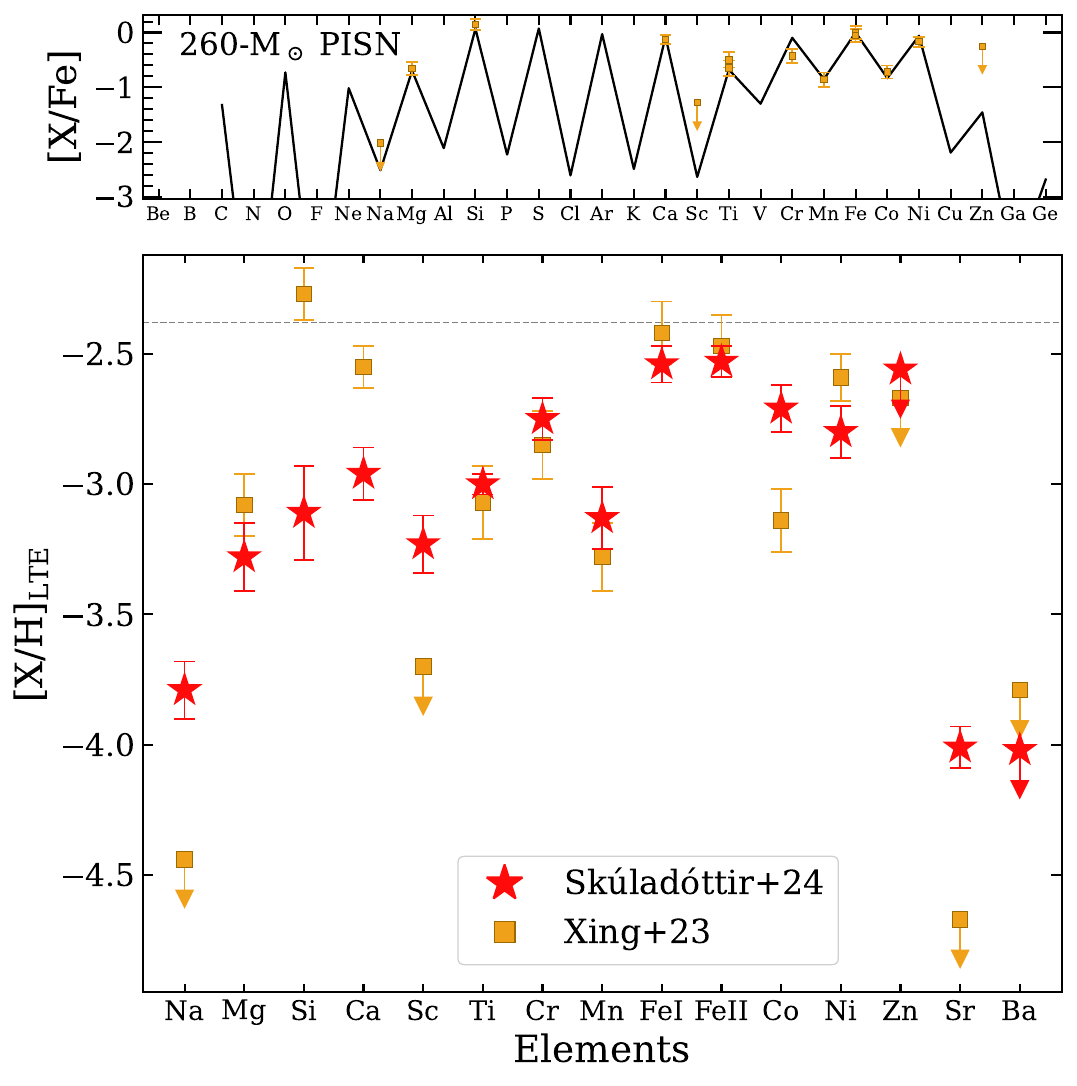}
        \caption{Chemical abundance pattern at the surface of the J1010+2358 star from the LAMOST survey \citep{Zhao+12}, as originally measured by \citet{Xing+23} (\textit{yellow squares} in the \textbf{top/bottom panels}), and from later, high-resolution observations from \citet{Skuladottir+24} (\textit{red stars}) (adapted from \citealt{Xing+23} and \citealt{Skuladottir+24}). The agreement with the top-panel $260 ~\Msun$ PISN model was not confirmed by the new measures.}
        \label{fig:PISN_signature_candidate}
    \end{minipage}
    \vspace{1.5em}
\end{figure}

Over twenty-five years of searches for the chemical signature of PISN enrichment in MW stars (e.g., looking for signs of strong elemental odd-even effect, and the absence of neutron-capture elements) also only yielded two candidate stars with chemical abundances consistent with PISN yields (J001820-093939, \citealt{Aoki+14}, and J1010+2358, \citealt{Xing+23, Jeena+23, JeenaBanerjee+24, Koutsouridou+24, Skuladottir+24, Thibodeaux+24}; other potential candidates have been discussed in \citealt{Salvadori+19, Aguado+23b}). The lack of clear PISN signatures has often been interpreted as evidence that the first stars were typically massive (enough to enable core-collapse events, Section~\ref{sec:CEMPs}), but probably not very massive ($< 140 ~\Msun$ on average), in order to explain an absence of PISNe signatures at low redshift. Particularly, the non-detection of PISNe descendants at $\mathrm{[Fe/H]} < -2.5$ in the SAGA catalog \citep{Suda+08, Suda+17} disfavors very top-heavy and flat IMFs (\citealt{Koutsouridou+24}, Figure~\ref{fig:PISN_non-detection_constraints}).

It has been suggested that the apparent lack of PISN enrichment may be due to an observational selection effect \citep[e.g.][]{Karlsson+08}: in fact, even a single PISN event might be able to enrich the neighboring gas to metallicities $\gg 10^{-3} ~\Zsun$ \citep[see also][]{Salvadori+19, Koutsouridou+23}, especially when metals are inhomogeneously mixed in the gas, potentially causing any second-generation stars forming out of this material to ``overshoot'' out of the EMP regime (also see e.g. \citealt{Greif+10, Wise+12, Pan+13, Ritter+15, Sluder+16, Magg+20, Tarumi+20, Gutcke+22}), and requiring searches of PISN enrichment over a much larger sample at higher metallicity. However, the prevalence of this ``overshoot'' scenario is debated, with the cosmological simulation from \citet{Magg+22} showing for example that only a few descendant stars forming from the gas enriched by Pop III PISNe are expected to form at such elevated metallicity, while the majority lie within the EMP range (also \citealt{Kordt+26}).

Based on the comparison with predictions from the \texttt{NEFERTITI} model, \citet{Koutsouridou+23} showed that the detection of even a single PISN descendant in the Galactic halo will place strong constraints on the steepest and most bottom-heavy Pop~III IMFs. For example, the blue area in Fig.~\ref{fig:PISN_non-detection_constraints} shows the constraints that would be enabled by a 
\begin{wrapfigure}[29]{r}{.5\textwidth}
    \centering
    \vspace{-1em}
    \includegraphics[width=\linewidth]{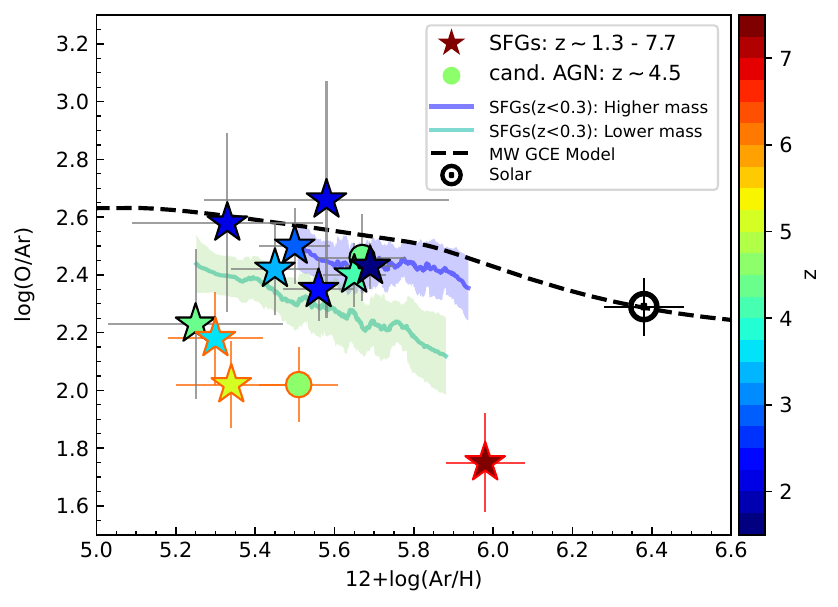}
    \caption{Log(O/Ar) vs $12 + \mathrm{Log(Ar/H)}$ for 11 star-forming galaxies at $z \approx 1.3 - 7.7$ (\textit{stars}) and two candidate AGN hosts at $z \approx 4.5$ (\textit{circles}), \textit{colored} by their redshift. The \textit{green} and \textit{blue lines} respectively show the sequence of mean values of low-redshift ($z < 0.3$) low-mass ($\langle \mathrm{Log}(M_\star / \Msun) \rangle = 7.23$) and higher-mass ($\langle \mathrm{Log}(M_\star / \Msun) \rangle = 9.41$) galaxies from the SDSS survey \citep{Bhattacharya+25b}, adopted from \citet{Bhattacharya+25a}. The MW solar neighborhood Galactic chemical evolution (GCE) model from \citet{Kobayashi+20} is also shown as a reference (\textit{black line}). Four of these objects at $z \sim 3.5 - 7.7$ (marked by \textit{orange/red edgecolors}) have Log(O/Ar) values below the MW GCE model, and even below the sequence traced by lower-mass, low-$z$ starbursts, possibly indicative of alternative enrichment scenarios, including a rapid PISN enrichment channel (particularly for the ERO 10612 at $z = 7.66$, \textit{red edgecolor}, with a low -- albeit uncertain -- estimated star-formation timescale of $4.37^{+4.13}_{-2.13}$~Myr).}
    \label{fig:PISNe_OtoAr}
\end{wrapfigure}
PISN-origin interpretation for the J1010+2358 LAMOST star (\citealt{Xing+23}, with an inferred $\mathrm{[Fe/H]} \approx -2.4$), whose chemical abundances were previously found to be in excellent agreement with the predicted yields of a massive $260 ~\Msun$ progenitor. Unfortunately, the measured abundances have not been confirmed by two independent follow-up studies \citep{Skuladottir+24, Thibodeaux+24}: both studies report mutually consistent measurements for the different chemical abundances, which differ significantly from those reported by \citet{Xing+23}, and therefore from the predicted chemical signature of a massive PISN (Figure~\ref{fig:PISN_signature_candidate}).

On the other hand, low-mass halos (with virial mass $\lesssim 10^7 ~\Msun$, e.g. \citealt{Mead+25}) may not be able to retain their metals after a SN event due to their low gravitational potential (also \citealt{CookeMadau14, Ji+15, Rey+25}, but see \citealt{Rey+25, Storck+26} for alternative predictions of high retention at all halo masses\footnote{The efficiency of metal retention and fallback depends sensitively on halo mass, explosion energy, and radiative pre-processing by the progenitor star, so that the exact mass threshold above which metals are typically retained is uncertain. See e.g. \citet{Whalen+08}, \citet{Whalen+13a}, \citet{Ritter+12}, and \citet{Johnson+13_SN} for other studies on Pop~III SN feedback in both minihalos and atomic-cooling halos.} down to $\sim  2 \times 10^6 ~\Msun$), so that minihalos hosting a PISN that were later incorporated in the MW halo may not preserve their chemical imprints.
For this reason, simulations and simpler semi-analytical models of early chemical enrichment \citep[e.g.][]{Jeon+15, Rossi2025} have suggested that PISN-produced material instead predominantly resides in the circumgalactic/intergalactic medium (CGM/IGM), which can be probed in absorbers along the line-of-sight of distant quasars (damped-Ly$\alpha$ systems or DLAs, e.g. \citealt{Simcoe+06, Kobayashi+11, Cooke+17, Welsh+19, Welsh+21, Welsh+22, Welsh+23, Welsh+24, Robert+22, Saccardi+23a, Saccardi+23b, Davies+23, Christensen+23, Zou+24, Sebastian+24, Sodini+24, Durocikova+25, Visbal+26, Higginson+26}), or even against the ultra-luminous afterglows of high-redshift gamma ray bursts (GRBs) \citep{Wang+12}.
However, a significant problem in the interpretation of the measured abundances in these systems comes from dust depletion (see e.g. \citealt{Chiaki+25} on C-grain production from Pop~III faint SNe, and \citealt{Schneider+04} for dust yields from PISNe; also \citealt{Otaki+26} for SN dust yields including rotation). A statistically significant study of the CGM through rare quasar sightlines is also challenging due to its highly multi-phase nature \citep[e.g.][]{Fumagalli+24}\footnote{However, see \citet{MirzaKhanlari+25} for a potential approach to overcome this problem by employing the diffuse ultra-violet (UV) extragalactic background as a background source across a much larger area in the HETDEX survey \citep{Gebhardt+21}.}.

Potential PISN signatures have been proposed to explain the large amount of Fe with low Mg measured in the broad-line region of a $z = 7.54$ quasar \citep{Yoshii+22}, and the abundance pattern (including odd-even abundance ratios [Mg,Si/Al], e.g. \citealt{Vanni+24}) in a $z \approx 3.4$ absorber from the \citet{Saccardi+23a} sample; a possible Fe-rich abundance pattern has also been reported for the GN-z11 galaxy at $z \approx 10.6$ \citep{Nakane+24}, although the interpretation of this pattern is degenerate with other solutions\footnote{Also see Section~\ref{sec:low-metallicity} and~\ref{sec:massive_halos} for a discussion of a Pop~III candidate identified through its \ion{He}{2} emission in the environment of GN-z11 \citep{Maiolino+24b, Maiolino+26, Ubler+26}.}.
A recent study of the [O/Ar] vs [Ar/H] relation (which can constrain galaxy chemical enrichment mechanisms at higher redshifts, $1 \lesssim z \lesssim 8$) has shown a parallel sequence of high-mass galaxies at low [O/Ar] values that may reveal different enrichment pathways at $z > 3.6$, possibly consistent with the Ar enhancement expected from PISNe enrichment \citep[see Figure~\ref{fig:PISNe_OtoAr}]{Bhattacharya+25a}.
Early PISN enrichment with elevated metal retention has also been invoked to explain the iron-metallicity plateau at the UFD regime of the local stellar mass-metallicity relation \citep{Rey+25}.
Finally, the dearth of CEMP stars in the Galactic bulge, which is predicted to host the oldest stars in the Galaxy \citep[e.g.][]{Tumlinson10, Salvadori+10}, has been proposed to indirectly probe the existence of PISNe, as the mean [C/Fe] of first-stars polluted environments is expected to decrease for an increasing contribution of PISNe \citep{Pagnini+23}.

\vspace{.6em}
\subsection{Very-massive/super-massive Pop~III stars? N-loud systems or N-emitters}
\label{sec:N-emitters}

It has been argued that Pop~III enrichment may be linked to the recently discovered nitrogen overabundance in some rare high-redshift sources.
Indeed, some galaxies with nitrogen emission lines of \ion{N}{4}] and \ion{N}{3}] in the UV have been revealed by JWST, first in GN-z11 at $z = 10.6$ \citep{Bunker+23, Cameron+23_Nemitter, Maiolino+24a} and soon thereafter in other high-$z$ systems \citep[e.g., by][]{Isobe2023, Marques-Chaves2024Extreme-N-emitt, Castellano+24, Arellano-Cordova2025CLASSY-XII:-nit,Schaerer2024Discovery-of-a-, Topping+24, Topping+25, Curti2025JADES:-The-star}. These objects are called N-emitters or N-loud systems in the literature \citep[see ][for a recent compilation]{Ji+26}. Including a few lensed $z \sim 3$ galaxies and very rare $z \sim 0$ objects, currently approximately $\sim 60$ such objects are known, after the recent systematic search of \citet{Morel2025Discovery-of-ne} using JWST archival data. Selecting for UV N-emission also yields a small fraction ($\sim 10$\%) of active galactic nuclei (AGN) and ``little red dots'' (LRDs)\footnote{LRDs are a class of compact, very red JWST-selected sources (often at $z\gtrsim 4$, \citealt{Kocevski+25}), with a characteristic red rest-frame optical continuum, blue/UV excess, and in many cases broad Balmer lines possibly suggestive of obscured AGN activity, although their nature remains debated \citep[e.g.][]{Kocevski+23, Matthee+24, Labbe+25}.} \citep{Morel2025Discovery-of-ne}, and conversely, a subsample of AGN and LRDs show \ion{N}{4}] or \ion{N}{3}] emission lines \citep[e.g.,][]{Ubler2023GA-NIFS:-A-mass}. This shows a small overlap between these categories, indicating likely similar conditions and possibly common physical mechanisms at play in these objects. 

The presence of nitrogen UV emission lines translates overall into a high N abundance, more specifically high (super-solar) N/O ratios \citep[see e.g.][]{Ji+26, Morel2025Discovery-of-ne}, as established by many independent studies considering also effects of a multi-component ISM, such as high densities, density gradients, uncertainties in the electron temperature, and others \citep[see, e.g,][]{Martinez+25, Morel2025Discovery-of-ne, Arellano-Cordova+26, Berg+26}.
The majority of N-emitters are found at relatively low metallicity ($\la (0.1-0.2)$ solar in O/H), although objects up nearly solar have also recently been found \citep{Morel2025Discovery-of-ne}.
Other abundances which can be derived in these objects include carbon, neon, and sometimes also helium, argon, sulfur, and iron 
\citep[see][]{Yanagisawa2024Strong-He-I-Emi, Watanabe2026Chemical-Abunda, Gimenez-Alcazar2026Shape-of-Direct}.
Overall, N-emitters stand out by their high (super-solar) N/O ratio, normal C/O and high N/C ratios, and normal Ne/O, compared to the bulk of star-forming galaxies at similar metallicities, both at $z \sim 0$ and up to $z \sim 3$ at least \citep{Morel2025Discovery-of-ne, Schaerer2026Nitrogen-abunda, Cataldi2025Tracing-Nitroge}.
Possible helium enhancements in N-emitters have been reported \citep[][]{Yanagisawa2024Strong-He-I-Emi, Gimenez-Alcazar2026Shape-of-Direct}, although accurate determinations of the He abundance are challenging.
Although the average N/O ratio of star-forming galaxies may increase beyond $z>4$ \citep[as, e.g.,][]{Cataldi2025Tracing-Nitroge, Rusakov2026Diverse-Histori}, average spectra of $z \sim 5-10$ galaxies do not show nitrogen UV lines \citep{Roberts-Borsani2024Between-the-Ext}, and N-emitters are a rare population that represents a fraction of approximately $\sim (1-3) \times 10^{-3}$ of emission-line galaxies at $z \sim 3$, which increases beyond $\ga 10$\% at $z \ga 10$ \citep{Morel2025Discovery-of-ne,Schaerer2026N-emitters-as-p}.

Many scenarios have been proposed to explain the high N/O abundances of N-emitters, including ejecta from Wolf-Rayet (WR) stars, very massive stars (VMS, $~ 10^2 - 10^3 ~\Msun$), extremely-massive stars (EMS, $10^3 - 10^4 ~\Msun$), supermassive stars (SMS, $> 10^4 ~\Msun$), AGB stars, or also extreme Pop~III stars (as detailed in the next paragraph). Other studies have suggested rapid enrichment due to intermittent star formation \citep{KobayashiFerrara24}, tidal disruption of stars in encounters with BHs, differential winds, or a top-heavy IMF, and no consensus has yet been reached \citep[see, e.g.,][]{Bunker+23, Cameron+23_Nemitter, Bekki2023A-model-for-GN-, Charbonnel2023N-enhancement-i, Vink2023Very-massive-st, Nagele2023Multiple-Channe,Pascale2023Nitrogen-enrich, Pascale2024A-Young-Super-S, Watanabe2023EMPRESS.-XIII.-, Watanabe2026Chemical-Abunda, KobayashiFerrara24, Nandal2024, Nandal2025, Gieles+25, DAntona2025Dating-N-loud-A, Berg+26}.
Empirically, it has been noted early on that the relative C, N, O and H abundances of N-emitters resemble those of globular cluster (GC) stars \citep{Charbonnel2023N-enhancement-i,Senchyna2024GN-z11-in-Conte,Ji+26}, which show peculiar abundance patterns, also in other elements \citep[][]{Bastian2018Multiple-Stella}; the link with GC has been further explored by \citet{Marques-Chaves2024Extreme-N-emitt} and by \citet{Schaerer2026N-emitters-as-p}, who suggest that N-emitters are signposts of GC in formation\footnote{Recently, possible connections between GC formation, SMS formation from star collisions within the dense clusters, N-emitters, and LRDs have also been discussed \citep[e.g.][]{Chisholm+26}, while WR-like spectral signatures identified in some LRDs may point to a population of massive stars capable of producing similarly extreme abundance patterns \citep{Perez-Gonzalez+26}.}.
The inferred chemical abundances may provide insights into the nucleosynthetic sites and origins of the observed elements.
From the measured N/C abundance ratios and lower limits, \citet{Morel2025Discovery-of-ne} conclude that the majority of N-emitters show abundances reflecting CNO burning, but no signs of He-burning products (carbon, in particular), over a wide metallicity range. \citet{Yanagisawa2024Strong-He-I-Emi} and other studies also support this conclusion. This is a strong constraint in favor of the proto-GC interpretation, with metal-enriched EMS being expected as a natural outcome of GC formation in the early universe and over a broad metallicity range \citep{Gieles+25}. On the other hand, the He constraint disfavors  chemical evolution models calling for the contribution of less extreme stars, as it requires the suppression of ejecta from core-collapse supernova ejecta and advanced phases of WR stars (WC stars), which might be present, depending on metallicity \citep[see, e.g.,][]{Charbonnel2023N-enhancement-i,Marques-Chaves2024Extreme-N-emitt,KobayashiFerrara24,Berg+26,Watanabe2026Chemical-Abunda}. A complete picture is, however, still lacking, and different explanations are being debated.

Let us briefly examine scenarios related to Pop~III enrichment in the present context. 
\citet{Nandal2024} proposed that Pop~III stars of a few thousand solar masses could explain the high N abundances of N-emitters. Using stellar evolution models of metal-free stars with initial masses $\sim 2000 - 9000 ~\Msun$, they computed their ejecta up to the end of He-burning. The cumulative ejecta are rich in He and N (from H-burning), and O (from He-burning) and correspond in particular to high (supersolar) N/O ratios, but also very high (highly super-solar) O/H. The former are comparable with the observations of N-emitters, but to bring the O/H abundance within the range of observations for normal and low-metallicity galaxies, the Pop~III ejecta have to be diluted within large amounts of gas. According to \citet{Nandal2024}, dilutions by $\sim 100$ times more gas than the ejecta are needed, and the gas needs to be primordial to maintain the high N/O ratio of the ejecta. Otherwise, mixing with gas of ``normal'' abundance ratios would lead to a rapid decrease of N/O and C/O, well below the observed values. The situation is similar in the study of \citet{Nandal2025}, who propose to explain one of the highest N/O ratios of N-emitters (in the AGN GS 3073) with primordial stars of masses $10^{2-4} ~\Msun$.

The amount of enriched ionized gas in N-emitters has been estimated in a few objects, and was found to be of the order of $\sim (1-2) \times 10^5 ~\Msun$ for GN-z11 and CEERS 1019, adopting electron densities $n_e \sim 10^5$ cm$^{-3}$ \citep{Charbonnel2023N-enhancement-i,Marques-Chaves2024Extreme-N-emitt}. Such amounts can, e.g., be obtained from the stellar wind ejecta of normal stellar populations of $\sim (0.3-1) \times 10^8 ~\Msun$ in stars \citep{Marques-Chaves2024Extreme-N-emitt,Berg+26}. By contrast, in the Pop~III SMS scenario discussed above, the same ionized-gas mass would require a SMS ejecta $\lesssim 10^3 ~\Msun$ to be mixed with $\sim 10^5 ~\Msun$ of nearly pristine gas, highlighting that the viability of this interpretation depends primarily on achieving the required dilution while preserving the extreme N/O ratios.

None of the described scenarios can be so far ruled out. However, while Pop~III SMS models might reproduce the observed abundance ratios after substantial dilution with pristine gas \citep{Nandal2024, Nandal2025}, we emphasize that such models remain highly idealized: current applications rely on stellar-evolution yields and prescribed dilution rather than a self-consistent model for SMS formation, ejecta release, and mixing into the observed ionized gas. Current abundance patterns also do not provide compelling evidence for a Pop~III origin and may be more naturally explained by dense-cluster or other non-Pop~III channels \citep[e.g.][]{Cameron+23}. Nonetheless, further studies and additional constraints will help test all these different pathways.
For example, as the nitrogen signatures are observed in nebular emission, probing recently enriched and actively ionized gas rather than fossil abundance records, in the SMS Pop~III scenario one could expect to find leftover undiluted metal-free gas close to the site of N-enrichment. Additionally, Pop~III spectral signatures (as discussed below, e.g. Section~\ref{sec:spectral-hardness}) from remaining Pop~III stars coeval to the SMS could be expected. 
Deep spatially-resolved observations of N-emitters could in principle spot these signatures from nearby gas/stellar populations. IFU observations should also be key to establish if the observed N-enhancement is localized (e.g., associated with compact star clusters if related to proto-GCs), measure the typical chemical composition of the host galaxy, and determine the ages of the stellar populations associated with N-enhancement. These will provide fundamental tests of the proposed scenarios, and should allow to understand the nature of N-emitters and the origin of the unusual abundances seen in these objects.

\vspace{.75em}
\section{Direct search for Pop~III stars at high and intermediate redshifts ($\lowercase{z} \gtrsim 3$)}
\label{sec:high-intermediate_zs}

So far we discussed indirect constraints on the masses, explosion mechanisms, and chemical yields of the first stars through archaeological studies, probing Pop~III activity through its long-lived descendants and enrichment products.
However, the advent of JWST has shifted the search for Pop~III stars from a purely indirect endeavor to one in which direct observational tests have become possible, allowing us to directly probe the metallicity, ionizing spectra, and stellar populations of galaxies at epochs where pristine gas may survive.
In this section, we review current observational efforts aimed at identifying ongoing or recent Pop~III star formation at high and intermediate redshifts, ranging from searches for extremely metal-poor star-forming complexes (Section~\ref{sec:low-metallicity}) and hard ionizing spectra (Sections~\ref{sec:spectral-hardness} and~\ref{sec:late_PopIII_obs}), including hybrid Pop~II/III systems (possibly within large-scale overdensities, Section~\ref{sec:massive_halos}), and even indications towards the search for individual, strongly lensed Pop~III stars (Section~\ref{sec:lensed_stars}).

\vspace{.6em}
\subsection{Search for extremely metal-poor complexes and evidence of efficient enrichment at high redshifts}
\label{sec:low-metallicity}

The transition from a top-heavy-IMF regime typical of the first stars (Section~\ref{sec:PopIII_IMF_theory}) to the bottom-heavy IMF observed in the local Universe is expected to occur once the gas reaches a critical level of chemical enrichment, above which cooling and fragmentation properties change significantly. This transition is directly relevant for observational searches, as it sets the expected metallicity range of systems where Pop~III-like star formation may still occur. However, the value of the critical metallicity for the transition is extremely uncertain, with studies focusing on the required carbon and oxygen abundances for metal-line cooling to become dominant over H$_2$ cooling typically placing it at $\sim 10^{-3} - 10^{-4} ~\Zsun$ \citep{Omukai00, Bromm+01a, BrommLoeb03, Frebel+07, ShardaKrumholz22}, while a scenario with a dust-driven transition has also been suggested at lower critical metallicities ($\sim 10^{-6} - 10^{-4} ~\Zsun$), provided that a critical dust-to-gas mass ratio of $\sim 4.4 \times 10^{-9}$ is reached in star-forming clouds \citep{Omukai+05, Schneider+12, Chiaki+13}.
While 3D simulations have confirmed that efficient cooling
by metals and dust grains \citep{TsuribeOmukai06, TsuribeOmukai08, Clark+08, Dopcke+11, Dopcke+13, Safranek-Shrader+16, Chiaki+16, ChiakiWise19} promotes fragmentation at small scales, it is hard to follow the subsequent accretion phase for sufficiently long times ($\sim 10^4 - 10^5$~yr, e.g. \citealt{ChiakiYoshida22}) to constrain the actual final shape of the stellar IMF.

Recent studies further show that this transition is likely to be gradual and dependent on additional environmental factors, such as the strength of the local radiation field. The metallicity-dependence of the IMF has been explored most recently through RHD simulations by \citet{Chon+21, Chon+22, Chon+24}, finding that, while the number of low-mass stars increases with metallicity above $\sim 10^{-5} ~\Zsun$, the stellar IMF only converges to a present-day IMF above $\sim 10^{-2} ~\Zsun$.  \citet{Chon+21} further indicated that the shape of the IMF at $\sim 10^{-2} - 10^{-1} ~\Zsun$ at $z \gtrsim 10$ is affected by
CMB heating, which suppresses fragmentation and reduces the number of low-mass stars (also see the study of \citealt{vanVeenen+25} with full MHD simulations).
In the case of strong Lyman–Werner (LW) irradiation ($J_{21} \sim 10^3$, in units of $10^{-21} ~\si{erg.s^{-1}.cm^{-2}.Hz^{-1}.sr^{-1}}$)\footnote{Also see Section~\ref{sec:massive_halos} for a discussion of the effect of LW radiation on the SFE of primordial clouds.}, \citet{ChonOmukai25}, and \citet{NandalChon26} showed that SMSs with masses above $\sim 10^4 ~\Msun$ can form at metallicities below $\sim 10^{-3} ~\Zsun$; above $\sim 10^{-2} ~\Zsun$, this star-formation channel is suppressed, leading instead to the formation of dense stellar clusters that may still host very massive stars of $\sim 10^3 ~\Msun$. 

Finally, once stars are formed, early studies such as that of \citet{CassisiCastellani93} also indicate that even trace amounts of CNO nuclei (already present at an initial metallicity of $10^{-6} ~\Zsun$ or lower) may be sufficient to cause departures from a typical zero-metallicity stellar evolution regime.

Homogeneous Galactic-chemical-evolution models predict rapid chemical enrichment, easily allowing near-solar (as e.g. in the luminous $z \approx 11$ CEERS galaxy from \citealt{Harikane+26}) or even super-solar metallicities to be reached within a few hundred Myr after the onset of star formation in a given galaxy \citep{KobayashiFerrara24}. 
Even faster local enrichment may occur when Pop~III enrichment is taken into account, as a single PISN episode from a massive Pop~III progenitor can eject tens of solar masses of metals; however, the metallicity reached by the surrounding gas depends sensitively on the dilution mass and on the efficiency of metal mixing (see e.g. the discussion in Section~\ref{sec:PISNe}). 
Therefore, observing truly pristine galaxies is extremely challenging even on timescale grounds alone.
Additionally, constraining very low metallicities through nebular emission from
gas photoionized by young Pop~III or extremely metal-poor stellar populations
requires probing metal lines in an extremely faint regime. \citet{Inoue11}, for instance, proposed [\ion{O}{3}]~$\lambda5007$/H$\beta < 0.1$ as a robust criterion for identifying galaxies below $\sim 10^{-3}~\Zsun$, while \citet{Katz+23} emphasized that robustly probing pristine or nearly pristine stellar populations may require to constrain [\ion{O}{3}]~$\lambda5007$ line luminosities $10^5$ times fainter than H$\beta$.

\begin{figure*}[!t]
    \centering

    \begin{minipage}{\textwidth}
        \centering

        \begin{minipage}[c]{0.73\textwidth}
            \centering
            \includegraphics[width=\linewidth]{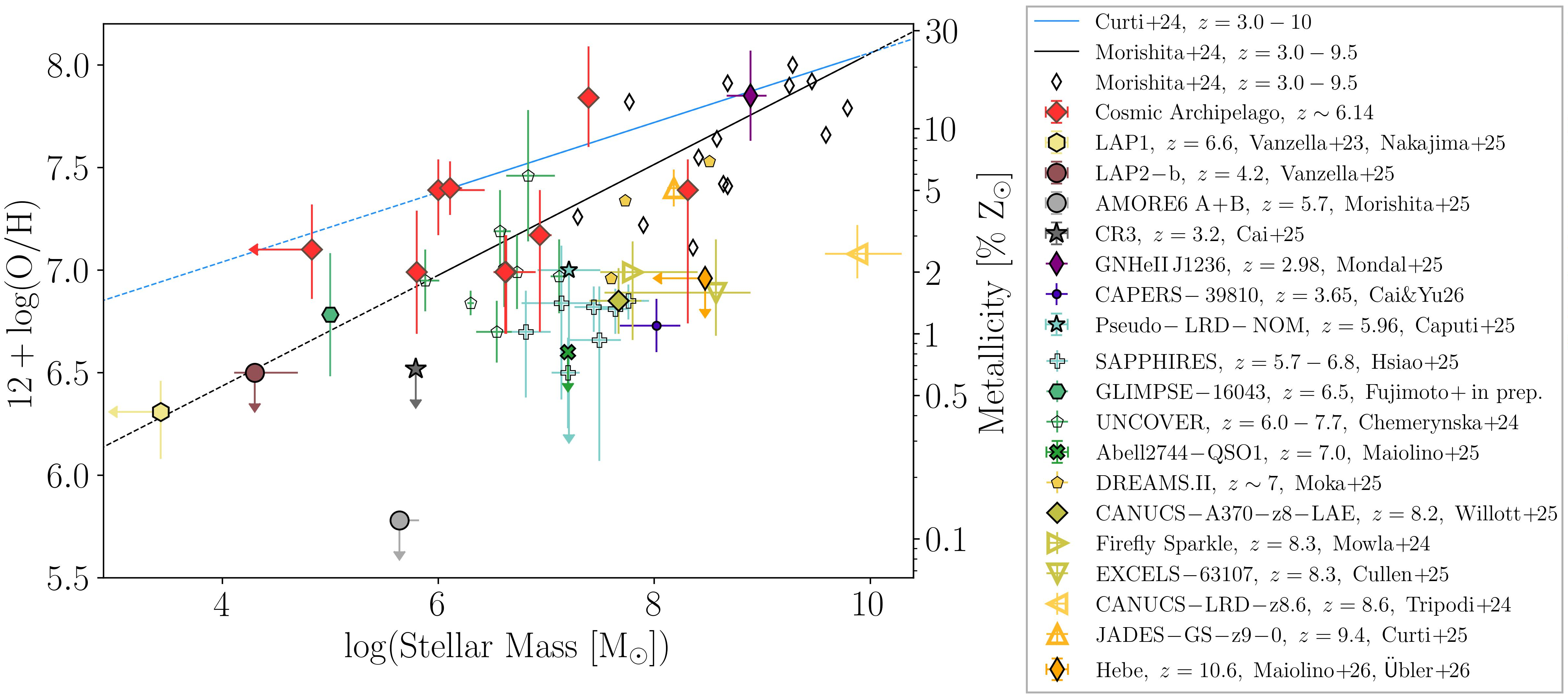}
        \end{minipage}
        \hfill
        \begin{minipage}[c]{0.25\textwidth}
            \caption{Mass-metallicity relation at $z > 3$ from \citet{Bolamperti+26}, including low-mass, star-forming complexes with undetected metal lines that indicate metallicities $\lesssim 8 \times 10^{-3} ~\Zsun$ (LAP2, \citealt{Vanzella+26}, AMORE6, \citealt{Morishita+25, Messa+26}, Hebe \citealt{Maiolino+26, Ubler+26} and MPG-CR3, \citealt{Cai+25} from Table~\ref{tab:PopIII_candidates_metal-poor}).}
            \label{fig:mass-metallicity+PopIII_candidates}
        \end{minipage}

    \end{minipage}

    \vspace{1.2em}

    \includegraphics[width=\linewidth]{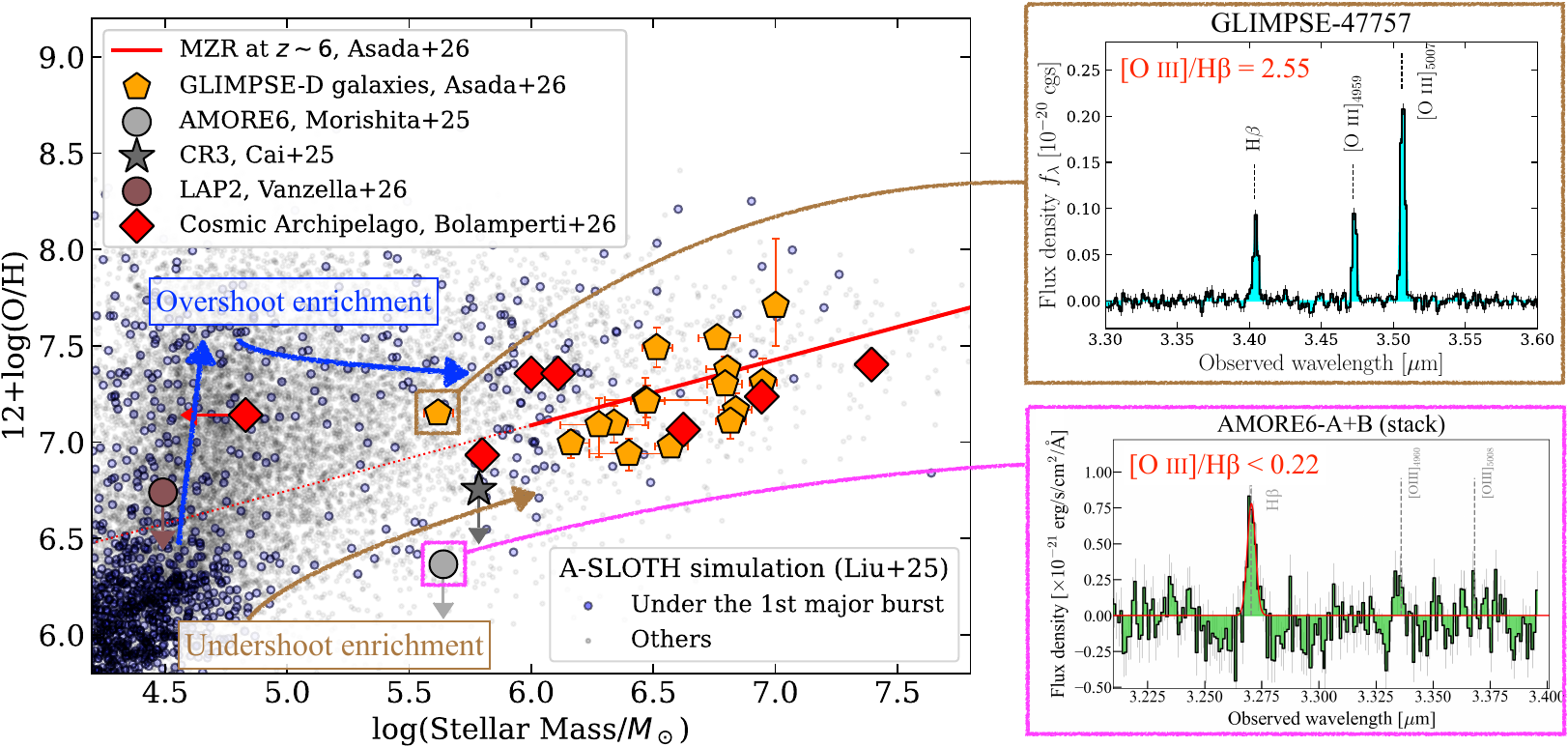}

    \caption{Mass-metallicity relation from deep observations behind the GLIMPSE-D cluster, compared with predictions from \texttt{A-SLOTH} simulations \citep{Liu+25}. The simulations indicate a dominant fast-enrichment (``overshoot'') channel immediately after the first Pop~III star-formation episode (\textit{blue arrow}), resulting in metal-rich galaxies at the low-mass end as in the example of the \textbf{top-right inset}; rare cases of slower (``undershoot'', \textit{brown arrow}) enrichment may however also result in EMP systems such as the AMORE6 object discovered by \citet{Morishita+25} (\textbf{bottom-right inset}, adapted from \citealt{Morishita+25}). Note that the same \citet{Nakajima+22} conversion from the R3 line ratio to metallicity has been adopted for all observations in this plot for consistency, resulting in metallicity estimates that do not always match the values presented in the respective discovery papers; particularly, a higher metallicity upper limit of $12 + \mathrm{Log (O/H)} < 6.37$ is shown for the AMORE6 candidate, with respect to Table~\ref{tab:PopIII_candidates_metal-poor}. Adapted from \citet{Asada+26}.}
    \vspace{3em}
    \label{fig:mass-metallicity_overshoot-undershoot}

\end{figure*}

\begin{table*}

\caption{Properties of observed low-mass, EMP candidate Pop~III systems.}

\centering
\scriptsize

\begin{adjustbox}{max width=\textwidth}
\begin{tabular}{lcccc}

\toprule
 & \textbf{MPG-CR3} & \textbf{LAP2}\textsuperscript{\textcolor{red}{(c)}} & \textbf{AMORE6}\textsuperscript{\textcolor{red}{(e)}} & \textbf{Hebe}\textsuperscript{\textcolor{red}{(g)}} \\
\midrule

$z$ & $3.193 \pm 0.016$ & $4.189 \pm 0.003$ & $5.7253 \pm 0.0001$ & $10.5862 \pm 0.0003$ \\

$M_\star$ [M$_\odot$] & $\approx 6.1 \times 10^5$ & $\approx (1.2 - 5.0) \times 10^4$ & $(4.37 {}^{+2.24}_{-0.73}) \times 10^{5}$ & $\approx (0.2 - 6) \times 10^5$\textsuperscript{\textcolor{red}{(h)}} \\

$Z$ [Z$_\odot$] & $< 0.008$ & $< 0.006$ & $< 0.0012$ & $< 0.019$ \\

12 + Log(O/H) & $< 6.52$ & $< 6.5$ & $< 5.78$ & $< 6.96$ \\

SFR [M$_\odot$ yr$^{-1}$] & — & — & $0.35 \pm 0.06$\textsuperscript{\textcolor{red}{(f)}} & — \\

$M_{\rm UV}$ & — & $\approx -12.2$ & $-14.52 {}^{+0.07}_{-0.08}$ & — \\

$\beta$-slope & — & — & $-2.77 {}^{+0.07}_{-0.09}$ & — \\

\midrule
\multicolumn{5}{c}{\textbf{Line fluxes [erg s$^{-1}$ cm$^{-2}$]}} \\
\midrule

Ly$\alpha$ & $(5.8 \pm 0.7) \times 10^{-17}$ & $(2.64 {}^{+0.37}_{-0.38}) \times 10^{-18}$\textsuperscript{\textcolor{red}{(d)}} & $(4.95 \pm 0.92) \times 10^{-19}$ & $< 1.5 \times 10^{-18}$\textsuperscript{\textcolor{red}{(i)}} \\

H$\alpha$ & $(4.2 \pm 0.6) \times 10^{-18}$\textsuperscript{\textcolor{red}{(a)}} & $(1.68 \pm 0.30) \times 10^{-19}$ & — & — \\

H$\beta$ & $(6.3 \pm 0.7) \times 10^{-19}$ & $(7.2 \pm 2.6) \times 10^{-20}$ & $(4.9 \pm 0.6) \times 10^{-20}$ & — \\

H$\gamma$ & — & — & — & $(4.1 \pm 0.7) \times 10^{-20}$ \\

\ion{He}{2}~$\lambda1640$ & —\textsuperscript{\textcolor{red}{(b)}} & — & — & $(1.11 \pm 0.17) \times 10^{-19}$ \\

[\ion{O}{3}]~$\lambda5007$ & $< 5.6 \times 10^{-19}$ & $< 5 \times 10^{-20}$ & $< 1.1 \times 10^{-20}$ & — \\

\midrule
\multicolumn{5}{c}{\textbf{EWs [Å]}} \\
\midrule

Ly$\alpha$ & $822 \pm 101$ & — & — & — \\

H$\alpha$ & $2814 \pm 327$ & $647 {}^{+302}_{-220}$ & — & — \\

H$\beta$ & — & — & $1594.7 \pm 206.9$ & — \\

H$\gamma$ & — & — & — & $> 350$ \\

\ion{He}{2}~$\lambda1640$ & — & $< 80$ & — & $> 47$ \\

\bottomrule

\textbf{References} & \citet{Cai+25} & \citet{Vanzella+26} & \citet{Morishita+25} & \makecell[c]{\citet{Maiolino+26} \\ \citet{Ubler+26} \\ \citet{Rusta+26}} \\
\bottomrule

\end{tabular}
\end{adjustbox}
\vspace{0.5em}
\begin{minipage}{\textwidth}
\tiny
\centering

\noindent\textsuperscript{\textcolor{red}{(a)}} Includes a rescaling to account for potential flux losses, as the source lies near the edge of the MSA shutter.\par

\noindent\textsuperscript{\textcolor{red}{(b)}} \ion{He}{2}~$\lambda1640$ line coinciding with strong OH skyline for this object.\par

\noindent\textsuperscript{\textcolor{red}{(c)}} The UV magnitude at 1700 Å for this source is de-lensed, while fluxes are not corrected for lensing. De-lensed fluxes were obtained by dividing by $\mu_\mathrm{tot}$, assuming a total magnification factor of $\mu_\mathrm{tot} = 50 \pm 5$.\par

\noindent\textsuperscript{\textcolor{red}{(d)}} From independent VLT/MUSE measure, $(5.50 \pm 0.58) \times 10^{-18} ~\mathrm{erg \, s^{-1} \, cm^{-2}}$.\par

\noindent\textsuperscript{\textcolor{red}{(e)}} Absolute UV magnitude and physical properties from SED fitting are reported for only one of the two lensed images (AMORE6-B), as the other one (AMORE6-A) suffers from large uncertainties, likely due to its location near bright galaxies as well as its smaller magnification factor ($\mu = 39.32_{-3.48}^{+3.73}$, vs $\mu = 77.69_{-5.92}^{+8.37}$ for AMORE6-B).\par

\noindent\textsuperscript{\textcolor{red}{(f)}} SFR inferred from H$\beta$, larger than the value inferred from rest-frame UV luminosity -- $(0.0186_{-0.0035}^{+0.0033}) ~\mathrm{M_\odot \, yr^{-1}}$ -- or from averaging over the last 100 Myr of the best-fit star-formation history -- $(0.0038_{-0.0007}^{+0.0017}) ~\mathrm{M_\odot \, yr^{-1}}$ --, supporting the presence of a very young starburst.\par

\noindent\textsuperscript{\textcolor{red}{(g)}} Note that this system is located $\sim 3$~kpc from the bright galaxy GN-z11 (with an estimated $M_\mathrm{UV} = -21.50 \pm 0.02$ and $\log M_\star / \Msun = 8.73 \pm 0.06$, \citealt{Bunker+23}), consistent with Pop~III formation in a pristine satellite of a massive halo (as discussed in Section~\ref{sec:massive_halos}), while its large distance from GN-z11 is incompatible with ionization from the central AGN proposed by \citet{Maiolino+24a}.\par

\noindent\textsuperscript{\textcolor{red}{(h)}} Pop~III mass estimate derived from \citet{Rusta+26} based on the \ion{He}{2}/H$\gamma$ ratio (which appears to exclude steep IMFs, favoring top-heavy distributions, especially for stellar ages $< 1$~Myr), combined with the \ion{He}{2} luminosity. \par

\noindent\textsuperscript{\textcolor{red}{(i)}} Measured fluxes need to be corrected for a small magnification factor of 1.42, which is not accounted for in the reported line fluxes; see table 1 of \citet{Maiolino+26} for the corresponding de-lensed line luminosities.\par

An up-to-date list of discovered Pop III candidates is maintained at the following page: \url{https://utcfc.github.io/pages/popiiicandidates.html}.

\end{minipage}

\label{tab:PopIII_candidates_metal-poor}
\end{table*}

Indirect metallicity probes from strong-line ratios \citep[e.g.][]{Curti+23, Sanders+24, Sanders+25, Chakraborty+25, Scholte+25, Korber+26} allow us to study the mass-metallicity relation at high redshifts (e.g. \citealt{Matthee+23, Nakajima+23, Curti+24, Sarkar+25, Faisst+25, Li+25, Stanton+26}, but also see \citealt{Morishita+24, Arellano-Cordova+26} for estimates solely based on the direct-$T_e$ method). 
Recently, \citet{Isobe+26} extended stack-based strong-line calibrations down to $12 + \mathrm{Log(O/H)} \approx 7.0$ using deep JWST/NIRSpec spectra, highlighting that metallicity estimates and upper limits for extremely metal-poor galaxies can depend sensitively on the adopted calibration and line-ratio threshold. Throughout this review, unless otherwise stated, we report the values quoted in the original discovery papers rather than homogenizing all measurements to a single calibration. 

While JWST has pushed the high-$z$ frontier to unprecedented depths (with the farthest spectroscopically confirmed galaxy, MoM-z14, currently lying at $z \approx 14.44$, \citealt{Naidu+26}), and various studies have attempted to extend the mass-metallicity relation to an extremely faint galaxy population of $\lesssim 10^6 ~\Msun$, often with the aid of gravitational lensing (e.g., \citealt{Vanzella+23, Vanzella+24, Chemerynska+24, Fujimoto+25a, Fujimoto+25b, Nakajima+25, Hsiao+25, Asada+26, Trussler+26}, also see Figure~\ref{fig:mass-metallicity+PopIII_candidates}), only a few candidate low-mass, star-forming complexes with undetected metal lines -- constraining metallicities $\lesssim 8 \times 10^{-3} ~\Zsun$ -- have been reported at $z \gtrsim 3$ (MPG-CR3, \citealt{Cai+25}, LAP2, \citealt{Vanzella+26}, AMORE6, \citealt{Morishita+25, Messa+26}, and Hebe, \citealt{Maiolino+24b, Maiolino+26, Ubler+26, Rusta+26}, see Table~\ref{tab:PopIII_candidates_metal-poor}). 
Notably, the current record holder for the lowest metallicity upper limit at $z > 3$, the AMORE6 galaxy at $z \approx 5.7$, with a metallicity $\lesssim 0.1\% ~\Zsun$ and a stellar mass of $\sim (4 - 7) \times 10^5 ~\Msun$ \citep{Morishita+25}, still lies one order of magnitude above the most metal-poor star observed in the local Universe, i.e. the red giant star SDSS J0715-7334 in the Large Magellanic Cloud, with a measured metallicity $\sim (0.42  - 1.1) \times 10^{-4} ~\Zsun$ \citep{Ji+26}. 
Deep spectroscopic observations of the strongly lensed LAP1 system at $z \approx 6.6$ \citep{Vanzella+20, Vanzella+23} also revealed weak but significant detections of [\ion{O}{3}] and \ion{C}{4} metal lines even fainter than the upper limit established for AMORE6 at the sensitivity of \citet{Morishita+25} study, constraining a gas-phase oxygen abundance $(4.2 \pm 1.8) \times 10^{-3}$ times lower than the solar value \citep{Nakajima+25}\footnote{On the other hand, the authors highlighted an exceptionally hard ionizing radiation field matching theoretical predictions for an extremely metal-deficient stellar population, as well as an elevated carbon-to-oxygen ratio consistent with Pop~III nucleosynthetic yields (Section~\ref{sec:CEMPs}). These observations establish LAP1 as a candidate ``self-enriched'' Pop~III system or ``fossil in the making'' (see the discussion in Section~\ref{sec:spectral-hardness}, and a summary of LAP1 observed properties in Table~\ref{tab:PopIII_candidates_hybrid}).}.

On the contrary, metal enrichment in high-$z$ galaxies appears to be pervasive up to the highest redshifts (see \citealt{Bunker+23, Castellano+24, DEugenio+24, Abdurrouf+24, Hsiao+24, Carniani+25, Helton+25, Witstok+26, Harikane+26, Alvarez-Marquez+26} for examples at $z > 10$)\footnote{Note however that -- among these candidates -- the case of JADES-GS-z12 at $z\simeq 12.5$ is of particular interest for Pop~III studies from an archaeological standpoint, as \citet{DEugenio+24} reported an elevated C/O ratio consistent with enrichment from Pop~III SNe (see Section~\ref{sec:CEMPs} for a discussion of elevated carbon in Pop~III chemical signatures).}.
These observations suggest a fast enrichment process in early galaxies. By analyzing deep NIRSpec observations behind the GLIMPSE-D cluster to probe the mass-metallicity relation down to $\sim 10^{5.6} ~\Msun$, \citet{Asada+26} reported strong [\ion{O}{3}] line detection even from the lowest-mass galaxy in the sample (GLIMPSE-47757). By comparing with predictions from \texttt{A-SLOTH} simulations at even lower masses from \citet{Liu+25}, they suggested two distinct pathways for early metal enrichment, with a dominant fast enrichment (``overshoot'') channel that occurs right after the first Pop~III star-formation episode, but also rare cases in which delayed enrichment (``undershoot'', e.g. due to inefficient mixing within the halos, see the discussion in Section~\ref{sec:PISNe}) can result in EMP systems such as the AMORE6 galaxy; this comparison is illustrated in Figure~\ref{fig:mass-metallicity_overshoot-undershoot}, together with the spectra of GLIMPSE-47757 and AMORE6. 

However, metallicity estimates in high-redshift galaxies remain subject to important observational and modeling biases. Robust determinations of the ionization, thermal and chemical properties of galaxies require multiphase gas models, while unresolved temperature fluctuations and uncertain density structure can lead to underestimation of the metallicity of emitting regions. Particularly, assuming uniform, low-density conditions in high-density environments has been shown to lead to underestimated metallicity from UV and optical nebular diagnostics \citep{Martinez+25, Moreschini+26}, and biases have been pointed out in $T_e$-based metallicity estimates due to unresolved temperature fluctuations, within individual \ion{H}{2} regions and across different \ion{H}{2} regions \citep[e.g.][]{Cameron+23, Rosales-Ortega2026The-DESIRED-str}. Joint JWST and ALMA observations also provide direct evidence for a multi-zone electron-density structures at high redshifts, clearly indicating the need for spatially resolved modeling of galaxy emission (\citealt{Usui+25, Harikane+25}, also see \citealt{Topping+25, Castellano+26}).

On the other hand, genuinely pristine or nearly pristine gas would contribute little to no oxygen-line emission, and therefore may be difficult to identify directly in integrated spectra. \citet{Lewis+25}, for example, examined potential selection bias effects favoring enriched, strong line emitters in observational studies of the mass-metallicity relation at $z \sim 3 - 6$ (also see \citealt{Kotiwale+26} for a study of selection effects in mass-metallicity-relation estimates at $z \sim 5 - 7$). 
Moreover, the metallicity of high-$z$ galaxies may be overestimated when faint, metal-poor regions in their outskirts are overlooked in the presence of negative metallicity gradients \citep[e.g.][]{Li+25}.
Thus, current data may both underestimate the metallicity of emitting gas and remain insensitive to faint/non-emitting pristine pockets, complicating the interpretation of galaxy-integrated metallicity measurements and assessments of pristine gas reservoirs across time. \citet{Pollock+26} recently identified extremely metal-poor neutral gas in three UV-bright galaxies at $z \approx 7.8 - 9.3$, with metallicities substantially below those inferred from nebular emission in the same systems, indicating that absorption spectroscopy may reveal diffuse, chemically primitive gas in the outer ISM and/or inner CGM that is missed by luminosity-weighted emission-line measurements.

\vspace{.6em}
\subsection{Residual Pop~III formation at late cosmic times traced by spectral-hardness signatures}
\label{sec:spectral-hardness}

\begin{figure*}[!t]
    \centering

    \begin{minipage}[t]{0.49\textwidth}
        \vspace{0pt}
        \centering

        \includegraphics[width=\linewidth]{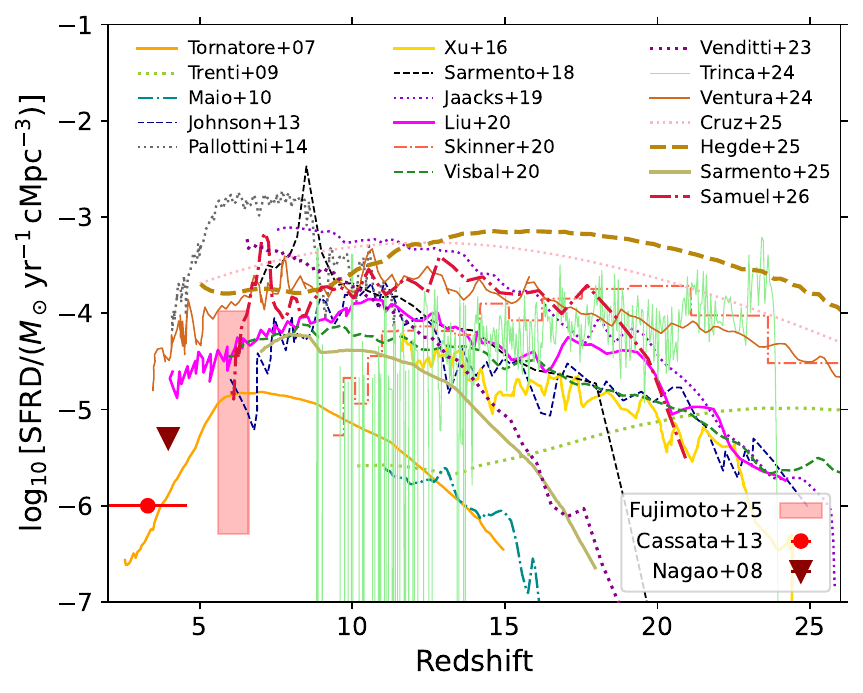}

        \vspace{-.5em}
        \caption{Pop~III SFRD evolution as a function of redshift according to various models and simulations (listed in Table~\ref{tab:PopIII_cosmological_simulations}). Tentative observational constraints from \citet{Nagao+08}, \citet{Cassata+13} and \citet{Fujimoto+25b} are also shown for comparison (see Section~\ref{sec:late_PopIII_obs} for further details). An interactive version of this figure with downloadable data is available at the following page: \url{https://utcfc.github.io/pages/popiiisfrd.html}.}
        \label{fig:PopIII_SFRD}
    \end{minipage}
    \hfill
    \begin{minipage}[t]{0.49\textwidth}
        \vspace{0pt}
        \vspace{1.1em}
        \centering

        \begingroup
        \setlength{\abovecaptionskip}{0pt}
        \setlength{\belowcaptionskip}{4pt}

        \captionof{table}{List of the cosmological simulations included in the Pop~III SFRD comparison of Figure~\ref{fig:PopIII_SFRD}, adapted and updated from \citet{Venditti+23}. Table columns show: model reference, code/simulation name and type of adopted numerical scheme, box size $L$~$[\si{cMpc}/h]$, and DM particle mass $m_\mathrm{DM}$~$[10^5 ~ \si{M_\odot}/h]$.}
        \label{tab:PopIII_cosmological_simulations}
        \endgroup

        \scriptsize
        \setlength{\tabcolsep}{3pt}
        \renewcommand{\arraystretch}{0.9}

        \begin{adjustbox}{max width=\linewidth}
        \begin{tabular}{>{\tiny}r|>{\scriptsize}c|c|c}
           {\footnotesize Reference} & {\footnotesize Method} & $L$ & $m_\mathrm{DM}$ \\
            \hline
            \citet{HegdeFurlanetto25} & \texttt{abcd} (SAM) & -- & -- \\
            \citet{Cruz+25} & \texttt{Zeus21} (SAM) & -- & -- \\
            \citet{Trinca+24} & \texttt{CAT} (SAM) & -- & -- \\
            \citet{TrentiStiavelli09} & SAM & -- & -- \\
            \citet{Venditti+23} & \texttt{dustyGadget} (SPH) & 50.0 & 353 \\
            \citet{SarmentoScannapieco25} & \texttt{RAMSES-RT} (AMR) & 24.0 & 7.81 \\
            \citet{Sarmento+18} & \texttt{RAMSES} (AMR) & 12.0 & 0.99 \\
            \citet{Tornatore+07} & \texttt{Gadget} (SPH) & 10.0 & 36.2 \\
            \citet{Pallottini+14} & \texttt{RAMSES} (AMR) & 10.0 & 4.70 \\
            \citet{Ventura+24} & \texttt{Meraxes} (SAM) & 10.0 & 0.10 \\
            \citet{Xu+16a} & \texttt{ENZO} (AMR) & 4.30 & 0.21 \\
            \citet{LiuBromm20} & \texttt{GIZMO} (MFM) & 4.00 & 0.36 \\
            \citet{Jaacks+19} & \texttt{GIZMO} (MFM) & 4.00 & 0.29 \\
            \citet{Samuel+26} & \texttt{CampFIRE-6k} (MFM) & 3.37 & 0.23 \\
            \citet{Johnson+13} & \texttt{Gadget-2} (SPH) & 2.84 & 0.04 \\
            \citet{Visbal+20} & SAM & 2.01 & 0.05 \\
            \citet{Maio+10} & \texttt{Gadget-2} (SPH) & 0.70 & 0.01 \\
            \citet{SkinnerWise20} & \texttt{ENZO} (AMR) & 0.67 & 0.01 \\
        \end{tabular}
        \end{adjustbox}

    \end{minipage}

\end{figure*}

While star formation at $z \lesssim 10$ is certainly dominated by metal-enriched, Pop~II stars, a host of cosmological numerical simulations \citep{Tornatore+07, Maio+10, Johnson+13, Pallottini+14, Xu+16a, Xu+16b, Sarmento+18, SarmentoScannapieco22, SarmentoScannapieco25, Jaacks+19, LiuBromm20, BennettSijacki20, SkinnerWise20, Venditti+23, Zier+25, Storck+26} and semi-analytical models \citep{TrentiStiavelli09, Visbal+20, Riaz+22, Trinca+24, Ventura+24, HegdeFurlanetto25} agree in finding a residual level of Pop~III star formation at late cosmic times, possibly down to the EoR, and even into the post-reionization era. Figure~\ref{fig:PopIII_SFRD} reports predictions for the Pop~III star formation rate density (SFRD) evolution down to $z \sim 3$ from a variety of models, listed in Table~\ref{tab:PopIII_cosmological_simulations}. In fact, although locally the ISM of high-$z$ galaxies is quickly polluted after the onset of star formation (Section~\ref{sec:low-metallicity}), metal enrichment is a highly inhomogeneous and globally inefficient process on large scales, that can leave patches of pristine gas capable of forming Pop~III stars well after Cosmic Dawn.

\begin{figure}
    \centering
    \includegraphics[width=\linewidth]{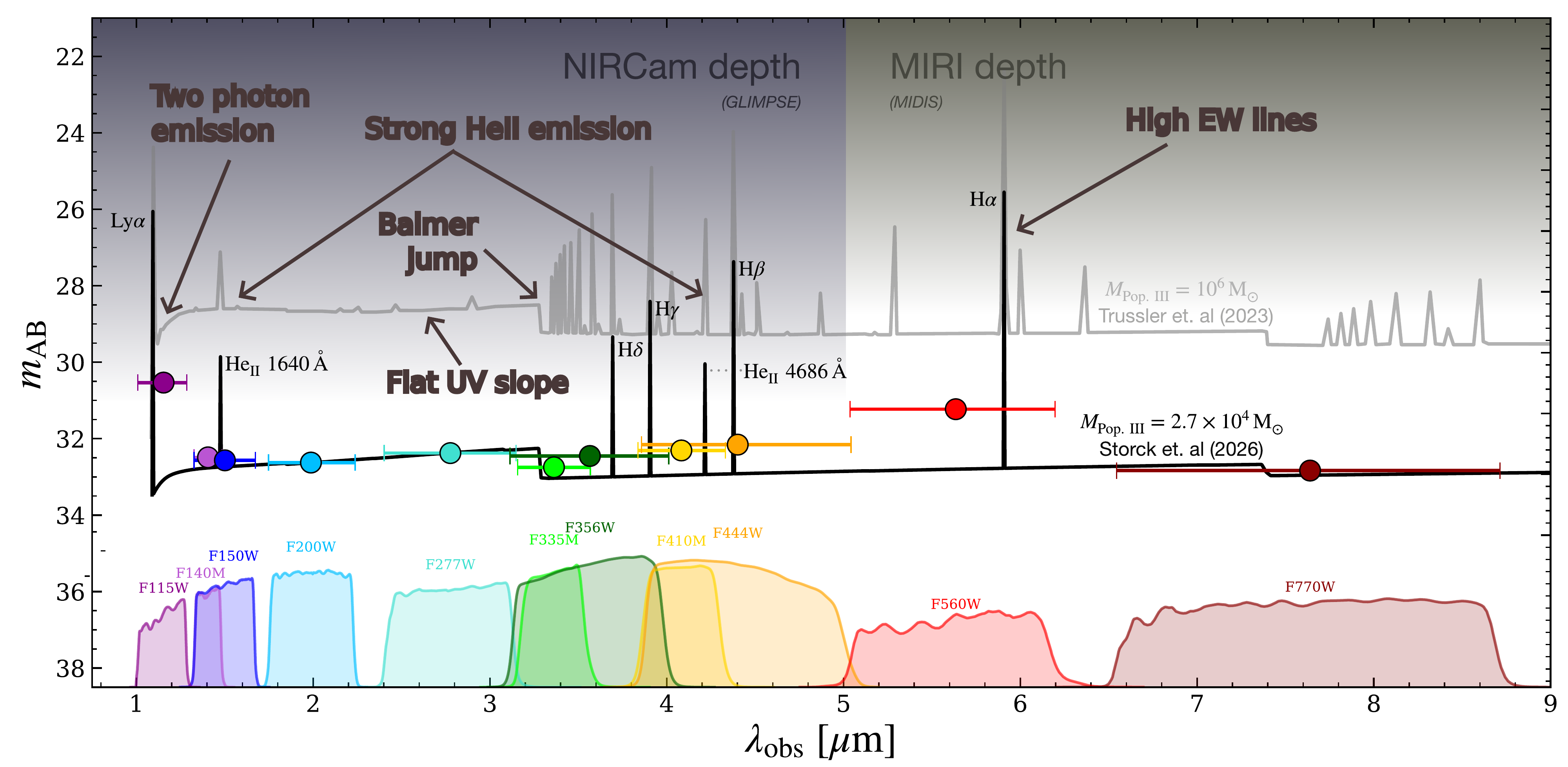}
    \caption{Example spectrum of a theoretical $z = 8$ Pop~III galaxy (\textit{grey}), assuming a top-heavy Salpeter-like IMF in the range $[50, 500] ~\Msun$ at the nominal stellar mass $M_\star = 10^6 ~\Msun$ from the \texttt{Yggdrasil} model \citep{Zackrisson+11}, observed 0.01~Myr after an instantaneous starburst \citep{Trussler+23}; typical spectral-hardness features associated with Pop~III-dominated spectra are highlighted in the figure (see text for further details). This is compared with the spectrum of the brightest Pop~III galaxy from the \texttt{MEGATRON} simulation suite at $z = 9.8$ (\textit{black}, redshifted at $z = 8$ for easier comparison with the \texttt{Yggdrasil} spectrum, and including filter throughputs and bandpass-averaged flux densities within JWST/NIRCam + MIRI bands), assuming a critical metallicity for Pop~III formation of $Z_\mathrm{crit} \approx 10^{-6} ~\Zsun$ and a top-heavy IMF with a characteristic mass of $100 ~\Msun$. Even in the most favorable case shown here, only a Pop~III mass up to of $2.7 \times 10^4 ~\Msun$ can form in the simulation, resulting in a much fainter spectrum, with UV magnitudes typically fainter than $\approx -11.5$, and an EW for H$\alpha$ and \ion{He}{2}~$\lambda1640$/$\lambda4686$ of 3870~\AA ~and 66~\AA ~respectively, as well as a redder UV $\beta$-slope due to the strong nebular-continuum contribution. Adapted from \citet{Storck+26}.} 
    \vspace{4em}
    \label{fig:PopIII_galaxy_example_spectrum}
\end{figure}

The search for direct signatures from this residual stellar population is extremely challenging. In Section~\ref{sec:low-metallicity} we emphasized the difficulty in identifying EMP or even pristine star complexes at high redshifts, due to sensitivity limits. Therefore, most studies aiming to spot an active Pop~III component in high-$z$ galaxies rather rely on a combination of low-metallicity and spectral-hardness diagnostics.
Massive, metal-poor stars should in fact produce intense ionizing radiation, resulting in exceptionally strong \ion{He}{2} recombination lines \citep[e.g.][]{TumlinsonShull00, Tumlinson+01, Bromm+01b, Oh+01, Schaerer02, Schaerer03, Raiter+10, Visbal+15, Venditti+24b, Venditti+26}, strong Ly$\alpha$ and Balmer \ion{H}{1} recombination lines \citep[e.g.][]{Inoue11, Mas-Ribas+16, NakajimaMaiolino22}, as well as steep Balmer jumps and flat UV slopes (redder than $~\sim -2.2$), due to powering of the nebular-continuum portion of the spectrum from hot, ionized gas in the vicinity of the stars \citep[e.g.][]{Raiter+10, Zackrisson+11, Trussler+23, Cameron+24} (see e.g. the example spectrum expected for a young Pop~III-dominated galaxy at $z = 8$ from the \texttt{Yggdrasil} model in Figure~\ref{fig:PopIII_galaxy_example_spectrum}, in grey).
However, all characteristic spectral-hardness features will decrease after a few Myr in instantaneous bursts, as the most massive stars evolve out of the main sequence \citep[e.g.][]{Schaerer02, Schaerer03, Katz+23, Venditti+26}, only allowing the identification of very young Pop~III populations. 
\citet{Katz+23} further emphasized that both the hard emission from massive Pop~III stars and the radiation emitted by their cooling surrounding nebular gas or supernova remnants are short-lived.

Note that, although the brightest stars in a Pop~III population are expected to be bluer than their non-Pop~III counterparts (owing to their hotter, more compact metal-free stellar atmospheres, e.g. \citealt{Schaerer02}), a greater contribution from the relatively redder nebular continuum has also been predicted to potentially result in a redder total spectrum for Pop~III-dominated galaxies, manifesting e.g. in less negative 1500~\AA \, UV $\beta$-slopes \citep{Schaerer02,Raiter+10,Trussler+23}. \citet{Storck+26}, for example, found values for $\beta$ between -2 and -1.5, redder than counterpart Pop~II-dominated galaxies, due to the strong contribution of nebular continuum (see e.g. the spectrum arising from the brightest Pop~III galaxy from the \texttt{MEGATRON} simulation suite in Figure~\ref{fig:PopIII_galaxy_example_spectrum}, in black).
\citet{Katz+25} also studied the dependence of the UV $\beta$-slope on stellar temperature and ISM conditions, showing that nebular gas properties can strongly impact this diagnostic, and that galaxy populations with very high ionizing efficiencies ($\xi_\mathrm{ion}$) may be difficult to detect in practice and lead to biases in observational measurements of $\xi_\mathrm{ion}$.
Furthermore, these predictions likely depend on the assumed ionizing photon escape fraction ($f_{\rm esc}$): in density-bounded systems, or in galaxies with very high $f_{\rm esc}$, fewer ionizing photons are reprocessed locally by the gas, weakening both recombination lines and nebular-continuum emission and potentially leading to bluer UV $\beta$-slopes \citep[e.g.][]{Zackrisson+13, Zackrisson+17, Topping+22, Saxena+26, Giovinazzo+26}.

The interpretation of hard spectral features is further complicated by a variety of confusing mechanisms/sources that can produce high enough temperatures to power e.g. strong \ion{He}{2} line emission, including X-ray binaries \citep{Schaerer+19, Saxena+20a, Saxena+20b, Senchyna+20, Cameron+24, Lecroq+24, Bray+25}, enriched VMS, WR stars and stripped He stars \citep{GrafenerVink+15, Kehrig+18, Saxena+20a, ShiraziBrinchmann12, Senchyna+21, Cameron+24, Martins+23, Tozzi+23, Wofford+23, Gomez-Gonzalez+24, Upadhyaya+24, Berg+26, Leitherer+25}, AGN \citep{Saxena+20a, Saxena+20b, ShiraziBrinchmann12, Tozzi+23, Liu+24, Topping+24} -- including direct-collapse black holes (DCBHs) \citep{Pacucci+15, Pacucci+16, Pacucci+26, Pallottini+15, Agarwal+16, NakajimaMaiolino22} --, shocked gas \citep{Kehrig+18, Lecroq+24, Flury+25}, and matter/density-bounded \ion{H}{2} regions\footnote{In ionization-bounded \ion{H}{2} regions, the gas cloud is optically thick to ionizing photons and the ionization front is fully contained within the nebula, so that most Lyman-continuum photons are absorbed locally and reprocessed into nebular emission. In matter-bounded, or density-bounded, regions, the gas column is insufficient to absorb all ionizing photons, allowing part of the ionizing radiation to escape and truncating the low-ionization outer layers of the nebula, which suppresses low-ionization nebular lines with respect to higher-ionization lines produced closer to the ionizing source, possibly enhancing high-ionization line ratios and mimicking harder ionizing conditions \citep{OsterbrockFerland06}.} \citep{Moreschini+26}.
Note that, while strong \ion{He}{2} lines from AGN \citep[e.g.][]{KollatschnyZetzl+13}, as well as VMS and WRs \citep[e.g.][]{Wofford+23, Schaerer+25, Berg+26}, may be distinguishable from Pop~III models thanks to larger expected line widths for these sources with respect to Pop~III stellar populations, \citet{GrafenerVink+15} showed that slow, strong winds of metal-poor ($< 0.1 ~\Zsun$) VMS may also produce narrower \ion{He}{2}~$\lambda1640$ lines with FWHMs of $\sim 300 - 500 ~\si{km.s^{-1}}$.

\begin{figure*}[!t]
    \centering
    \includegraphics[width=\linewidth]{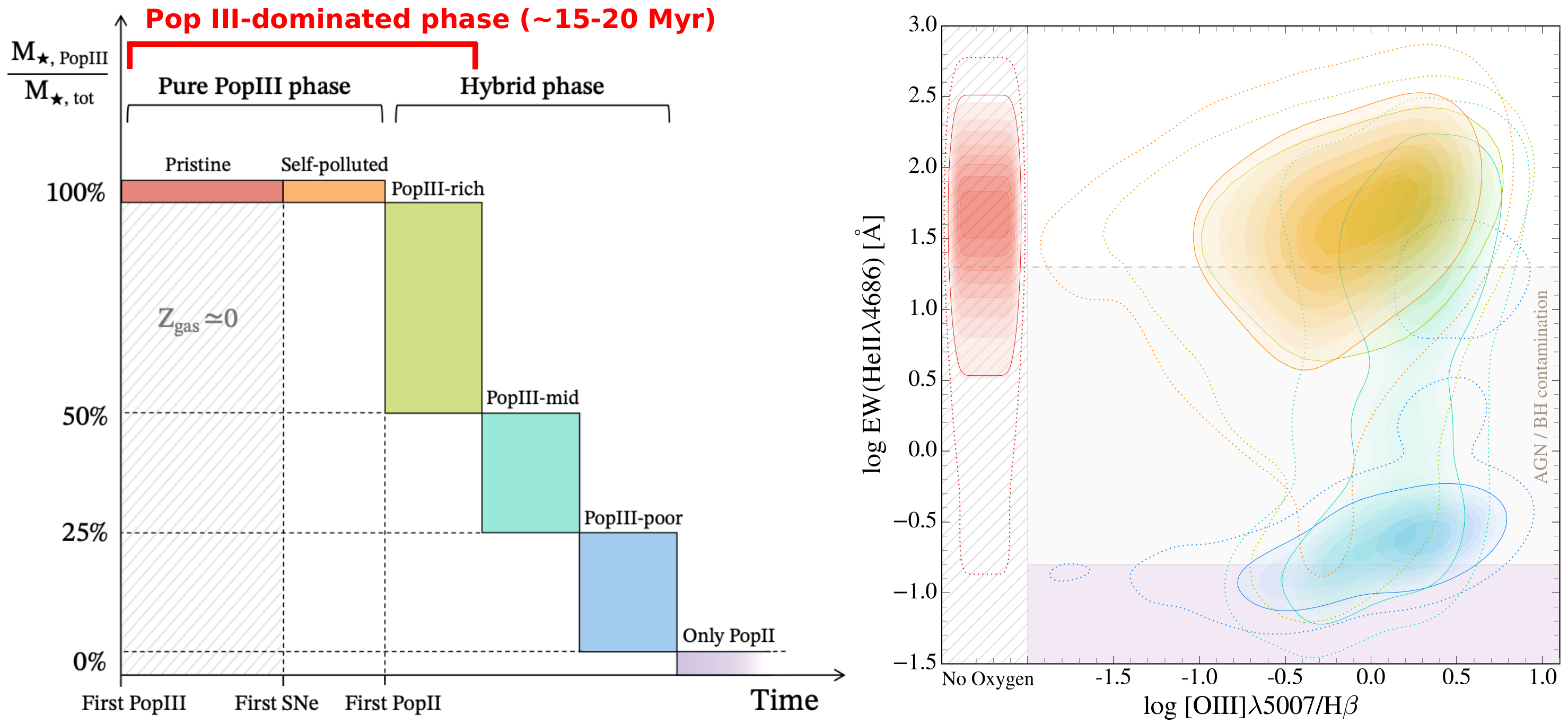}
    \caption{\textbf{Left:} schematic representation of the evolutionary stages of a Pop~III-hosting galaxy, based on the ratio between Pop~III ($M_\mathrm{\star,PopIII}$) and total ($M_\mathrm{\star,tot}$) stellar mass, with the Pop~III-dominated phase ($M_\mathrm{\star,PopIII} > 50\%  M_\mathrm{\star,tot}$, encompassing a brief, ``truly pristine'' stage, a ``self-polluted, Pop~III-pure'' stage, and a ``Pop~III-rich'' stage in hybrid galaxies that also host a coeval Pop~II component) lasting about $\sim 15 - 20$~Myr when a Larson-type IMF \citep{Larson+85} in the range $[0.8, 1000] ~\Msun$ with a characteristic mass of $10 ~\Msun$ is assumed. 
    \textbf{Right:} density distributions of Pop~III-hosting galaxies predicted by the \texttt{NEFERTITI} model on the EW(\ion{He}{2}~$\lambda4686$) versus [\ion{O}{3}]~$\lambda5007$/H$\beta$ diagram (assuming an ionization parameter $\mathrm{Log} U = -1$), with different \textit{colors} matching the evolutionary stages in the left-side scheme, and \textit{dotted} (\textit{solid}) lines including 95\% (68\%) of the population. The \textit{horizontal dashed line} marks the lower limit of the Pop~III diagnostics from \citet{NakajimaMaiolino22}, with the \textit{lilac shaded area} below highlighting the AGN and DCBH contamination zone. Adapted from \citet{Rusta+25}.}
    \vspace{2.5em}
    \label{fig:self-polluted+hybrid_PopIII_galaxies}
\end{figure*}

Finally, the semi-analytic study of \citet{Rusta+25} showed that a Pop~III component may be found in moderately enriched, low-mass ($M_\star \lesssim 10^{5} ~\Msun$, $M_\mathrm{vir} \lesssim 10^{7.5} ~\Msun$) galaxy environments, due to a surviving tail of massive Pop~III stars still powering hard emission, even after the most massive stars formed in the current (or previous) burst have exploded as SNe (``self-polluted'' pure Pop~III phase), or in the presence of coeval Pop~II formation (truly ``hybrid'' phase)\footnote{Note that numerical simulations have found recovery times of tens to several hundreds of Myr for the ejected gas to recollapse and be able to form stars again after the first Pop~III SN explosion \citep[e.g.][]{Jeon+14, ChiakiWise19, Chiaki+20}. This may limit the prevalence of scenarios in which massive Pop~III stars from a previous star-formation episode are still alive when Pop~II star formation is triggered (as in the ``Pop~III-rich'' phase suggested by \citealt{Rusta+25}), especially after energetic explosions or in shallow potential wells. However, Pop~III star formation may also be possible in more massive halos, where the deeper potential well and larger reservoir of dense gas can prevent complete gas evacuation after the first SN explosions (see e.g. the discussion in Section~\ref{sec:massive_halos}).}.
Strong metal-line emission (e.g. [\ion{O}{3}]~$\lambda5007$) may therefore be observed from pure-Pop~III galaxies after the first Pop~III starburst, before the onset of metal-enriched star formation. Conversely, bright \ion{He}{2} lines could still mark a dominant ($> 50\%$ in mass) Pop~III stellar population in the presence of active Pop~II stars (Figure~\ref{fig:self-polluted+hybrid_PopIII_galaxies}). Building on these considerations, the authors proposed new diagnostics for identifying Pop~III-dominated galaxy candidates, based, e.g., on \ion{C}{4}/\ion{C}{3}] versus \ion{C}{3}]/\ion{He}{2} line ratios; see also \citet{Cleri+23} for diagnostics relying on the [\ion{Ne}{5}]/[\ion{Ne}{3}] line ratio.

\vspace{.6em}
\subsection{Observational constraints on late Pop~III formation}
\label{sec:late_PopIII_obs}

Despite all these challenges, a handful of observational campaigns have been conducted to constrain the abundance of striking ``pure'' Pop~III systems at low and intermediate redshifts.
\citet{Nagao+08} first attempted to place some bounds on the Pop~III SFRD at $z \approx 4$, by applying a photometric selection criterion to search for strong Ly$\alpha$ and \ion{He}{2}~$\lambda1640$ line emitters with no associated metal-line detection in the Subaru Deep Field \citep{Kashikawa+04}. Based on the non-detection of convincing pure Pop~III candidates, they inferred an upper limit on the Pop~III SFRD at this redshift of $\sim 5 \times 10^{-6} ~\si{\Msun.yr^{-1}.cMpc^{-1}}$. 
More recently, \citet{Fujimoto+25a} proposed a NIRCam-based selection method for prominent H$\alpha$ and deficient [\ion{O}{3}] emitters at $5.5 \lesssim z \lesssim 6.5$, further refined by \citet{Fujimoto+25b} to exclude potential exotic low-metallicity (but not genuinely metal-free) sources with an unusually strong Balmer jump, such as the previously identified GLIMPSE-16043 system at $z \approx 6.5$. By applying this selection to the five legacy JWST survey fields CEERS \citep{Finkelstein+24}, PRIMER-UDS/COSMOS \citep{Donnan+24}, JOF
\begin{wrapfigure}[19]{r}{.5\textwidth}
    \centering
    \vspace{-2em}
    \includegraphics[width=\linewidth]{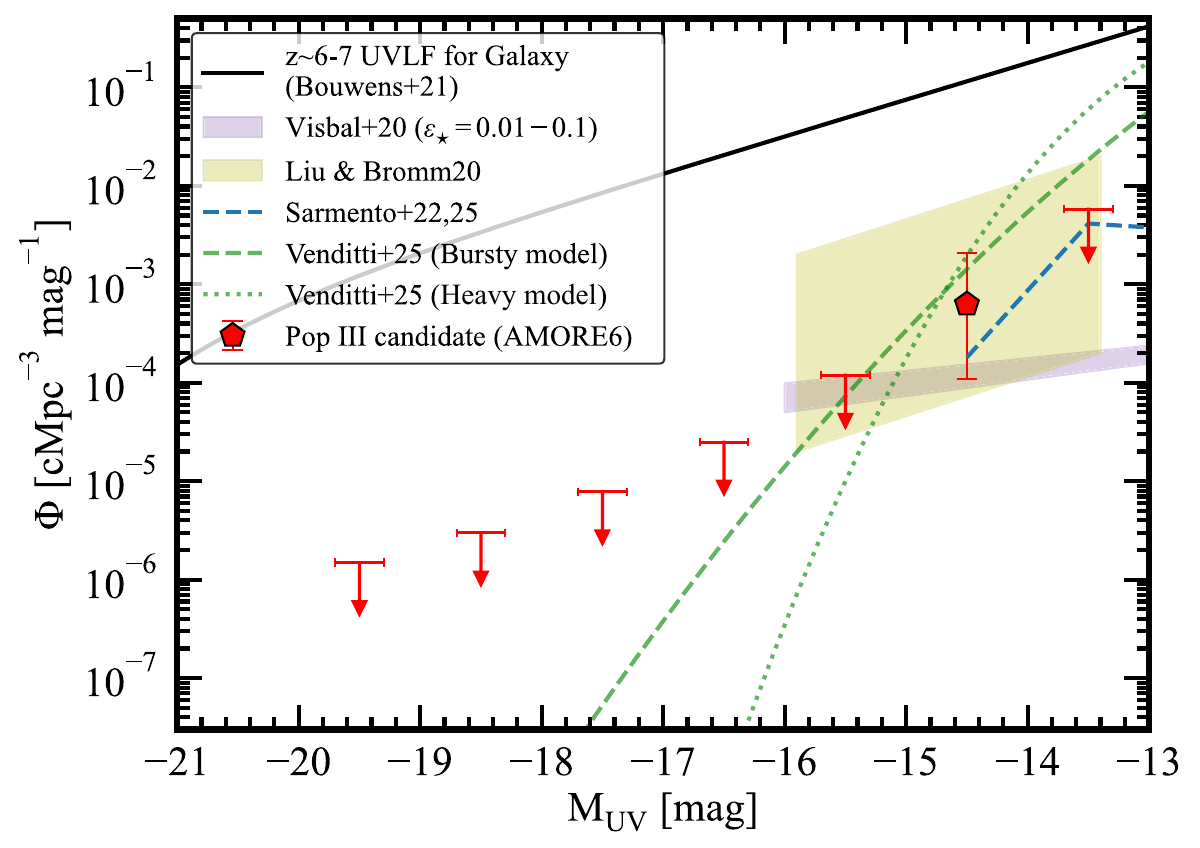}
    \caption{Pop~III UVLF observational constraints at $5.5 \lesssim z \lesssim 6.5$, based on the detection of a single Pop~III galaxy candidate (the AMORE6 galaxy at $z \approx 5.7$, \citealt{Morishita+24}) when applying a NIRCam-based selection criterion (aimed at identifying ``pure'' Pop~III systems through prominent-H$\alpha$/deficient-[\ion{O}{3}] line diagnostics) to ten JWST legacy fields. Reproduced from \citet{Fujimoto+25b}.}
    \label{fig:PopIII_UVLF_obs}
\end{wrapfigure}
\citep{Robertson+24, Eisenstein+25}, UNCOVER+MegaScience \citep{Bezanson+24, Suess+24}, and GLIMPSE \citep{Atek+25}, as well as five additional lensing cluster fields publicly released by the CANUCS team \citep{Sarrouh+26}, they were able to infer a constraint of $\sim 10^{-6} - 10^{-4} ~\si{\Msun.yr^{-1}.cMpc^{-1}}$ on the Pop~III SFRD, and some preliminary constraints on the Pop~III UV luminosity function (UVLF, Figure~\ref{fig:PopIII_UVLF_obs}) at these redshifts, based on the detection of the AMORE6 candidate at $z \approx 5.7$ (\citealt{Morishita+24}, see Section~\ref{sec:low-metallicity} and Table~\ref{tab:PopIII_candidates_metal-poor}).
Another faint ($M_\mathrm{UV} \lesssim -14.3$) compact Ly$\alpha$ emitter at $z \approx 6.2$ (A370-z6LAE-2) was found to partially satisfy the updated selection criteria in \citet{Fujimoto+25b}, but it exhibited a S/N of only $\sim 1.4$ in the F410M filter, not robustly excluding contaminants exhibiting a significant Balmer jump as seen in GLIMPSE-16043.

\begin{table*}
\vspace{-3em}
\centering

\caption{Properties of observed ``hybrid'' and ``self-enriched'' candidate Pop~III-hosting systems.}

\scriptsize

\begin{adjustbox}{max width=\textwidth}
\begin{tabular}{lccccc}

\toprule
 & \textbf{GNHeII J1236+6215} & \textbf{CAPERS-UDS-32520}\textsuperscript{\textcolor{red}{(c)}} & \textbf{LAP1} & \textbf{RX J2129-z8HeII} & \textbf{EXCELS-63107} \\
\midrule

$z$ & $2.9803 \pm 0.0010$ & $5.1240 \pm 0.0002$ & $6.625 \pm 0.001$ & $8.1623 \pm 0.0007$ & $8.271$ \\

$M_\star$ [M$_\odot$] & $(7.8 \pm 3.1) \times 10^{8}$ & $\sim 10^9$ & $\lesssim 3300$ & $(5.6 {}^{+0.8}_{-0.7}) \times 10^{7}$ & $(3.72 {}^{+4.05}_{-3.36}) \times 10^{8}$\textsuperscript{\textcolor{red}{(h)}} \\

$M_\mathrm{III}$ [M$_\odot$] & — & — & — & $(7.8 \pm 1.4) \times 10^{5}$ & — \\

$Z$ [Z$_\odot$] & $ \sim 0.15$ & $\sim 0.15$ & $(4.2 \pm 1.8) \times 10^{-3}$\noindent\textsuperscript{\textcolor{red}{(e)}} & $\sim 0.1$ & $0.016$ \\

12 + Log(O/H) & $7.85 \pm 0.22$ & $7.88 \pm 0.07$ & $6.31 {}^{+0.15}_{-0.23}$ & $7.63 {}^{+0.14}_{-0.09}$ & $6.89 {}^{+0.26}_{-0.21}$ \\

[\ion{O}{3}]/H$\beta$ & $5.45 \pm 0.32$\textsuperscript{\textcolor{red}{(a)}} & — & $0.69 \pm 0.28$ & $5.5 \pm 0.8$ & $3.61 \pm 0.30$ \\

SFR [M$_\odot$yr$^{-1}$] & $12.2 \pm 2.0$\textsuperscript{\textcolor{red}{(b)}} & — & — & $9.56 {}^{+4.51}_{-1.70}$ & $7.8 \pm 0.6$ \\

$M_{\rm UV}$ & $-22.09 \pm 0.02$ & $-20.95 \pm 0.05$ & $> -10.4$ & $-19.58 {}^{+0.03}_{-0.02}$ & $-19.9 \pm 0.1$ \\

$\beta$-slope & $-2.18 \pm 0.06$ & $-1.68 \pm 0.05$ & — & $-2.53 {}^{+0.06}_{-0.07}$ & $-3.3 \pm 0.3$ \\

\midrule
\multicolumn{5}{c}{\textbf{Flux [erg s$^{-1}$ cm$^{-2}$]}} \\
\midrule

Ly$\alpha$ & $(2.30 \pm 0.54) \times 10^{-17}$ & — & $(6.08 \pm 1.70) \times 10^{-19}$ & — & — \\

H$\alpha$ & $(1.646 \pm 0.032) \times 10^{-17}$ & $(1.38 \pm 0.02) \times 10^{-17}$ & $(2.07 \pm 0.25) \times 10^{-19}$ & — & — \\

H$\beta$ & $(5.44 \pm 0.37) \times 10^{-18}$ & $(4.3 \pm 0.3) \times 10^{-18}$ & $(7.3 \pm 2.0) \times 10^{-20}$ & $(7.1 \pm 1.0) \times 10^{-19}$\textsuperscript{\textcolor{red}{(g)}} & $(1.069 \pm 0.084) \times 10^{-18}$ \\

H$\gamma$ & — & $(2.8 \pm 0.3) \times 10^{-18}$ & —  & — & — \\

\ion{He}{2}~$\lambda1640$ & $(8.8 \pm 1.8) \times 10^{-18}$ & — & $\lesssim 1.85 \times 10^{-19}$\noindent\textsuperscript{\textcolor{red}{(f)}} & $(1.20 \pm 0.22) \times 10^{-18}$ & — \\

\ion{He}{2}~$\lambda4686$ & — & — & $< 1.6 \times 10^{-19}$ & — \\

[\ion{O}{3}]~$\lambda5007$ & $(5.96 \pm 0.22) \times 10^{-17}$\textsuperscript{\textcolor{red}{(a)}} & $(2.96 \pm 0.04) \times 10^{-17}$\textsuperscript{\textcolor{red}{(d)}} & $(5.0 \pm 1.5) \times 10^{-20}$ & $(3.90 \pm 0.10) \times 10^{-18}$ & $(3.854 \pm 0.120) \times 10^{-18}$ \\

\midrule
\multicolumn{5}{c}{\textbf{EW [Å]}} \\
\midrule

Ly$\alpha$ & $19.2$ & — & $> 250$ & — & — \\

H$\alpha$ & $166.5$ & $734 \pm 33$ & $> 1800$ & — & — \\

H$\beta$ & $26.6$ & $137 \pm 13$ & $> 340$ & $202 \pm 34$ & — \\

H$\gamma$ & — & $35 \pm 4$ & — & — & — \\

\ion{He}{2}~$\lambda1640$ & $8.3$ & — & — & $21 \pm 4$ & — \\

\ion{He}{2}~$\lambda4686$ & — & — & — & $< 49$ & — \\

[\ion{O}{3}]~$\lambda5007$ & $248.8$ & $946 \pm 46$\textsuperscript{\textcolor{red}{(d)}} & — & $1015 \pm 83$ & — \\

\midrule
\multicolumn{5}{c}{\textbf{FWHM [km s$^{-1}$]}} \\
\midrule

Ly$\alpha$ & $758 \pm 90$ & — & — & — & — \\
H$\alpha$ & $268 \pm 41$ & — & — & — & — \\
H$\beta$ & $320 \pm 137$ & — & — & — & — \\
\ion{He}{2}~$\lambda1640$ & $573 \pm 191$ & — & — & — & — \\

\bottomrule
\textbf{References} & \citet{Mondal+25} & \citet{Reumert+26} & \citet{Nakajima+25} & \citet{Wang+24} & \citet{Cullen+25} \\
\bottomrule
\end{tabular}
\end{adjustbox}

\begin{minipage}{\textwidth}
\tiny
\centering

\noindent\textsuperscript{\textcolor{red}{(a)}} [\ion{O}{3}]~$\lambda5007$ line flux from Keck/MOSFIRE, around two times higher than measured with NIRSpec, $(2.986 \pm 0.052) \times 10^{-17} ~\si{erg.s^{-1}.cm^{-2}}$, possibly due to incomplete MSA coverage; the corresponding [\ion{O}{3}]/H$\beta$ ratio is computed consistently with the NIRSpec/MSA flux measures.

\noindent\textsuperscript{\textcolor{red}{(b)}} From photometric analysis, averaged over the last 10 Myr; from spectroscopic analysis: $\mathrm{SFR}_\mathrm{UV} = (9.8 \pm 0.1) ~\mathrm{M_\odot yr^{-1}}$, $\mathrm{SFR}_\mathrm{H\beta} = (7.6 \pm 0.4) ~\mathrm{M_\odot yr^{-1}}$, $\mathrm{SFR}_\mathrm{H\alpha} = (7.5 \pm 0.1) ~\mathrm{M_\odot yr^{-1}}$, $\mathrm{SFR}_\mathrm{Pa\beta} = (6.40 \pm 0.03) ~\mathrm{M_\odot yr^{-1}}$.

\noindent\textsuperscript{\textcolor{red}{(c)}} The authors also examined a companion that serendipitously lies across the slit mask (therefore dubbing the whole system ``banana and blueberry'' due to its morphology, possibly hinting to a disturbed, merger-like origin), finding clear signatures of H$\alpha$ emission, as well as potential \ion{He}{1} and \ion{He}{2} emission lines; however, they caution that photometric data for the ``blueberry'' system strongly favor a lower-redshift solution at $z \sim 0.7$.

\noindent\textsuperscript{\textcolor{red}{(d)}} Total flux of the [\ion{O}{3}]$~\lambda \lambda$4959,5007 doublet.

\noindent\textsuperscript{\textcolor{red}{(e)}} The authors further highlighted an elevated carbon-to-oxygen ratio of $\sim 1 - 2$ times higher than the solar value, consistent with Pop~III nucleosynthetic yields (Section~\ref{sec:CEMPs}). This observation, together with the exceptionally hard ionizing radiation field, inconsistent with chemically enriched stellar populations or accreting black holes, establishes LAP1 as a candidate ``self-enriched'' Pop~III system or ``fossil in the making'' (Section~\ref{sec:spectral-hardness}).

\noindent\textsuperscript{\textcolor{red}{(f)}} \citet{Vanzella+23} previously estimated a line flux for \ion{He}{2}~$\lambda1640$ of $(7.96 \pm 2.07) \times 10^{-19} ~\mathrm{erg \, s^{-1} \, cm^{-2}}$; however, the reliability of the \ion{He}{2} line detection was hampered by the presence of a small blueshift relative to the Balmer lines, and by the extreme required EW ($\gtrsim 200$ Å), therefore the authors safely considered the line undetected, placing a $1\sigma$ upper limit of $2.7 \times 10^{-19} ~\mathrm{erg \, s^{-1} \, cm^{-2}}$. While \citet{Nakajima+25} also report no clear detection of the line, they provide an upper limit that remains within the range expected for zero-metallicity stellar populations. Also note that the stellar continuum is undetected for this source.

\noindent\textsuperscript{\textcolor{red}{(g)}} Intrinsic line fluxes and upper limits are reported after applying corrections for lensing magnification and dust extinction, adopting $A_\mathrm{V} = 0.12 \pm 0.04$ from their spectro-photometric analysis.\par

\noindent\textsuperscript{\textcolor{red}{(h)}} Assuming an extended star-formation history in which the stellar mass is built up steadily, but with a recent ~3 Myr burst of star formation, forming a mass of $(2.24_{-2.13}^{+1.48}) \times 10^7 ~\mathrm{M_\odot}$.\par

An up-to-date list of discovered Pop III candidates is maintained at the following page: \url{https://utcfc.github.io/pages/popiiicandidates.html}.

\end{minipage}

\label{tab:PopIII_candidates_hybrid}
\end{table*}

On the other hand, while no systematic search for ``hybrid'' Pop~III systems has been performed so far, the early analysis of \citet{JimenezHaiman06} suggested that a fraction of $\sim 20 - 30\%$ of Pop~III stars could explain the \ion{He}{2}~$\lambda1640$ and Ly$\alpha$ emission observed in a stacked galaxy sample at $z \sim 3$ \citep{Steidel+01}. 
\citet{Cassata+13} identified a population of narrow \ion{He}{2} emitters at $2 < z < 4.6$ that are incompatible with expectations from WR and AGN models, while they may be explained by a Pop~III SFR of $\sim 0.1 - 10 ~\si{\Msun.yr^{-1}}$, resulting in a tentative constraint on the Pop~III SFRD of $\sim 10^{-6} ~\si{\Msun.yr^{-1}.cMpc^{-1}}$ at $z \sim 3$.
First claims of Pop~III detection \citep{Sobral+15} in a luminous $z \approx 6.6$ Ly$\alpha$ emitter, CR7, were not corroborated by later scrutiny \citep{Bowler+17, Shibuya+18, Sobral+19, Matthee+20, Marconcini+25}. Other early candidates with unusually strong Ly$\alpha$ emission \citep{MalhotraRhoads02, Yamada+05} and very blue UV slopes at $z \sim 7$ \citep{Bouwens+10} -- possibly indicative of very metal-poor stellar populations and/or unusual IMFs -- have also been later explained with young, dust/metal-poor stellar populations and nebular effects, without requiring metal-free stars or non-standard IMF assumptions \citep{SchaererdeBarros10, Finkelstein+10}.

Recent JWST observations have renewed interest in this search. Evidence for unusually hard stellar populations, possibly associated with Pop~III-like components even in the presence of metal enrichment, has been reported for several high-$z$ galaxies, e.g. the RX J2129-z8HeII \citep{Wang+24} and EXCELS-63107 \citep{Cullen+25} systems at $z \approx 8.1$, the CAPERS-UDS-32520 system at $z \approx 5.1$ \citep{Reumert+26}, and the GNHeII J1236+6215 system at $z \approx 3$ \citep{Mondal+25}; the properties of these candidates are summarized in Table~\ref{tab:PopIII_candidates_hybrid}. However, alternative explanations for the hard emission in these sources are also possible (including X-ray binaries, VMSs, WR/stripped He stars, AGN and shocked gas, see e.g. the discussion in Section~\ref{sec:spectral-hardness}).

\vspace{.6em}
\subsection{Efficient Pop~III formation in the halo of massive galaxies within large-scale overdensities}
\label{sec:massive_halos}

{
\sidecaptionvpos{figure}{c}
\begin{SCfigure}
    \centering
    \includegraphics[width=.55\linewidth]{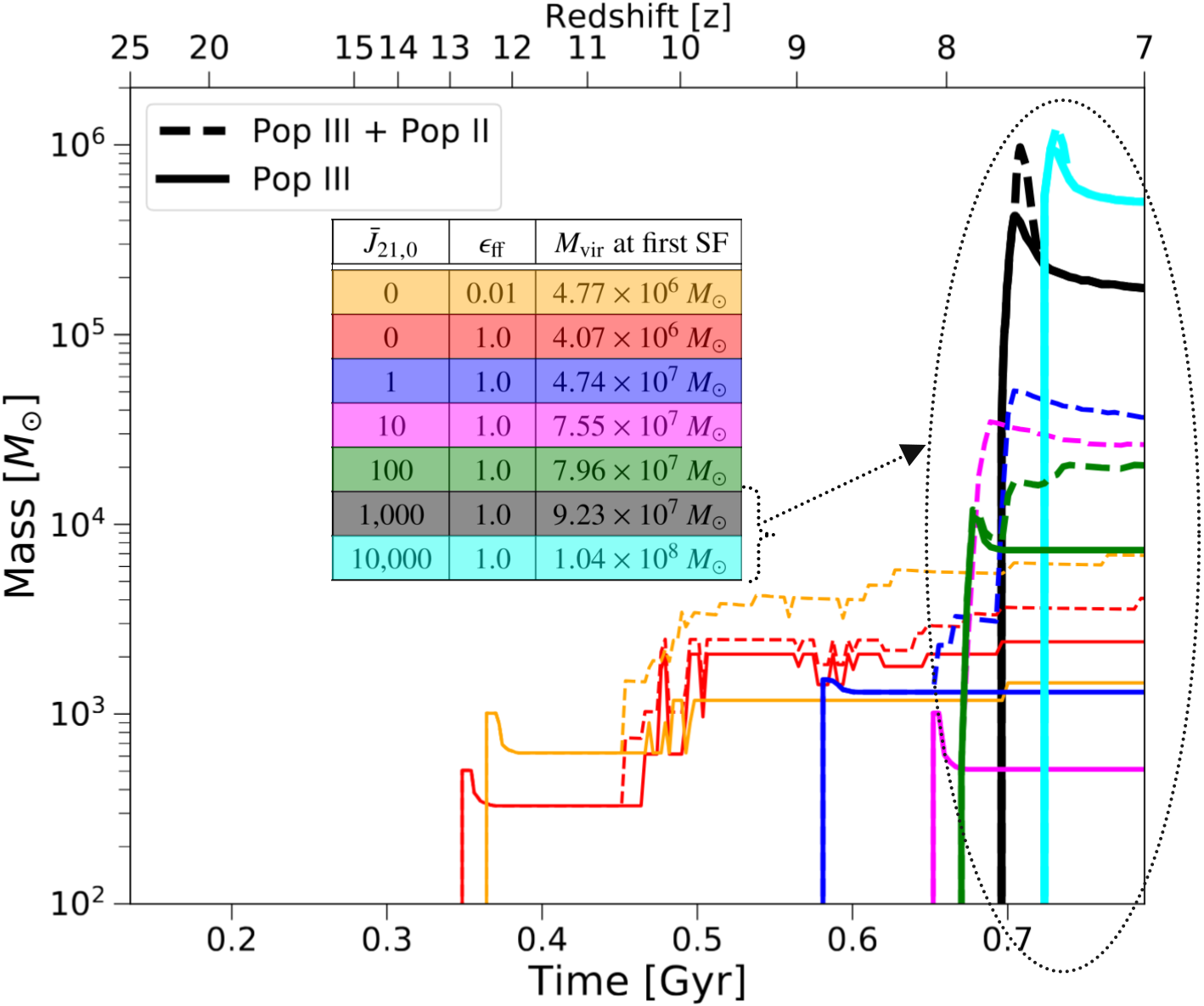}
    \caption{Stellar mass evolution (including Pop~III -- \textit{solid} -- and total -- \textit{dashed} -- star formation) within the virial radius of the same halo exposed to different levels of LW radiation (with an average intensity $\overline{J}_{21,0}$, in units of $10^{-21} ~\si{erg.s^{-1}.cm^{-2}.Hz^{-1}.sr^{-1}}$) and different assumed star-formation efficiencies ($\epsilon_\mathrm{ff}$, roughly quantifying how fast gas particles are converted into stars after they satisfy the star-formation criterion, relative to the free-fall timescale), highlighted in different \textit{colors}. When the halo is immersed in a strong LW background ($\sim 10^3 - 10^4 ~\si{J_{21}}$), star formation is delayed until the halo reaches a virial mass of $M_\mathrm{vir} \sim 10^8 ~\Msun$ and it becomes self-shielded to the radiation, resulting in massive late Pop~III starbursts up to $\sim 10^6 ~\Msun$. A fast transition to metal-enriched star formation is also observed mid-burst in these cases, demonstrating efficient self-pollution of the halo. Adapted from \citet{Jeong+26}.}
    \label{fig:late_PopIII_starburst_ACH}
    \vspace{1em}
\end{SCfigure}
}

While cosmic voids have long remained the main focus of searches for late Pop~III formation, due to their slower enrichment \citep[e.g.][]{Rowntree+24, Rodriguez-Medrano25}, recent studies suggest that Pop~III stars at late times could form in halos much more massive than the first minihalos \citep{Kulkarni+19, LiuBromm20, BennettSijacki20, Riaz+22, Storck+26}, including globally enriched halos within large-scale overdensities \citep{Venditti+23, HegdeFurlanetto25}.

Although intrinsically rare and likely subdominant in number, these systems may still be compelling due to their particular environmental conditions favoring efficient Pop~III star formation.
In fact, theoretical studies suggest that atomic-cooling halos immersed in a strong LW background (produced by a large concentration of nearby sources, e.g. \citealt{Trinca+26}) could experience a delayed onset of star formation, potentially leading to more massive starbursts and correspondingly stronger observational signatures than typical minihalos, due to efficient HD cooling (\citealt{GreifBromm06, Greif+08, Bromm+09, Sugimura+24, Jeong+26}, see e.g. Figure~\ref{fig:late_PopIII_starburst_ACH}).
\citet{Venditti+23} further indicated that the relative incidence of Pop~III activity may actually be higher in massive halos, despite their rarity: in fact, when pristine gas survives within overdense large-scale environments, dynamical interactions with nearby structures and satellites could promote gas compression, thereby triggering star formation within the pristine clumps \citep[e.g.][]{CorreaMagnus+24}.

\begin{figure*}
    \centering
    \includegraphics[width=\linewidth]{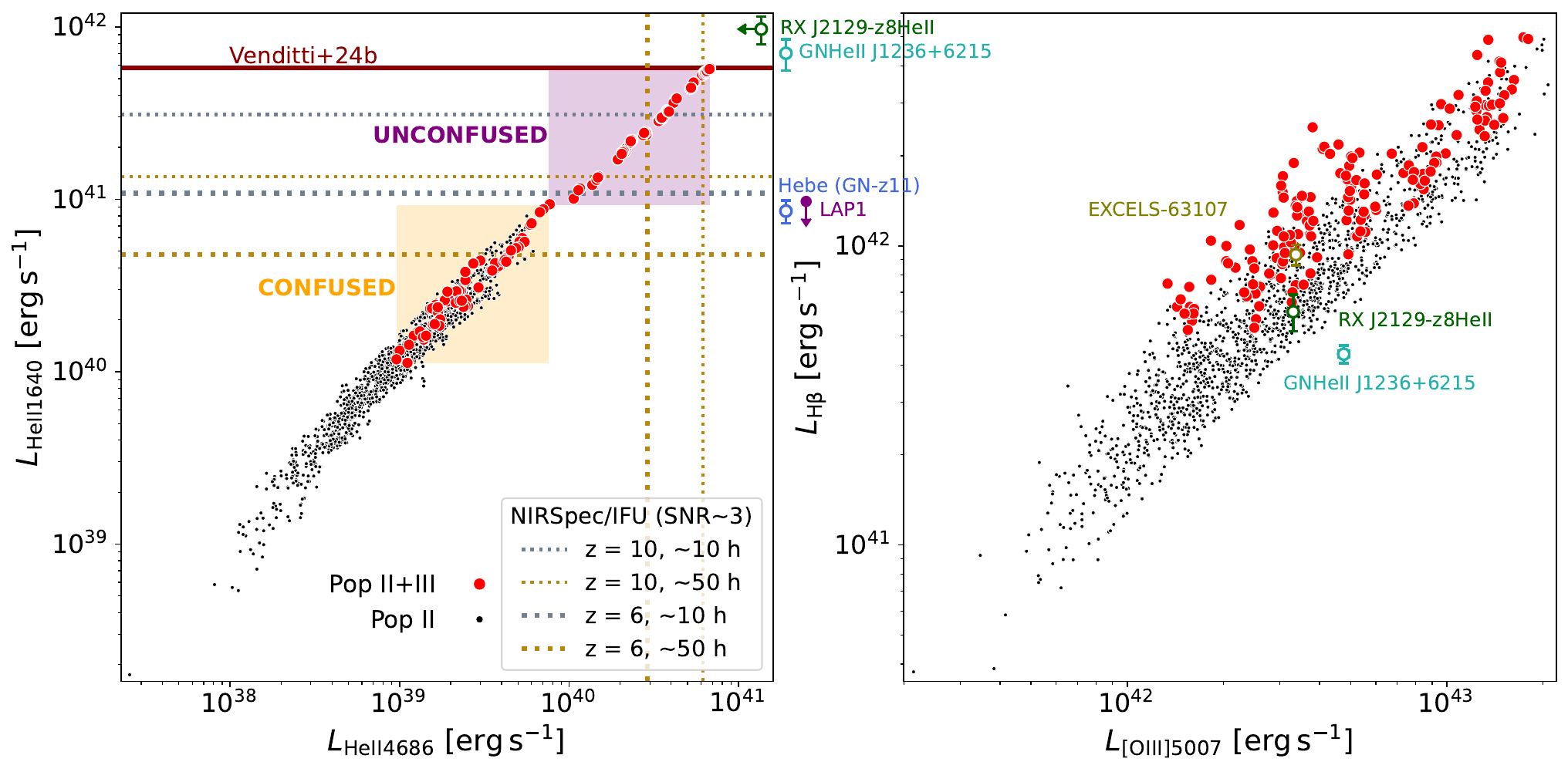}
    \caption{\textbf{Left:} Integrated \ion{He}{2} line luminosity at 1640~\AA ~($L_\mathrm{HeII1640}$) vs 4686~\AA ~($L_\mathrm{HeII4686}$) arising from massive halos ($M_\mathrm{vir} \gtrsim 10^{11} ~\Msun$, $M_\star \gtrsim 10^9 ~\Msun$) at $z \approx 6.5 - 9$ that host a sub-dominant Pop~III component, from the \texttt{dustyGadget} simulation suite \citep{Graziani+20, DiCesare+23, Venditti+23} (\textit{red circles}), compared with contributions from the dominant Pop~II component in these halos, as well as benchmark Pop~II-only halos (\textit{black dots}).
    Young ($\lesssim 1$ Myr), massive ($M_\mathrm{III} \sim 6 \times 10^5 ~\Msun$) Pop~III clusters in these environments are associated with strong \ion{He}{2}~$\lambda1640$ line emission ($L_\mathrm{HeII1640} \gtrsim 10^{41} ~\si{erg.s^{-1}}$), which cannot be produced by standard Pop~II populations alone (``unconfused'' \textit{purple shaded area}); older Pop~III populations may instead be confused with Pop~IIs (\textit{orange shaded area}).
    \textbf{Right:} same halos in the $L_\mathrm{H\beta} - L_\mathrm{[OIII]5007}$ plane, demonstrating that these systems are bright oxygen line emitters despite hosting active Pop~III formation, due to the dominant Pop~II component in the environment.
    Adapted from \citet{Venditti+26}. A comparison with observed ``pure'' (Table~\ref{tab:PopIII_candidates_metal-poor}) and ``hybrid'' (Table~\ref{tab:PopIII_candidates_hybrid}) Pop~III galaxy candidates is provided; note that for GNHeII-J1236+6215, Hebe and LAP1, no constraints are available on the \ion{He}{2}~$\lambda4686$ line, therefore measures of the \ion{He}{2}~$\lambda1640$ line luminosity are shown out of the $x$-axis boundaries of the first plot to avoid confusion.}
    \vspace{4em}
    \label{fig:HeII-Hbeta-OIII_luminosity_massive_PopIII_halos}
\end{figure*}

The enhanced signal from massive Pop~III starbursts could remain detectable at lower sensitivities, and be comparatively easier to identify -- even more so if occurring at intermediate redshifts. \citet{Venditti+26} recently investigated the detectability of a subdominant Pop~III component in/around massive ($M_\star \gtrsim 10^9 ~\Msun$) galaxies at $6.5 \lesssim z \lesssim 9$ from the \texttt{dustyGadget} cosmological simulation suite \citep{Graziani+20, DiCesare+23, Venditti+23}, finding that young ($\lesssim 1$ Myr), massive ($M_\mathrm{III} \sim 6 \times 10^5 ~\Msun$) Pop~III clusters in these environments are associated with strong \ion{He}{2} line emission that cannot be produced by standard Pop~II populations alone (see the left panel of Figure~\ref{fig:HeII-Hbeta-OIII_luminosity_massive_PopIII_halos})\footnote{Note that, with respect to the ``Pop~III-poor'' galaxies analyzed by \citet{Rusta+25} (Figure~\ref{fig:self-polluted+hybrid_PopIII_galaxies}), in which the subdominant, aging Pop~III populations can no longer power bright \ion{He}{2} emission, here Pop~III stellar particles can actively form in metal-poor pockets within the halos even after the onset of metal-enriched star formation. These young populations can raise the overall \ion{He}{2} brightness substantially, despite their low relative contribution in mass.}. Even brighter \ion{He}{2} luminosities may be achieved in the case of chemically homogeneous stellar evolution, as predicted for fast rotating Pop~III stars \citep{Sibony+22, Wasserman+26}. However, the role of more exotic hard sources that may be confused with Pop~IIIs (listed in Section~\ref{sec:spectral-hardness}) remains to be examined, as well as the effect of dust absorption and scattering through the ISM. A strong variability in the dust columns crossed by different lines-of-sight to the Pop~III sources was found in \citet{Venditti+23}, spanning values from $\Sigma_\mathrm{dust} \sim 10^{-3} ~\si{\Msun.kpc^{-2}}$ up to $\Sigma_\mathrm{dust} \sim 10^6 ~\si{\Msun.kpc^{-2}}$ even within a single simulated galaxy. For Pop~III clumps at significant distances from the central massive galaxy (up to $\sim 20$~kpc), dust attenuation may preferentially affect stellar populations lying in the most polluted regions, reducing their confusing \ion{He}{2} signal. Some of the most isolated Pop~III-forming pockets may be even detectable in a single spaxel of NIRSpec/IFU, further aiding their identification \citep{Venditti+26}.

On the other hand, a dominant Pop~II component within massive ``hybrid'' Pop~III hosts is expected to power strong metal line emission (right panel of Figure~\ref{fig:HeII-Hbeta-OIII_luminosity_massive_PopIII_halos}, also \citealt{Cleri+23, Rusta+25} in a lower-mass regime, see Section~\ref{sec:spectral-hardness}). Confusion from metal-enriched stellar populations may be even more important in massive halos hosting intense Pop~III starbursts, as e.g. \citet{Jeong+26} predicted a fast transition to Pop~II formation that happens mid-starburst, demonstrating efficient self-pollution in this scenario (Figure~\ref{fig:late_PopIII_starburst_ACH})\footnote{Note that, under these conditions, the first star-formation episode occurs when the halo has reached a mass of $\sim 10^8 ~\Msun$, so that the deeper potential well and the large reservoir of cold dense gas allow part of the gas to survive the first SN explosions, enabling a faster transition with respect to typical minihalo scenarios \citep[e.g.][]{Jeon+14, ChiakiWise19, Chiaki+20}.}. 
While these results may be affected by the limited mass resolution\footnote{Particularly, pre-enrichment of massive halos and their associated pristine pockets, caused by an unresolved population of minihalos at earlier epochs, may reduce the number of these systems that can preserve their metal-free nature at late cosmic epochs (see e.g. \citealt{TrentiStiavelli09, Johnson+10, Regan+20, Hicks+21, Hicks+24}).},
they further indicate that the detection of metal lines alone cannot exclude the presence of Pop~III stars in high-$z$ galaxy environments, and that identifying ``hybrid'' or ``self-polluted'' Pop~III hosts is essential for a complete census of Pop~III star formation across cosmic time.

We note that the exact rate of Pop~III star formation at $z \lesssim 10$ remains largely uncertain, with predictions spanning more than three orders of magnitude, from $\sim 10^{-6} ~\si{\Msun.yr^{-1}.cMpc^{-3}}$ to $\sim 10^{-3} ~\si{\Msun.yr^{-1}.cMpc^{-3}}$ (Figure~\ref{fig:PopIII_SFRD}). 
The pace and topology of metal enrichment are also poorly understood. Recent results from the \texttt{AGORA} collaboration \citep{Kim+26} showed that, when simulating the same massive halo at $z \approx 10$ starting from analogous initial conditions, and with sub-grid physics models calibrated to converge on global star-formation metrics, different hydrodynamical codes produce vastly different predictions in terms of the spatial distribution of dust and metals, reflected in the observed colors and morphology of mock images.
Accordingly, the very existence of pristine gas pockets in late, massive hosts remains far from settled. High-resolution, zoom-in studies from the \texttt{THESAN-ZOOM}\footnote{Note that Pop~III star formation and feedback are not explicitly modeled in the \texttt{THESAN-ZOOM}, but the Pop~III identification criterion relies instead on star-formation happening close to the imposed metallicity floor ($\sim 10^{-7} ~\Zsun$). Therefore, these predictions should be interpreted with caution.} \citep{Zier+25} and \texttt{MEGATRON} \citep{Storck+26} simulation suites suggest that, on the contrary, efficient metal mixing may suppress Pop~III formation in the vicinity of bright galaxies (but also see \citealt{Pan+13} for an implementation of sub-grid turbulent mixing in \texttt{RAMSES}, designed for applications in lower-resolution, larger simulated volumes, e.g. \citealt{Sarmento+18}). 
Evidence of a pervasive enrichment at the ISM scale has been reported up to the highest redshifts (Section~\ref{sec:low-metallicity}). On the other hand, the chemical nature of the CGM and IGM at high redshifts remains much more elusive, and mainly explored through metal-line absorption along rare quasar sightlines (Section~\ref{sec:PISNe}).
In this uncertain landscape, identifying isolated regions with ongoing Pop~III star formation would provide a direct way to locate pockets of pristine gas, and constrain how efficiently metals are mixed into the CGM/IGM around early galaxies.

\begin{figure}
    \centering
    \includegraphics[width=\linewidth]{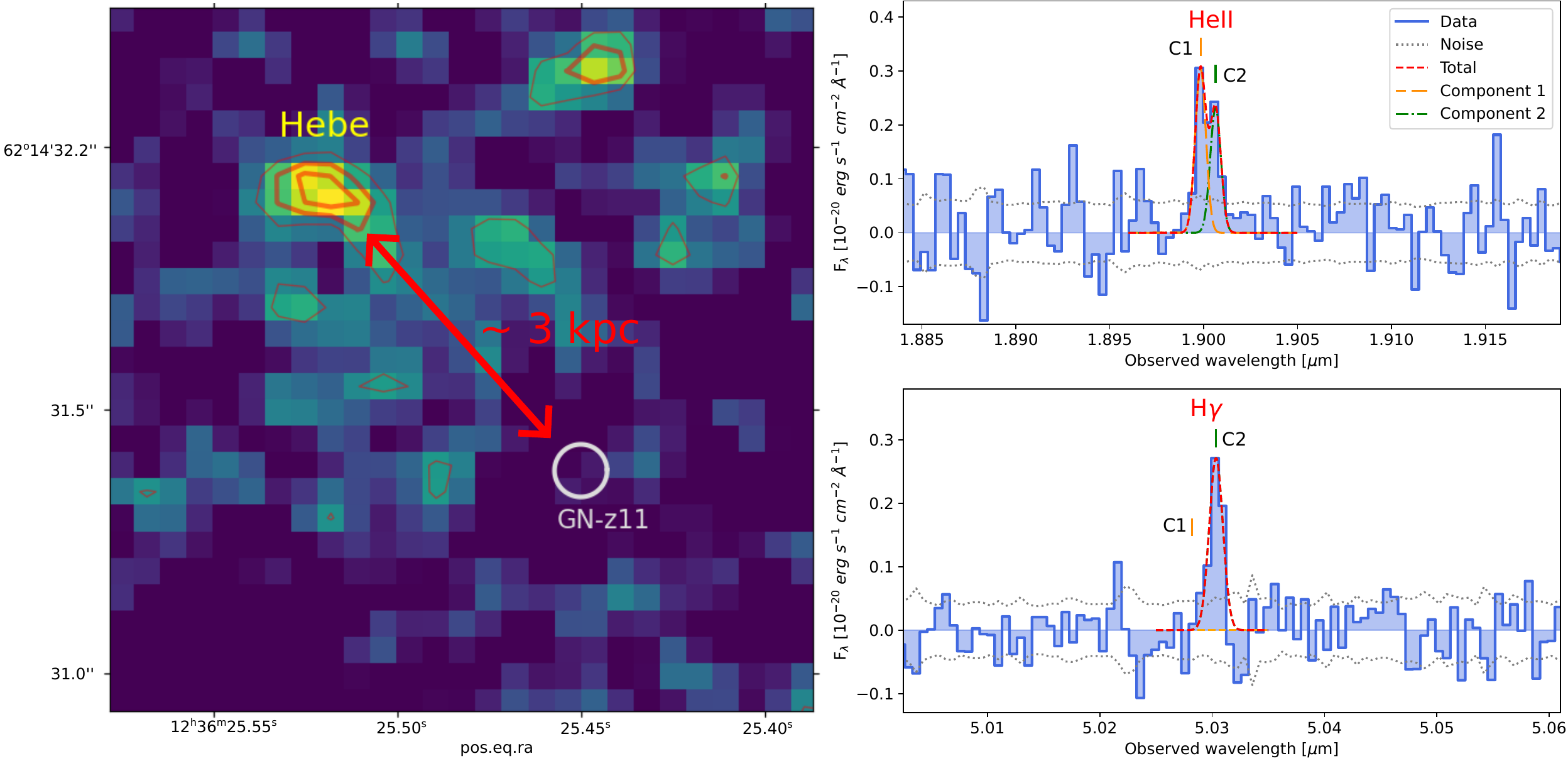}
    \caption{\textbf{Left:} continuum-subtracted map of the \ion{He}{2} emission at the redshift of Hebe in the halo of GN-z11, with contours indicating 3$\sigma$ (\textit{thin}), 4$\sigma$ and 5$\sigma$ (\textit{thick}) levels, and the \textit{white circle} the location of GN-z11. 
    \textbf{Right:} Hebe’s spectrum around the wavelength of the \ion{He}{2}~$\lambda1640$ (\textbf{top}) and H$\gamma$ lines (\textbf{bottom}, from \citealt{Ubler+26}), with \textit{colored lines} showing the simultaneous fit of two spectrally resolved components for the \ion{He}{2} line and their total (flux scale not corrected for aperture losses or lensing magnification). Adapted from \citet{Maiolino+26}.}
    \label{fig:Hebe+GN-z11}
    \vspace{2em}
\end{figure}

A number of systems possibly consistent with Pop~III formation within massive, enriched hosts have been reported in the literature. Very recently, the pristine nature of a strong and metal-poor \ion{He}{2} emitter identified by \citet{Maiolino+24b} has been further corroborated (\citealt{Maiolino+26, Ubler+26, Rusta+26}, also see Table~\ref{tab:PopIII_candidates_metal-poor} for its properties). This system, named Hebe, has been identified at $\sim 3$~kpc from the bright galaxy GN-z11 at $z \approx 10.6$ \citep{Bunker+23}, and it exhibits strong H$\gamma$ emission with no detected metal lines (see table 1 from \citealt{Maiolino+26}, as well as Figure~\ref{fig:Hebe+GN-z11} for a continuum-subtracted map of this system and the spectrum extracted around the wavelengths of the \ion{He}{2}~$\lambda1640$ and H$\gamma$ lines). Theoretical models by \citet{Rusta+26} and \citet{Jeon+26b} show that Hebe is consistent with a Pop~III mass of $\sim 2 \times 10^4 - 6 \times 10^5 ~\Msun$ formed in a pristine satellite of a massive halo, while its large distance from GN-z11 is incompatible with ionization from the central AGN proposed by \citet{Maiolino+24a} for this source.

A large Pop~III mass of $(7.8 \pm 1.4) \times 10^5 ~\Msun$ has also been suggested to explain the strong \ion{He}{2} emission observed in the ``hybrid'' Pop~III candidate RXJ2129-z8HeII at $z \approx 8.2$ (\citealt{Wang+24}, see Table~\ref{tab:PopIII_candidates_hybrid}), which however is not consistent 
with the predicted properties of hybrid Pop~III systems by \citet{Rusta+25}.
Another tentative Pop~III candidate (A370-z6LAE-1c) was presented by \citet{Fujimoto+25b} within the strongly lensed Ly$\alpha$ emitter A370-z6LAE-1 at $z \approx 5.9$ \citep{Claeyssens+22}, only partially satisfying the selection criterion proposed by \citet{Fujimoto+25a, Fujimoto+25b}: this object is consistent with Pop~III model tracks of intermediate age ($\approx 10 - 30$~Myr), with diminished H$\alpha$ strength, while still presenting some degeneracy with metal-enriched galaxies, and it is best fit by a low-mass $\approx 10^{6.6} ~\Msun$ stellar cluster hosted by a $\approx 10^8 ~\Msun$ galaxy, with very weak [\ion{O}{3}] emission. The host system exhibits an intrinsically elongated morphology spanning $\sim 4$~kpc, with A370-z6LAE-1c lying between the brightest core clump and two tail clumps (at only $\sim 1$~kpc separation from enriched neighbors), consistent with being either a metal-poor pocket within the host galaxy of A370-z6LAE-1 or an infalling metal-poor satellite.

The study of \citet{Venditti+25} indicates that efficient Pop~III formation -- possibly within heavy, atomic-cooling halos -- may be required to explain bright Pop~III candidates such as the AMORE6 galaxy at $z \approx 5.7$, based on preliminary constraints on the Pop~III UVLF at $z \sim 6$ (\citealt{Fujimoto+25a, Fujimoto+25b}, Figure~\ref{fig:PopIII_UVLF_obs}).
\citet{Koller+26} identified three low-metallicity satellites across seven low-metallicity galaxies at $z \approx 7.2 - 9.5$, which (while not metal-free) exhibit significant deviations from the local fundamental metallicity relation, possibly indicating strong accretion of pristine gas. They also observed a large scatter in radial metallicity gradients in these systems, further suggesting that early galaxies grow through structurally diverse and episodic processes, with a complex interplay between feedback/interaction-driven mixing processes driving the observed gradients.
Finally, \citet{Li+26} recently associated a Pop~III-like absorber at $z \approx 5.9$ with an unusually metal-poor galaxy overdensity. The absorber-galaxy cross-correlation suggests that the Pop~III-enriched gas traces the outskirts of a relatively massive atomic-cooling halo ($\gtrsim 10^9~\Msun$), and its fossil abundance pattern may originate either from delayed Pop~III formation in pristine regions of the halo, or from earlier Pop~III enrichment in a low-mass satellite or minihalo subsequently incorporated into the larger system.

These findings highlight the potential of spatially resolved searches, possibly aided by gravitational lensing\footnote{Also see Section~\ref{sec:lensed_stars} for a discussion of the detectability of single, lensed Pop~III stars.}. On the other hand, large distances from the central galaxy may cause Pop~III complexes to be overlooked in observations with a limited field-of-view (FoV) centered around bright, massive galaxies. \citet{Venditti+24b} predicted that $\sim 90\%$ or more of the Pop~III clusters forming in hybrid systems would be missed by NIRSpec/MOS pointings (with a narrow microshutter width of 0.2~arcsec) towards the central, massive galaxy; this would have been the case -- with an unlucky orientation of the instrument -- for the Hebe system, which is instead entirely within reach of NIRSpec/IFU pointings. Even larger areas will be covered through the HARMONI spectrograph of the Extremely Large Telescope (ELT), with planned integral-field modes spanning spatial samplings from a few to several tens of milliarcseconds and fields of view up to several arcseconds\footnote{\url{https://elt.eso.org/instrument/HARMONI/}}.

\vspace{.6em}
\subsection{Towards the discovery of individual lensed Pop~III stars}
\label{sec:lensed_stars}

Gravitational lensing can render individual stars detectable beyond the local Universe, and candidates for such lensed stars have currently been detected up to $z \approx 5 - 6$ \citep{Welch22a,Welch22b,Meena23,Furtak24}. While the very high magnifications required ($\mu\sim 10^2 - 10^4$) can be reached along many different sightlines to the high-redshift Universe \citep[e.g.][]{Zackrisson15}, cluster-lens fields are convenient targets since the critical curve on the plane of the sky is long \citep[$\sim 100$ arcsec;][]{Windhorst18}, usually well-separated from bright galaxies within the lens, and contained within an area in the sky that can be covered in a small number of HST or JWST pointings. Factors that affect the probability of having individual Pop~III stars lensed to detectable flux levels in such fields include the Pop~III SFRD($z$), the Pop~III IMF, and the intrinsic brightness of these stars in different evolutionary stages \citep{Rydberg13,Zackrisson24}. 

At evolutionary stages close to the Zero Age Main Sequence (ZAMS), where the effective temperature ($T_\mathrm{eff}$) may be as high as $\sim 10^5$~K for a star $> 100 ~\Msun$ \citep[e.g.][]{Schaerer02}, the magnifications required to lens a single Pop~III star at $z>6$ into a range accessible with JWST  ($\lesssim 30 - 31$~AB mag for very deep imaging) are estimated at $\mu \gtrsim 1000$ for masses $\sim 10^2 - 10^3 ~\Msun$, and $\mu \gtrsim 10^4$ for Pop~III stars $<100 ~\Msun$ \citep[e.g.][]{Larkin23}\footnote{We note that rapidly accreting SMSs of $\sim10^4 - 10^5 ~\Msun$ (see e.g. Section~\ref{sec:N-emitters}) are predicted to develop highly inflated, luminous envelopes, potentially making them directly detectable by JWST without strong gravitational lensing \citep{Surace+18, Surace+19}. However, their poorly constrained formation rate and short lifetimes complicate predictions on the number of observable SMSs, which remain highly uncertain.}.
At later stages of evolution, where the Pop~III star may have evolved to much lower $T_\mathrm{eff}$, the star becomes brighter \citep[when measured in $f_\nu$ units, like AB magnitudes;][]{Windhorst18,Zackrisson24,Hassan25}  and requires correspondingly lower magnifications ($\mu<1000$ for a Pop~III star of $\sim 100 ~\Msun$ at $z=6$).
While the lower-$T_\mathrm{eff}$ states of Pop~III stars on the late main sequence and beyond are very short-lived, the magnification probability distribution at the high-$\mu$ end (typically $P(>\mu)\propto \mu^{-2}$, although microlensing may alter this behaviour; \citealt{Palencia24}) can lead to a lensing bias which renders stars in these short-lived, low-$T_\mathrm{eff}$ states more likely to be detected, despite being intrinsically less common in a given cosmological volume \citep{Zackrisson24}.

Detailed calculations of the detectability of lensed Pop~III stars in different evolutionary states suggest that the probability of a $z\gtrsim 6$ Pop~III star $\leq 500 ~\Msun$ being lensed to brighter than $\sim 28$ AB mag in a single cluster-lens field is, at best, at the \%-level \citep{Zackrisson24}. However, in a multi-cluster survey like the Cycle 4 JWST program VENUS (\citealt{Fujimoto+25_VENUS}, including 60 cluster-lens fields imaged to 28 AB mag), the probability of achieving $\sim 1$ detection within the whole survey can become substantial. Based on the Pop~III models used in \citet{Zackrisson24}, the most probable properties of a star detected this way would be $z\sim 6$, mass $>200\ ~\Msun$ and $T_\mathrm{eff}<15000$ K.  This object would have a very red SED across the JWST/NIRCam filters, and a peak brightness close to the 28 AB mag threshold. The object is expected to have a magnification in the hundreds, and should appear as two resolvable mirror images, which may have slightly different brightness (and may hence not both be detectable) that would moreover vary independently in brightness due to microlensing in the cluster lens. Follow-up spectroscopy and deeper imaging would likely be required to distinguish such a candidate from an LRD or some other very red point-like source \citep[for examples of JWST/NIRSpec data on F444W$\sim 28$ AB mag LRDs, see][]{deGraaff25}. If the lensed star is part of a star cluster, then this cluster may be visible as an extended object with a bluer SED around the lensed red star \citep[for an example SED of the compound object, see][]{Zackrisson26}. A nebula powered by hotter surrounding stars could produce detectable emission lines, which would help pin down the exact redshift and set limits on the gaseous metallicity. 

It must be stressed, however, that probability estimates of this type are highly sensitive to the assumed evolution of Pop~III stars across the HR-diagram, which is affected by rotation, binary evolution and the treatment of convection within these stars. Estimates of the magnification needed to achieve detection are also dependent on the assumed lensing model, with uncertainties that increase for increasing magnifications. Even if a candidate object is found, confirming it as a Pop~III star will be very difficult, since spectroscopic follow-up observations with JWST of a $\sim 28$ AB mag point source with $T_\mathrm{eff} < 15000$~K at $z \sim 6$ may not be able to detect many absorption lines, and such stars would be exceedingly faint at the wavelengths where ELT spectroscopy would perform better \citep{Lundqvist24}.
Alternatively, very deep imaging of a single cluster-lens field, like that provided by the GLIMPSE survey \citep[reaching $\approx 30.7$ ~AB mag in F444W;][]{Atek+25}, $\sim 10$ times deeper than VENUS, would make the single-field detection probability comparable to that of multi-field survey like VENUS and bring intrinsically faint stellar populations within reach \citep[e.g.,][]{Furtak26}.
The downside of this approach is that the lensed star would then be expected to appear at much fainter apparent brightness (close to the 30.7 AB mag limit), rendering follow-up spectroscopy near-impossible. The star would also have a 10 times lower magnification, making its identification as a lensed-star candidate much harder, and blending with nearby objects in the source plane more severe.

\vspace{.75em}
\section{Alternative searches for Pop~III stars at ultra-high redshifts ($\lowercase{z} > 10$)}
\label{sec:ultra-high_zs}

Finding the very first star-formation sites at Cosmic Dawn is extremely challenging, due to the intrinsic faintness of the first minihalos \citep[e.g.,][]{Schauer+20}.
While Hebe is currently the only Pop~III candidate proposed at $z \gtrsim 10$ (Section~\ref{sec:massive_halos}), a potential top-heavy IMF component -- typical of Pop~III stellar populations -- has been proposed to alleviate the tension between pre-JWST galaxy formation models and the observed abundance of UV-bright galaxies at $z \gtrsim 10$ \citep{Inayoshi+22, Finkelstein+23, Harikane+23, Harikane+24, Yung+24, Trinca+24, Ventura+24, Cueto+24, Hutter+25, Lu+25, Harvey+25, Jeong+25, Mauerhofer+25, Perez-Gonzalez+25}; we refer the reader to Section~\ref{sec:low-metallicity} for a more in-depth discussion of the metallicity dependence of the IMF in early galaxies. On the other hand, a variety of other solutions have been proposed to solve this apparent tension, including burstier star-formation histories \citep[e.g.,][]{Mason+23, Sun+23, Shen+23}, feedback-free starbursts \citep[e.g.,][]{Dekel+23}, density-modulated star-formation models \citep{Somerville+25}, and attenuation-free models \citep[e.g.,][]{Ferrara+23}; see e.g. the review from Fontana et al. in preparation from the same conference proceeding series for further details.

Time-domain searches provide a complementary route to search for Pop III stars at ultra-high redshifts, as their explosions may be substantially brighter than their stellar progenitors.
Massive PISNe may in principle remain detectable with JWST beyond $z \sim 10$, although the expected survey yield is limited by their low rates and by the small area covered by sufficiently deep, multi-epoch observations. Their intrinsic rates are also extremely uncertain: the number of stars forming in
the $\sim 140 - 260 ~\Msun$ interval conventionally associated with PISN
progenitors depends strongly on the assumed IMF, so that uncertainties in the Pop
III IMF (Section~\ref{sec:PopIII_IMF_theory}) compound those in
the Pop III star formation history (Section~\ref{sec:spectral-hardness}). Nonetheless, the fractional contribution of Pop III PISNe to the transient population should increase at high redshifts, where Pop III stars become more prevalent. Owing to the long duration of their prompt-emission light curve (hundreds of days in the rest-frame for the long-lasting emission associated with the release of thermal energy deposited by the shock, and with the following radioactive decay of synthesized $^{56}$Ni, e.g. \citealt{Kasen+11}), cosmological time dilation may extend their visibility to several years or even decades in the observer frame at high redshifts -- while also making their transient nature more difficult to establish over a limited observational baseline. 

By compiling predictions of \citet{Hummel+12}, \citet{Johnson+13} and \citet{Magg+16}, \citet{Hartwig+18b} reported largely different PISN intrinsic rates per JWST/NIRCam FoV ($4.84 ~\si{arcmin^2}$) at $z \leq 7.5$, varying from $6 \times 10^{-6} ~\si{yr^{-1}}$ to $2 \times 10^{-4} ~\si{yr^{-1}}$, and more consistent rates of $3 - 7 \times 10^{-4} ~\si{yr^{-1}}$ at $7.5 < z \leq 12$, yielding a detection rate up to $\sim 2-3 \times 10^{-5} ~\si{yr^{-1}.FoV^{-1}}$ from each of these redshift ranges with their proposed optimized 2-filter strategy (F200W-F356W with 600~s exposure for $z \leq 7.5$, and F200W-F277W with 12~ks exposure for $7.5 < z \leq 12$, respectively).
On the other hand, a substantially higher intrinsic rate of $\approx 26$ PISN events per year within a $\sim 1 ~\si{deg^{2}}$ survey area at $z > 8$ (which would translate to $\sim 3.5 \times 10^{-2} ~\si{yr^{-1}}$ in the considered JWST FoV) was found with the most favorable IMF considered by \citet{Venditti+24a} using the optimistic Pop III SFRD from \citet{Jaacks+19}, and variations up to a factor $\sim 60$ were predicted within a broad range of possible Pop III IMFs.
The much larger FoV of the Roman Wide Field Instrument (WFI) ($0.281~\si{deg^2}$) will enable substantially more efficient sampling of
rare transients, and a survey strategy combining F158 and F213 down to limiting magnitudes of 27.0 and 26.5 respectively (capable of discriminating between classical type Ia/II SNe at $z > 1.5$ and PISNe/super-luminous SNe at $z > 6$) was proposed by \citet{Moriya+22a}. 
However, this dedicated F213 strategy is not included in the currently adopted High-Latitude Time-Domain Survey, whose reddest imaging band is F184 and whose extended component is expected to discover and characterize long-duration super-luminous and pair-instability SNe possibly out to $z \sim 5$
\citep{RomanObservationsTimeAllocationCommittee+25}. On the other hand, the greater depth and longer-wavelength coverage of JWST may in principle allow exceptionally luminous PISNe to be detected even at $z \sim 30$. 
\citet{Jeon+26a} recently suggested a promising strategy to reveal early star-forming minihalos, by capturing PISNe in overdense fields with an accelerated star-formation history, in which bright $250 ~\Msun$ PISNe could reach $m_\mathrm{AB} \sim 28 - 29$ even at $z \sim 30$ and -- owing to cosmological time dilation -- remain bright for approximately two decades in the observer frame\footnote{By comparison, \citet{Venditti+24a} examined the visibility of various PISN progenitors at $z \sim 6$ across the best combination of JWST/NIRCam and Roman/WFI filters suggested by \citet{Hartwig+18b} and \citet{Moriya+22a} (considering existing JWST surveys with different limiting magnitudes, and the sensitivity for the Roman survey recommended by \citealt{Moriya+22a}, respectively), finding a visibility time window ranging from $\sim 2$~months up to $\sim 9$~yr for progenitors $\gtrsim 200 ~\Msun$.}.
A possible PISN interpretation (with a $250 - 260 ~\Msun$ metal-free progenitor, exploding at $z \approx 15$) has been initially discussed for the ``Capotauro'' source, an F356W-dropout identified in the CEERS survey as potentially consistent with a luminous galaxy at a best-fit redshift of $\sim 30$ \citep{Gandolfi+26a}, for which possible signs of variability over an $\sim 800$~day baseline have been subsequently revealed \citep{Ferrara+26}. 
However, alternative explanations, including low-$z$ interlopers (i.e. dusty galaxies at intermediate redshifts), a cool local brown dwarf, or a contaminated grism spectrum, have been presented as viable alternatives \citep{Gandolfi+26b, Bradac+26}, and recently, \citet{Liu+26} reported a non-zero proper motion for this source consistent with it being a foreground brown dwarf, leaving no high-$z$ PISN detection securely confirmed so far. The reader may also refer to Section~\ref{sec:PISNe} for other low-redshift PISN candidates proposed in the literature, and for a discussion of archaeological constraints.

PISNe could be revealed through their afterglow on much longer timescales, up to a hundred times the duration of their prompt emission. However, \citet{Hartwig+18b} predict that the PISN afterglow is not accessible through current observational facilities. Other non-PISN channels linked to Pop III explosions that may in principle be observable at ultra-high redshifts include interaction-powered core-collapse SNe, which can become super-luminous due to the interaction of the ejecta with a dense circumstellar shell \citep{Whalen+13c}, jet-driven SNe and gamma-ray bursts (GRBs) from rapidly rotating Pop III VMSs, predicted respectively for tracks that directly collapse into BHs, and for progenitors subject to pulsational pair instability \citep{Toma+16, LazarBromm22, Volpato+24}, and Pop III SMS explosions triggered by general-relativistic instability \citep{Jockel+26}.
Unfortunately, while JWST surveys have identified tens of SNe up to $z \sim 5$ \citep{Fox+26}, none of these has a Pop III interpretation. The highest-redshift SN candidate is the likely SN component following GRB 250314A at $z \approx 7.3$ \citep{Levan+25}, whereas the strongly lensed Eos SN at $z = 5.13$ currently represents the highest-redshift spectroscopically confirmed supernova \citep{Coulter+26}, consistent with a Type IIp SN exploded in a metal-poor pocket ($< 1 \% ~\Zsun$), albeit not metal-free \citep{Asada+26}. 
Bright GRB afterglows may remain detectable in the near-infrared and radio to $z \gtrsim 20$, and early absorption-line spectroscopy could probe the surrounding metal enrichment and potentially distinguish PISN from core-collapse SN yields \citep{Wang+12}. Their initial discovery, however, requires wide-field high-energy facilities. \citet{Wei+26} derive Pop III GRB upper limits at $z > 6$ of $< 0.06 ~\si{yr^{-1}}$ for Einstein Probe/WXT and $< 0.13 ~\si{yr^{-1}}$ for SVOM/ECLAIRs for a $100 - 500 ~\Msun$ Pop III IMF, and further predict that, although Pop III GRBs remain subdominant below $z \sim 10$, their contribution should increase towards $z > 16$, where any detection would represent a particularly strong Pop III candidate (albeit with a very low absolute event rate).

Another promising way to determine the properties of the first stars at extremely high redshifts is through 21-cm and other line-intensity maps (LIM). The basic idea is to search for their combined effect on the intergalactic (21-cm) and galactic (typical nebular emission lines) gas, rather than attempting to find individual objects. 
This has the advantage of reaching earlier times and accessing the integrated emission of all galaxies, including the very faint ones where Pop~III star formation may be more prevalent. 
The 21-cm signal traces the thermal, ionization, and excitation state of neutral hydrogen in the IGM. As such, it is sensitive to the radiation emitted by all star formation, whether it is Pop~II or~III~\citep{Furlanetto+06}.
Theoretical work predicts that at $z\gtrsim 15$ these radiative fields are dominated by Pop~III rather than metal-enriched systems~\citep{Ventura+23, Cruz+25}, making the epochs of Lyman-$\alpha$ coupling and X-ray heating promising avenues to find Pop~III star formation~\citep{Mirocha+18, Munoz+22}.
While Pop~III star formation is expected to continue at lower redshifts ($z\lesssim 15$, Section~\ref{sec:spectral-hardness}), it will be more difficult to disentangle. A potential avenue is given by supersonic streaming velocities between dark matter and baryons, which strongly affect the formation of Pop~III stars in minihalos~\citep{Visbal+12}, leaving a distinct acoustic signature in the 21-cm signal~\citep{Munoz+19b}. 
Beyond 21-cm, LIM of star-forming emission lines can provide a handle on the integrated galaxy population beyond the reach of individual detections~\citep[as reviewed in, e.g.,][]{BernalKovetz22}. This can be turned into a powerful Pop~III search through for instance the \ion{He}{2} line, which acts as a differentiator between Pop~II and Pop~III-driven star formation, allowing for statistical searches of metal-free stars~\citep{Visbal+15, Parsons+22}.

Finally, a potential avenue for constraining early Pop~III star formation comes from the local Cosmic Infrared Background (CIB), which contains the integrated signature of emission accumulated over the entire history of our Universe, including sources inaccessible to individual telescopic studies. An excess in the NIR band ($\sim 1 - 10 ~\si{\mu m}$) has been reported with respect to the contribution from known resolved galaxies, possibly revealing redshifted UV emission from unresolved galaxy populations at $z \gtrsim 10$, up to the first stars era (see e.g. the review of \citealt{Kashlinsky+18}). When inferred from measurements of the mean CIB intensity, this excess is subject to substantial uncertainties, especially due to the challenges associated with foreground subtraction. A particularly promising tool for the characterization of faint populations below the sensitivity/angular-resolution threshold of our instruments comes therefore from the study of the power spectrum of source-subtracted CIB fluctuations \citep[e.g.][]{Kashlinsky+02, Kashlinski+07, Kashlinsky+25, Thompson+07, Helgason+12, Zemcov+14, Mitchell-Wynne+15, Arendt+16, Kaminsky+26}. The expected contribution from early Pop~III star formation to both the average CIB excess and its fluctuations has been explored for example by \citet{Santos+02}, \citet{SalvaterraFerrara03}, \citet{Yue+13}, \citet{Helgason+16}, and \citet{Sun+21}. Pop~III remnants at high redshift may also be among the sources of the cosmic soft X-ray background, which has been found to yield a statistically significant cross-correlation with the CIB (\citealt{Cappelluti+13}, but for a more complete discussion on the signature of Pop
III black hole remnants refer e.g. to the review of \citealt{Hammerle+20}).

\vspace{.75em}
\section{Conclusions}

In this review, we have summarized the current status of the search for Pop~III stars, from indirect constraints in the local Universe to the first direct searches enabled by JWST at high redshift. A broad theoretical consensus has emerged that primordial star formation was characterized by an IMF more top-heavy than that of present-day stars. However, the exact shape, characteristic mass, and high-mass cutoff of the Pop~III IMF remain poorly constrained, with important implications for the nature of the first supernovae, the timing of the transition to Pop~II star formation, and the early chemical enrichment of the ISM, CGM and IGM.

Cosmic archaeology provides some of the strongest indirect constraints on the Pop~III IMF. The abundance patterns of extremely metal-poor and CEMP-no stars indicate enrichment by massive, short-lived Pop~III progenitors, likely including faint and normal core-collapse supernovae. At the same time, the absence of confirmed metal-free low-mass survivors provides evidence against a dominant population of long-lived Pop~III stars. Searches for PISN signatures have so far produced only a small number of candidates, suggesting either that very massive Pop~III progenitors were rare, or that their chemical signatures are difficult to preserve in second-generation stars. Particularly, if metals produced by the first stars were not typically retained by low-mass hosts, they may be more efficiently traced in the CGM/IGM around higher-$z$ galaxies (e.g. from absorption studies along quasar sightlines) than in the stellar halo of the Milky Way or in nearby dwarfs.

At high redshift, JWST has revealed that metal enrichment is already widespread in early galaxies formed within the first billion year of cosmic history, making the identification of truly pristine systems challenging. Nevertheless, extremely metal-poor candidates and/or galaxies with unusually hard ionizing spectra provide promising laboratories for testing different Pop~III scenarios. A key lesson from both observations and simulations is that the presence of metal lines does not necessarily rule out ongoing Pop~III star formation. Hybrid and self-polluted Pop~III systems may thus represent an important pathway to detect Pop~III activity, especially at later cosmic epochs.

The most promising searches will likely combine multiple approaches. Spatially resolved spectroscopy is essential to separate metal-poor clumps from nearby enriched systems, while gravitational lensing can push observations to intrinsically faint stellar complexes, and even individual stars. Time-domain surveys offer an independent route through Pop~III supernovae, especially bright PISNe, which may be identified up to Cosmic Dawn. At the same time, improved modeling of stellar evolution, binary interactions, rotation, radiative feedback, metal mixing, dust attenuation, and nebular emission is required to distinguish genuine Pop~III signatures from other spectrally hard contaminants such as AGN, WR stars, very massive Pop~II stars, X-ray binaries, and shocked gas. For the highest-$z$ observed candidates ($z \gtrsim 15$), further studies are also needed to fully characterize these systems and their possible confusion with low-$z$ interlopers.

Overall, the search for Pop~III stars is entering a new phase. Rather than relying on a single, decisive observable, future progress will likely come from the convergence of near-field abundance constraints, high-redshift spectroscopy, transient searches, lensing-aided observations, and 21-cm studies aiming to characterize neutral hydrogen in the first star-forming sites, with interpretations informed by comparisons with cosmological simulations. 
The growing population of candidates is also already providing vital clues on the IMF, SFE, and feedback of the first stars.


\begin{acknowledgments}
We thank the Committee, the Secretary Gabriella Deconi, and the Local Organizing team of the Sexten Center for Astrophysics ``Riccardo Giacconi''. 
AV acknowledges funding from the Cosmic Frontier Center and the University of Texas at Austin’s College of Natural Sciences.
EZ acknowledges project grant 2022-03804 from the Swedish Research Council.
YA and SF acknowledge funding to the Dunlap Institute through an endowment established by the David Dunlap family and the University of Toronto.
SS acknowledges funding from the ERC Starting grant NEFERTITI H2020/804240 (PI: Salvadori).
EV acknowledges financial support through grants INAF GO Grant 2024 ``Mapping Star Cluster Feedback in a Galaxy 450 Myr after the Big Bang'' and the European Union – NextGenerationEU within PRIN 2022 project n.20229YBSAN - Globular clusters in cosmological simulations and lensed fields: from their birth to the present epoch.
JBM was supported by NSF Grants AST-2307354 and AST-2408637, and by the NSF-Simons AI Institute for Cosmic Origins.
AJB acknowledges funding from the ``FirstGalaxies'' Advanced Grant from the European Research Council (ERC) under the European Union’s Horizon 2020 research and innovation programme (Grant agreement No. 789056).
PGP-G acknowledges support from grant PID2022-139567NB-I00 funded by Spanish Ministerio de Ciencia, Innovaci\'on y Universidades MCIU/AEI/10.13039/501100011033,
FEDER ``Una manera de hacer
Europa''.
CC acknowledges support from the Swiss National Science Foundation
(SNF; Project 200020-192039).
RS acknowledge support from the PRIN 2022 MUR project 2022CB3PJ3 - First Light And Galaxy aSsembly (FLAGS) funded by the European Union – Next Generation EU.
HA acknowledges support from CNES, focused on the JWST mission, and the French National Research Agency (ANR) under grant ANR-21-CE31-0838.
LC acknowledges funding from the French government under the France 2030 investment plan, as part of the Initiative d’Excellence d’Aix-Marseille Université – A*MIDEX AMX-22-RE-AB-101.
PD acknowledges support from an NSERC discovery grant (RGPIN-2025-06182).
CK acknowledges funding from the UK Science and Technology Facilities Council through grant ST/Y001443/1.
MC acknowledges financial support from the INAF RF2024 GO Grant ``Revealing the nature of bright galaxies at cosmic dawn with deep JWST spectroscopy''.
PS acknowledges financial support from INAF RF2024 Large Grant ``UNDUST: UNveiling the Dawn of the Universe with JWST''.
\end{acknowledgments}

\bibliography{main}{}

@ARTICLE{BarkanaLoeb01,
       author = {{Barkana}, R. and {Loeb}, A.},
        title = "{In the beginning: the first sources of light and the reionization of the universe}",
      journal = {\physrep},
         year = 2001,
        month = jul,
       volume = {349},
       number = {2},
        pages = {125-238},
          doi = {10.1016/S0370-1573(01)00019-9},
archivePrefix = {arXiv},
       eprint = {astro-ph/0010468},
 primaryClass = {astro-ph},
       adsurl = {https://ui.adsabs.harvard.edu/abs/2001PhR...349..125B}
}

@ARTICLE{BrommLarson04,
       author = {{Bromm}, Volker and {Larson}, Richard B.},
        title = "{The First Stars}",
      journal = {\araa},
         year = 2004,
        month = sep,
       volume = {42},
       number = {1},
        pages = {79-118},
          doi = {10.1146/annurev.astro.42.053102.134034},
archivePrefix = {arXiv},
       eprint = {astro-ph/0311019},
 primaryClass = {astro-ph},
       adsurl = {https://ui.adsabs.harvard.edu/abs/2004ARA&A..42...79B}
}

@ARTICLE{Bromm13,
       author = {{Bromm}, Volker},
        title = "{Formation of the first stars}",
      journal = {Reports on Progress in Physics},
         year = 2013,
        month = nov,
       volume = {76},
       number = {11},
          eid = {112901},
        pages = {112901},
          doi = {10.1088/0034-4885/76/11/112901},
archivePrefix = {arXiv},
       eprint = {1305.5178},
 primaryClass = {astro-ph.CO},
       adsurl = {https://ui.adsabs.harvard.edu/abs/2013RPPh...76k2901B}
}

@ARTICLE{KlessenGlover23,
       author = {{Klessen}, Ralf S. and {Glover}, Simon C.~O.},
        title = "{The First Stars: Formation, Properties, and Impact}",
      journal = {\araa},
         year = 2023,
        month = aug,
       volume = {61},
        pages = {65-130},
          doi = {10.1146/annurev-astro-071221-053453},
archivePrefix = {arXiv},
       eprint = {2303.12500},
 primaryClass = {astro-ph.CO},
       adsurl = {https://ui.adsabs.harvard.edu/abs/2023ARA&A..61...65K}
}

@ARTICLE{Chabrier03,
       author = {{Chabrier}, Gilles},
        title = "{Galactic Stellar and Substellar Initial Mass Function}",
      journal = {\pasp},
         year = 2003,
        month = jul,
       volume = {115},
       number = {809},
        pages = {763-795},
          doi = {10.1086/376392},
archivePrefix = {arXiv},
       eprint = {astro-ph/0304382},
 primaryClass = {astro-ph},
       adsurl = {https://ui.adsabs.harvard.edu/abs/2003PASP..115..763C}
}

@ARTICLE{Bate+95,
       author = {{Bate}, Matthew R. and {Bonnell}, Ian A. and {Price}, Nigel M.},
        title = "{Modelling accretion in protobinary systems}",
      journal = {\mnras},
         year = 1995,
        month = nov,
       volume = {277},
       number = {2},
        pages = {362-376},
          doi = {10.1093/mnras/277.2.362},
archivePrefix = {arXiv},
       eprint = {astro-ph/9510149},
 primaryClass = {astro-ph},
       adsurl = {https://ui.adsabs.harvard.edu/abs/1995MNRAS.277..362B}
}

@ARTICLE{Federrath+10,
       author = {{Federrath}, Christoph and {Banerjee}, Robi and {Clark}, Paul C. and {Klessen}, Ralf S.},
        title = "{Modeling Collapse and Accretion in Turbulent Gas Clouds: Implementation and Comparison of Sink Particles in AMR and SPH}",
      journal = {\apj},
         year = 2010,
        month = apr,
       volume = {713},
       number = {1},
        pages = {269-290},
          doi = {10.1088/0004-637X/713/1/269},
archivePrefix = {arXiv},
       eprint = {1001.4456},
 primaryClass = {astro-ph.SR},
       adsurl = {https://ui.adsabs.harvard.edu/abs/2010ApJ...713..269F}
}

@ARTICLE{Haiman+96,
       author = {{Haiman}, Zoltan and {Thoul}, Anne A. and {Loeb}, Abraham},
        title = "{Cosmological Formation of Low-Mass Objects}",
      journal = {\apj},
         year = 1996,
        month = jun,
       volume = {464},
        pages = {523},
          doi = {10.1086/177343},
archivePrefix = {arXiv},
       eprint = {astro-ph/9507111},
 primaryClass = {astro-ph},
       adsurl = {https://ui.adsabs.harvard.edu/abs/1996ApJ...464..523H}
}

@ARTICLE{Tegmark+97,
       author = {{Tegmark}, Max and {Silk}, Joseph and {Rees}, Martin J. and {Blanchard}, Alain and {Abel}, Tom and {Palla}, Francesco},
        title = "{How Small Were the First Cosmological Objects?}",
      journal = {\apj},
         year = 1997,
        month = jan,
       volume = {474},
        pages = {1},
          doi = {10.1086/303434},
archivePrefix = {arXiv},
       eprint = {astro-ph/9603007},
 primaryClass = {astro-ph},
       adsurl = {https://ui.adsabs.harvard.edu/abs/1997ApJ...474....1T}
}

@ARTICLE{Yoshida+03,
       author = {{Yoshida}, Naoki and {Abel}, Tom and {Hernquist}, Lars and {Sugiyama}, Naoshi},
        title = "{Simulations of Early Structure Formation: Primordial Gas Clouds}",
      journal = {\apj},
         year = 2003,
        month = aug,
       volume = {592},
       number = {2},
        pages = {645-663},
          doi = {10.1086/375810},
archivePrefix = {arXiv},
       eprint = {astro-ph/0301645},
 primaryClass = {astro-ph},
       adsurl = {https://ui.adsabs.harvard.edu/abs/2003ApJ...592..645Y}
}

@ARTICLE{Abel02,
       author = {{Abel}, Tom},
        title = "{The Basic Building Blocks of Galaxies}",
      journal = {\apss},
         year = 2002,
        month = jul,
       volume = {281},
       number = {1},
        pages = {471-473},
          doi = {10.1023/A:1019509211972},
       adsurl = {https://ui.adsabs.harvard.edu/abs/2002Ap&SS.281..471A}
}

@ARTICLE{Abel+02,
       author = {{Abel}, Tom and {Bryan}, Greg L. and {Norman}, Michael L.},
        title = "{The Formation of the First Star in the Universe}",
      journal = {Science},
         year = 2002,
        month = jan,
       volume = {295},
       number = {5552},
        pages = {93-98},
          doi = {10.1126/science.1063991},
archivePrefix = {arXiv},
       eprint = {astro-ph/0112088},
 primaryClass = {astro-ph},
       adsurl = {https://ui.adsabs.harvard.edu/abs/2002Sci...295...93A}
}

@ARTICLE{Bromm+02,
       author = {{Bromm}, Volker and {Coppi}, Paolo S. and {Larson}, Richard B.},
        title = "{The Formation of the First Stars. I. The Primordial Star-forming Cloud}",
      journal = {\apj},
         year = 2002,
        month = jan,
       volume = {564},
       number = {1},
        pages = {23-51},
          doi = {10.1086/323947},
archivePrefix = {arXiv},
       eprint = {astro-ph/0102503},
 primaryClass = {astro-ph},
       adsurl = {https://ui.adsabs.harvard.edu/abs/2002ApJ...564...23B}
}

@ARTICLE{Yoshida+08,
       author = {{Yoshida}, Naoki and {Omukai}, Kazuyuki and {Hernquist}, Lars},
        title = "{Protostar Formation in the Early Universe}",
      journal = {Science},
         year = 2008,
        month = aug,
       volume = {321},
       number = {5889},
        pages = {669},
          doi = {10.1126/science.1160259},
archivePrefix = {arXiv},
       eprint = {0807.4928},
 primaryClass = {astro-ph},
       adsurl = {https://ui.adsabs.harvard.edu/abs/2008Sci...321..669Y}
}

@ARTICLE{Clark+08,
       author = {{Clark}, Paul C. and {Glover}, Simon C.~O. and {Klessen}, Ralf S.},
        title = "{The First Stellar Cluster}",
      journal = {\apj},
         year = 2008,
        month = jan,
       volume = {672},
       number = {2},
        pages = {757-764},
          doi = {10.1086/524187},
       adsurl = {https://ui.adsabs.harvard.edu/abs/2008ApJ...672..757C}
}

@ARTICLE{Clark+11a,
       author = {{Clark}, Paul C. and {Glover}, Simon C.~O. and {Klessen}, Ralf S. and {Bromm}, Volker},
        title = "{Gravitational Fragmentation in Turbulent Primordial Gas and the Initial Mass Function of Population III Stars}",
      journal = {\apj},
         year = 2011,
        month = feb,
       volume = {727},
       number = {2},
          eid = {110},
        pages = {110},
          doi = {10.1088/0004-637X/727/2/110},
archivePrefix = {arXiv},
       eprint = {1006.1508},
 primaryClass = {astro-ph.GA},
       adsurl = {https://ui.adsabs.harvard.edu/abs/2011ApJ...727..110C}
}

@ARTICLE{Clark+11b,
       author = {{Clark}, Paul C. and {Glover}, Simon C.~O. and {Smith}, Rowan J. and {Greif}, Thomas H. and {Klessen}, Ralf S. and {Bromm}, Volker},
        title = "{The Formation and Fragmentation of Disks Around Primordial Protostars}",
      journal = {Science},
         year = 2011,
        month = feb,
       volume = {331},
       number = {6020},
        pages = {1040},
          doi = {10.1126/science.1198027},
archivePrefix = {arXiv},
       eprint = {1101.5284},
 primaryClass = {astro-ph.CO},
       adsurl = {https://ui.adsabs.harvard.edu/abs/2011Sci...331.1040C}
}

@ARTICLE{Greif+11,
       author = {{Greif}, Thomas H. and {Springel}, Volker and {White}, Simon D.~M. and {Glover}, Simon C.~O. and {Clark}, Paul C. and {Smith}, Rowan J. and {Klessen}, Ralf S. and {Bromm}, Volker},
        title = "{Simulations on a Moving Mesh: The Clustered Formation of Population III Protostars}",
      journal = {\apj},
         year = 2011,
        month = aug,
       volume = {737},
       number = {2},
          eid = {75},
        pages = {75},
          doi = {10.1088/0004-637X/737/2/75},
archivePrefix = {arXiv},
       eprint = {1101.5491},
 primaryClass = {astro-ph.CO},
       adsurl = {https://ui.adsabs.harvard.edu/abs/2011ApJ...737...75G}
}

@ARTICLE{Hosokawa+11,
       author = {{Hosokawa}, Takashi and {Omukai}, Kazuyuki and {Yoshida}, Naoki and {Yorke}, Harold W.},
        title = "{Protostellar Feedback Halts the Growth of the First Stars in the Universe}",
      journal = {Science},
         year = 2011,
        month = dec,
       volume = {334},
       number = {6060},
        pages = {1250},
          doi = {10.1126/science.1207433},
archivePrefix = {arXiv},
       eprint = {1111.3649},
 primaryClass = {astro-ph.CO},
       adsurl = {https://ui.adsabs.harvard.edu/abs/2011Sci...334.1250H}
}

@ARTICLE{Susa+19,
       author = {{Susa}, Hajime},
        title = "{Merge or Survive: Number of Population III Stars per Minihalo}",
      journal = {\apj},
         year = 2019,
        month = jun,
       volume = {877},
       number = {2},
          eid = {99},
        pages = {99},
          doi = {10.3847/1538-4357/ab1b6f},
archivePrefix = {arXiv},
       eprint = {1904.09731},
 primaryClass = {astro-ph.GA},
       adsurl = {https://ui.adsabs.harvard.edu/abs/2019ApJ...877...99S}
}

@ARTICLE{Sharda+25,
       author = {{Sharda}, Piyush and {Menon}, Shyam H. and {Gerasimov}, Roman and {Bromm}, Volker and {Burkhart}, Blakesley and {Haemmerl{\'e}}, Lionel and {van Veenen}, Lisanne and {Wibking}, Benjamin D.},
        title = "{Magnetic fields limit the mass of Population III stars even before the onset of protostellar radiation feedback}",
      journal = {\mnras},
         year = 2025,
        month = jul,
       volume = {541},
       number = {1},
        pages = {L1-L7},
          doi = {10.1093/mnrasl/slaf043},
archivePrefix = {arXiv},
       eprint = {2501.12734},
 primaryClass = {astro-ph.GA},
       adsurl = {https://ui.adsabs.harvard.edu/abs/2025MNRAS.541L...1S}
}

@ARTICLE{ShardaMenon25,
       author = {{Sharda}, Piyush and {Menon}, Shyam H.},
        title = "{Population III star formation in the presence of turbulence, magnetic fields, and ionizing radiation feedback}",
      journal = {\mnras},
         year = 2025,
        month = jun,
       volume = {540},
       number = {2},
        pages = {1745-1764},
          doi = {10.1093/mnras/staf803},
archivePrefix = {arXiv},
       eprint = {2405.18265},
 primaryClass = {astro-ph.GA},
       adsurl = {https://ui.adsabs.harvard.edu/abs/2025MNRAS.540.1745S}
}

@ARTICLE{Sharda+20,
       author = {{Sharda}, Piyush and {Federrath}, Christoph and {Krumholz}, Mark R.},
        title = "{The importance of magnetic fields for the initial mass function of the first stars}",
      journal = {\mnras},
         year = 2020,
        month = sep,
       volume = {497},
       number = {1},
        pages = {336-351},
          doi = {10.1093/mnras/staa1926},
archivePrefix = {arXiv},
       eprint = {2002.11502},
 primaryClass = {astro-ph.GA},
       adsurl = {https://ui.adsabs.harvard.edu/abs/2020MNRAS.497..336S}
}

@ARTICLE{Sharda+21,
       author = {{Sharda}, Piyush and {Federrath}, Christoph and {Krumholz}, Mark R. and {Schleicher}, Dominik R.~G.},
        title = "{Magnetic field amplification in accretion discs around the first stars: implications for the primordial IMF}",
      journal = {\mnras},
         year = 2021,
        month = may,
       volume = {503},
       number = {2},
        pages = {2014-2032},
          doi = {10.1093/mnras/stab531},
archivePrefix = {arXiv},
       eprint = {2007.02678},
 primaryClass = {astro-ph.GA},
       adsurl = {https://ui.adsabs.harvard.edu/abs/2021MNRAS.503.2014S}
}

@ARTICLE{vanVeenen+25,
       author = {{van Veenen}, Lisanne and {Sharda}, Piyush and {Viti}, Serena and {Menon}, Shyam H.},
        title = "{Radiation magnetohydrodynamics simulations of Population III star formation during the epoch of reionization}",
      journal = {\aap},
         year = 2026,
        month = mar,
       volume = {708},
          eid = {A61},
        pages = {A61},
          doi = {10.1051/0004-6361/202558114},
archivePrefix = {arXiv},
       eprint = {2511.11314},
 primaryClass = {astro-ph.GA},
       adsurl = {https://ui.adsabs.harvard.edu/abs/2026A&A...708A..61V}
}

@ARTICLE{Sadanari+21,
       author = {{Sadanari}, Kenji Eric and {Omukai}, Kazuyuki and {Sugimura}, Kazuyuki and {Matsumoto}, Tomoaki and {Tomida}, Kengo},
        title = "{Magnetohydrodynamic effect on first star formation: pre-stellar core collapse and protostar formation}",
      journal = {\mnras},
         year = 2021,
        month = aug,
       volume = {505},
       number = {3},
        pages = {4197-4214},
          doi = {10.1093/mnras/stab1330},
archivePrefix = {arXiv},
       eprint = {2105.03430},
 primaryClass = {astro-ph.GA},
       adsurl = {https://ui.adsabs.harvard.edu/abs/2021MNRAS.505.4197S}
}

@ARTICLE{Sadanari+24,
       author = {{Sadanari}, Kenji Eric and {Omukai}, Kazuyuki and {Sugimura}, Kazuyuki and {Matsumoto}, Tomoaki and {Tomida}, Kengo},
        title = "{Impact of turbulent magnetic fields on disk formation and fragmentation in first star formation}",
      journal = {\pasj},
         year = 2024,
        month = aug,
       volume = {76},
       number = {4},
        pages = {823-840},
          doi = {10.1093/pasj/psae051},
archivePrefix = {arXiv},
       eprint = {2405.15045},
 primaryClass = {astro-ph.GA},
       adsurl = {https://ui.adsabs.harvard.edu/abs/2024PASJ...76..823S}
}

@ARTICLE{Prole+22,
       author = {{Prole}, Lewis R. and {Clark}, Paul C. and {Klessen}, Ralf S. and {Glover}, Simon C.~O. and {Pakmor}, R{\"u}diger},
        title = "{Primordial magnetic fields in Population III star formation: a magnetized resolution study}",
      journal = {\mnras},
         year = 2022,
        month = aug,
       volume = {516},
       number = {2},
        pages = {2223-2234},
          doi = {10.1093/mnras/stac2327},
archivePrefix = {arXiv},
       eprint = {2206.11919},
 primaryClass = {astro-ph.GA},
       adsurl = {https://ui.adsabs.harvard.edu/abs/2022MNRAS.516.2223P}
}

@ARTICLE{StacyBromm13,
       author = {{Stacy}, Athena and {Bromm}, Volker},
        title = "{Constraining the statistics of Population III binaries}",
      journal = {\mnras},
         year = 2013,
        month = aug,
       volume = {433},
       number = {2},
        pages = {1094-1107},
          doi = {10.1093/mnras/stt789},
archivePrefix = {arXiv},
       eprint = {1211.1889},
 primaryClass = {astro-ph.CO},
       adsurl = {https://ui.adsabs.harvard.edu/abs/2013MNRAS.433.1094S}
}

@ARTICLE{Stacy+16,
       author = {{Stacy}, Athena and {Bromm}, Volker and {Lee}, Aaron T.},
        title = "{Building up the Population III initial mass function from cosmological initial conditions}",
      journal = {\mnras},
         year = 2016,
        month = oct,
       volume = {462},
       number = {2},
        pages = {1307-1328},
          doi = {10.1093/mnras/stw1728},
archivePrefix = {arXiv},
       eprint = {1603.09475},
 primaryClass = {astro-ph.GA},
       adsurl = {https://ui.adsabs.harvard.edu/abs/2016MNRAS.462.1307S}
}

@ARTICLE{Susa+14,
       author = {{Susa}, Hajime and {Hasegawa}, Kenji and {Tominaga}, Nozomu},
        title = "{The Mass Spectrum of the First Stars}",
      journal = {\apj},
         year = 2014,
        month = sep,
       volume = {792},
       number = {1},
          eid = {32},
        pages = {32},
          doi = {10.1088/0004-637X/792/1/32},
archivePrefix = {arXiv},
       eprint = {1407.1374},
 primaryClass = {astro-ph.GA},
       adsurl = {https://ui.adsabs.harvard.edu/abs/2014ApJ...792...32S}
}

@ARTICLE{Safranek-Shrader+14,
       author = {{Safranek-Shrader}, Chalence and {Milosavljevi{\'c}}, Milo{\v{s}} and {Bromm}, Volker},
        title = "{Star formation in the first galaxies - II. Clustered star formation and the influence of metal line cooling}",
      journal = {\mnras},
         year = 2014,
        month = feb,
       volume = {438},
       number = {2},
        pages = {1669-1685},
          doi = {10.1093/mnras/stt2307},
archivePrefix = {arXiv},
       eprint = {1307.1982},
 primaryClass = {astro-ph.CO},
       adsurl = {https://ui.adsabs.harvard.edu/abs/2014MNRAS.438.1669S}
}

@ARTICLE{Sugimura+20,
       author = {{Sugimura}, Kazuyuki and {Matsumoto}, Tomoaki and {Hosokawa}, Takashi and {Hirano}, Shingo and {Omukai}, Kazuyuki},
        title = "{The Birth of a Massive First-star Binary}",
      journal = {\apjl},
         year = 2020,
        month = mar,
       volume = {892},
       number = {1},
          eid = {L14},
        pages = {L14},
          doi = {10.3847/2041-8213/ab7d37},
archivePrefix = {arXiv},
       eprint = {2002.00012},
 primaryClass = {astro-ph.GA},
       adsurl = {https://ui.adsabs.harvard.edu/abs/2020ApJ...892L..14S}
}

@ARTICLE{Wollenberg+20,
       author = {{Wollenberg}, Katharina M.~J. and {Glover}, Simon C.~O. and {Clark}, Paul C. and {Klessen}, Ralf S.},
        title = "{Formation sites of Population III star formation: The effects of different levels of rotation and turbulence on the fragmentation behaviour of primordial gas}",
      journal = {\mnras},
         year = 2020,
        month = may,
       volume = {494},
       number = {2},
        pages = {1871-1893},
          doi = {10.1093/mnras/staa289},
archivePrefix = {arXiv},
       eprint = {1912.06377},
 primaryClass = {astro-ph.GA},
       adsurl = {https://ui.adsabs.harvard.edu/abs/2020MNRAS.494.1871W}
}

@ARTICLE{Latif+22,
       author = {{Latif}, Muhammad A. and {Whalen}, Daniel and {Khochfar}, Sadegh},
        title = "{The Birth Mass Function of Population III Stars}",
      journal = {\apj},
         year = 2022,
        month = jan,
       volume = {925},
       number = {1},
          eid = {28},
        pages = {28},
          doi = {10.3847/1538-4357/ac3916},
archivePrefix = {arXiv},
       eprint = {2109.10655},
 primaryClass = {astro-ph.GA},
       adsurl = {https://ui.adsabs.harvard.edu/abs/2022ApJ...925...28L}
}

@ARTICLE{Jaura+22,
       author = {{Jaura}, Ondrej and {Glover}, Simon C.~O. and {Wollenberg}, Katharina M.~J. and {Klessen}, Ralf S. and {Geen}, Sam and {Haemmerl{\'e}}, Lionel},
        title = "{Trapping of H II regions in Population III star formation}",
      journal = {\mnras},
         year = 2022,
        month = may,
       volume = {512},
       number = {1},
        pages = {116-136},
          doi = {10.1093/mnras/stac487},
archivePrefix = {arXiv},
       eprint = {2202.09803},
 primaryClass = {astro-ph.SR},
       adsurl = {https://ui.adsabs.harvard.edu/abs/2022MNRAS.512..116J}
}

@ARTICLE{TangChen24,
       author = {{Tang}, Ching-Yao and {Chen}, Ke-Jung},
        title = "{Clumpy structures within the turbulent primordial cloud}",
      journal = {\mnras},
         year = 2024,
        month = apr,
       volume = {529},
       number = {4},
        pages = {4248-4261},
          doi = {10.1093/mnras/stae764},
archivePrefix = {arXiv},
       eprint = {2303.00751},
 primaryClass = {astro-ph.GA},
       adsurl = {https://ui.adsabs.harvard.edu/abs/2024MNRAS.529.4248T}
}

@ARTICLE{ItoOmukai+24,
       author = {{Ito}, Mana and {Omukai}, Kazuyuki},
        title = "{First star formation in extremely early epochs}",
      journal = {\pasj},
         year = 2024,
        month = aug,
       volume = {76},
       number = {4},
        pages = {850-862},
          doi = {10.1093/pasj/psae054},
archivePrefix = {arXiv},
       eprint = {2405.10073},
 primaryClass = {astro-ph.GA},
       adsurl = {https://ui.adsabs.harvard.edu/abs/2024PASJ...76..850I}
}

@ARTICLE{Hirano+14,
       author = {{Hirano}, Shingo and {Hosokawa}, Takashi and {Yoshida}, Naoki and {Umeda}, Hideyuki and {Omukai}, Kazuyuki and {Chiaki}, Gen and {Yorke}, Harold W.},
        title = "{One Hundred First Stars: Protostellar Evolution and the Final Masses}",
      journal = {\apj},
         year = 2014,
        month = feb,
       volume = {781},
       number = {2},
          eid = {60},
        pages = {60},
          doi = {10.1088/0004-637X/781/2/60},
archivePrefix = {arXiv},
       eprint = {1308.4456},
 primaryClass = {astro-ph.CO},
       adsurl = {https://ui.adsabs.harvard.edu/abs/2014ApJ...781...60H}
}

@ARTICLE{Hirano+15,
       author = {{Hirano}, S. and {Hosokawa}, T. and {Yoshida}, N. and {Omukai}, K. and {Yorke}, H.~W.},
        title = "{Primordial star formation under the influence of far ultraviolet radiation: 1540 cosmological haloes and the stellar mass distribution}",
      journal = {\mnras},
         year = 2015,
        month = mar,
       volume = {448},
       number = {1},
        pages = {568-587},
          doi = {10.1093/mnras/stv044},
archivePrefix = {arXiv},
       eprint = {1501.01630},
 primaryClass = {astro-ph.GA},
       adsurl = {https://ui.adsabs.harvard.edu/abs/2015MNRAS.448..568H}
}

@ARTICLE{HiranoBromm17,
       author = {{Hirano}, Shingo and {Bromm}, Volker},
        title = "{Formation and survival of Population III stellar systems}",
      journal = {\mnras},
         year = 2017,
        month = sep,
       volume = {470},
       number = {1},
        pages = {898-914},
          doi = {10.1093/mnras/stx1220},
archivePrefix = {arXiv},
       eprint = {1612.06387},
 primaryClass = {astro-ph.GA},
       adsurl = {https://ui.adsabs.harvard.edu/abs/2017MNRAS.470..898H}
}

@ARTICLE{Hosokawa+16,
       author = {{Hosokawa}, Takashi and {Hirano}, Shingo and {Kuiper}, Rolf and {Yorke}, Harold W. and {Omukai}, Kazuyuki and {Yoshida}, Naoki},
        title = "{Formation of Massive Primordial Stars: Intermittent UV Feedback with Episodic Mass Accretion}",
      journal = {\apj},
         year = 2016,
        month = jun,
       volume = {824},
       number = {2},
          eid = {119},
        pages = {119},
          doi = {10.3847/0004-637X/824/2/119},
archivePrefix = {arXiv},
       eprint = {1510.01407},
 primaryClass = {astro-ph.GA},
       adsurl = {https://ui.adsabs.harvard.edu/abs/2016ApJ...824..119H}
}

@ARTICLE{Chon+18,
       author = {{Chon}, Sunmyon and {Hosokawa}, Takashi and {Yoshida}, Naoki},
        title = "{Radiation hydrodynamics simulations of the formation of direct-collapse supermassive stellar systems}",
      journal = {\mnras},
         year = 2018,
        month = apr,
       volume = {475},
       number = {3},
        pages = {4104-4121},
          doi = {10.1093/mnras/sty086},
archivePrefix = {arXiv},
       eprint = {1711.05262},
 primaryClass = {astro-ph.GA},
       adsurl = {https://ui.adsabs.harvard.edu/abs/2018MNRAS.475.4104C}
}

@ARTICLE{Regan+20,
       author = {{Regan}, John A. and {Wise}, John H. and {Woods}, Tyrone E. and {Downes}, Turlough P. and {O'Shea}, Brian W. and {Norman}, Michael L.},
        title = "{The Formation of Very Massive Stars in Early Galaxies and Implications for Intermediate Mass Black Holes}",
      journal = {The Open Journal of Astrophysics},
         year = 2020,
        month = dec,
       volume = {3},
       number = {1},
          eid = {15},
        pages = {15},
          doi = {10.21105/astro.2008.08090},
archivePrefix = {arXiv},
       eprint = {2008.08090},
 primaryClass = {astro-ph.GA},
       adsurl = {https://ui.adsabs.harvard.edu/abs/2020OJAp....3E..15R}
}

@ARTICLE{BeersChristlieb05,
       author = {{Beers}, Timothy C. and {Christlieb}, Norbert},
        title = "{The Discovery and Analysis of Very Metal-Poor Stars in the Galaxy}",
      journal = {\araa},
         year = 2005,
        month = sep,
       volume = {43},
       number = {1},
        pages = {531-580},
          doi = {10.1146/annurev.astro.42.053102.134057},
       adsurl = {https://ui.adsabs.harvard.edu/abs/2005ARA&A..43..531B}
}

@ARTICLE{Ricotti+10,
       author = {{Ricotti}, Massimo},
        title = "{The First Galaxies and the Likely Discovery of Their Fossils in the Local Group}",
      journal = {Advances in Astronomy},
         year = 2010,
        month = jan,
       volume = {2010},
          eid = {271592},
        pages = {271592},
          doi = {10.1155/2010/271592},
archivePrefix = {arXiv},
       eprint = {0911.2792},
 primaryClass = {astro-ph.CO},
       adsurl = {https://ui.adsabs.harvard.edu/abs/2010AdAst2010E..33R}
}

@ARTICLE{FrebelNorris15,
       author = {{Frebel}, Anna and {Norris}, John E.},
        title = "{Near-Field Cosmology with Extremely Metal-Poor Stars}",
      journal = {\araa},
         year = 2015,
        month = aug,
       volume = {53},
        pages = {631-688},
          doi = {10.1146/annurev-astro-082214-122423},
archivePrefix = {arXiv},
       eprint = {1501.06921},
 primaryClass = {astro-ph.SR},
       adsurl = {https://ui.adsabs.harvard.edu/abs/2015ARA&A..53..631F}
}

@ARTICLE{Bonifacio25,
       author = {{Bonifacio}, Piercarlo and {Caffau}, Elisabetta and {Fran{\c{c}}ois}, Patrick and {Spite}, Monique},
        title = "{The most metal-poor stars}",
      journal = {\aapr},
         year = 2025,
        month = jul,
       volume = {33},
       number = {1},
          eid = {2},
        pages = {2},
          doi = {10.1007/s00159-025-00159-2},
archivePrefix = {arXiv},
       eprint = {2504.06335},
 primaryClass = {astro-ph.GA},
       adsurl = {https://ui.adsabs.harvard.edu/abs/2025A&ARv..33....2B}
}

@ARTICLE{UmedaNomoto02,
       author = {{Umeda}, Hideyuki and {Nomoto}, Ken'ichi},
        title = "{Nucleosynthesis of Zinc and Iron Peak Elements in Population III Type II Supernovae: Comparison with Abundances of Very Metal Poor Halo Stars}",
      journal = {\apj},
         year = 2002,
        month = jan,
       volume = {565},
       number = {1},
        pages = {385-404},
          doi = {10.1086/323946},
archivePrefix = {arXiv},
       eprint = {astro-ph/0103241},
 primaryClass = {astro-ph},
       adsurl = {https://ui.adsabs.harvard.edu/abs/2002ApJ...565..385U}
}

@ARTICLE{HegerWoosley02,
       author = {{Heger}, A. and {Woosley}, S.~E.},
        title = "{The Nucleosynthetic Signature of Population III}",
      journal = {\apj},
         year = 2002,
        month = mar,
       volume = {567},
       number = {1},
        pages = {532-543},
          doi = {10.1086/338487},
archivePrefix = {arXiv},
       eprint = {astro-ph/0107037},
 primaryClass = {astro-ph},
       adsurl = {https://ui.adsabs.harvard.edu/abs/2002ApJ...567..532H}
}

@ARTICLE{HegerWoosley+10,
       author = {{Heger}, Alexander and {Woosley}, S.~E.},
        title = "{Nucleosynthesis and Evolution of Massive Metal-free Stars}",
      journal = {\apj},
         year = 2010,
        month = nov,
       volume = {724},
       number = {1},
        pages = {341-373},
          doi = {10.1088/0004-637X/724/1/341},
archivePrefix = {arXiv},
       eprint = {0803.3161},
 primaryClass = {astro-ph},
       adsurl = {https://ui.adsabs.harvard.edu/abs/2010ApJ...724..341H}
}

@ARTICLE{MeynetMaeder02,
       author = {{Meynet}, G. and {Maeder}, A.},
        title = "{Stellar evolution with rotation. VIII. Models at Z = 10$^{-5}$ and CNO yields for early galactic evolution}",
      journal = {\aap},
         year = 2002,
        month = aug,
       volume = {390},
        pages = {561-583},
          doi = {10.1051/0004-6361:20020755},
archivePrefix = {arXiv},
       eprint = {astro-ph/0205370},
 primaryClass = {astro-ph},
       adsurl = {https://ui.adsabs.harvard.edu/abs/2002A&A...390..561M}
}

@ARTICLE{Meynet+06,
       author = {{Meynet}, G. and {Ekstr{\"o}m}, S. and {Maeder}, A.},
        title = "{The early star generations: the dominant effect of rotation on the CNO yields}",
      journal = {\aap},
         year = 2006,
        month = feb,
       volume = {447},
       number = {2},
        pages = {623-639},
          doi = {10.1051/0004-6361:20053070},
archivePrefix = {arXiv},
       eprint = {astro-ph/0510560},
 primaryClass = {astro-ph},
       adsurl = {https://ui.adsabs.harvard.edu/abs/2006A&A...447..623M}
}

@ARTICLE{Nomoto+06,
       author = {{Nomoto}, Ken'ichi and {Tominaga}, Nozomu and {Umeda}, Hideyuki and {Kobayashi}, Chiaki and {Maeda}, Keiichi},
        title = "{Nucleosynthesis yields of core-collapse supernovae and hypernovae, and galactic chemical evolution}",
      journal = {\nphysa},
         year = 2006,
        month = oct,
       volume = {777},
        pages = {424-458},
          doi = {10.1016/j.nuclphysa.2006.05.008},
archivePrefix = {arXiv},
       eprint = {astro-ph/0605725},
 primaryClass = {astro-ph},
       adsurl = {https://ui.adsabs.harvard.edu/abs/2006NuPhA.777..424N}
}

@ARTICLE{Ekstrom+08,
       author = {{Ekstr{\"o}m}, S. and {Meynet}, G. and {Chiappini}, C. and {Hirschi}, R. and {Maeder}, A.},
        title = "{Effects of rotation on the evolution of primordial stars}",
      journal = {\aap},
         year = 2008,
        month = oct,
       volume = {489},
       number = {2},
        pages = {685-698},
          doi = {10.1051/0004-6361:200809633},
archivePrefix = {arXiv},
       eprint = {0807.0573},
 primaryClass = {astro-ph},
       adsurl = {https://ui.adsabs.harvard.edu/abs/2008A&A...489..685E}
}

@ARTICLE{ChatzopoulosWheeler12,
       author = {{Chatzopoulos}, E. and {Wheeler}, J. Craig},
        title = "{Effects of Rotation on the Minimum Mass of Primordial Progenitors of Pair-instability Supernovae}",
      journal = {\apj},
         year = 2012,
        month = mar,
       volume = {748},
       number = {1},
          eid = {42},
        pages = {42},
          doi = {10.1088/0004-637X/748/1/42},
archivePrefix = {arXiv},
       eprint = {1201.1328},
 primaryClass = {astro-ph.HE},
       adsurl = {https://ui.adsabs.harvard.edu/abs/2012ApJ...748...42C}
}

@ARTICLE{Yoon+12,
       author = {{Yoon}, S.-C. and {Dierks}, A. and {Langer}, N.},
        title = "{Evolution of massive Population III stars with rotation and magnetic fields}",
      journal = {\aap},
         year = 2012,
        month = jun,
       volume = {542},
          eid = {A113},
        pages = {A113},
          doi = {10.1051/0004-6361/201117769},
archivePrefix = {arXiv},
       eprint = {1201.2364},
 primaryClass = {astro-ph.SR},
       adsurl = {https://ui.adsabs.harvard.edu/abs/2012A&A...542A.113Y}
}

@ARTICLE{LimongiChieffi18,
       author = {{Limongi}, Marco and {Chieffi}, Alessandro},
        title = "{Presupernova Evolution and Explosive Nucleosynthesis of Rotating Massive Stars in the Metallicity Range -3 {\ensuremath{\leq}} [Fe/H] {\ensuremath{\leq}} 0}",
      journal = {\apjs},
         year = 2018,
        month = jul,
       volume = {237},
       number = {1},
          eid = {13},
        pages = {13},
          doi = {10.3847/1538-4365/aacb24},
archivePrefix = {arXiv},
       eprint = {1805.09640},
 primaryClass = {astro-ph.SR},
       adsurl = {https://ui.adsabs.harvard.edu/abs/2018ApJS..237...13L}
}

@ARTICLE{Murphy+21,
       author = {{Murphy}, Laura J. and {Groh}, Jose H. and {Ekstr{\"o}m}, Sylvia and {Meynet}, Georges and {Pezzotti}, Camila and {Georgy}, Cyril and {Choplin}, Arthur and {Eggenberger}, Patrick and {Farrell}, Eoin and {Haemmerl{\'e}}, Lionel and {Hirschi}, Raphael and {Maeder}, Andr{\'e} and {Martinet}, S{\'e}bastien},
        title = "{Grids of stellar models with rotation - V. Models from 1.7 to 120 M$_{☉}$ at zero metallicity}",
      journal = {\mnras},
         year = 2021,
        month = feb,
       volume = {501},
       number = {2},
        pages = {2745-2763},
          doi = {10.1093/mnras/staa3803},
archivePrefix = {arXiv},
       eprint = {2012.07420},
 primaryClass = {astro-ph.SR},
       adsurl = {https://ui.adsabs.harvard.edu/abs/2021MNRAS.501.2745M}
}

@ARTICLE{Jeena+23,
       author = {{Jeena}, S.~K. and {Banerjee}, Projjwal and {Chiaki}, Gen and {Heger}, Alexander},
        title = "{Rapidly rotating massive Population III stars: a solution for high carbon enrichment in CEMP-no stars}",
      journal = {\mnras},
         year = 2023,
        month = dec,
       volume = {526},
       number = {3},
        pages = {4467-4483},
          doi = {10.1093/mnras/stad3028},
archivePrefix = {arXiv},
       eprint = {2306.06433},
 primaryClass = {astro-ph.SR},
       adsurl = {https://ui.adsabs.harvard.edu/abs/2023MNRAS.526.4467J}
}

@ARTICLE{Martinet+23,
       author = {{Martinet}, S{\'e}bastien and {Meynet}, Georges and {Ekstr{\"o}m}, Sylvia and {Georgy}, Cyril and {Hirschi}, Raphael},
        title = "{Very massive star models. I. Impact of rotation and metallicity and comparisons with observations}",
      journal = {\aap},
         year = 2023,
        month = nov,
       volume = {679},
          eid = {A137},
        pages = {A137},
          doi = {10.1051/0004-6361/202347514},
archivePrefix = {arXiv},
       eprint = {2309.00062},
 primaryClass = {astro-ph.SR},
       adsurl = {https://ui.adsabs.harvard.edu/abs/2023A&A...679A.137M}
}

@ARTICLE{Roberti+24,
       author = {{Roberti}, Lorenzo and {Limongi}, Marco and {Chieffi}, Alessandro},
        title = "{Zero and Extremely Low-metallicity Rotating Massive Stars: Evolution, Explosion, and Nucleosynthesis Up to the Heaviest Nuclei}",
      journal = {\apjs},
         year = 2024,
        month = feb,
       volume = {270},
       number = {2},
          eid = {28},
        pages = {28},
          doi = {10.3847/1538-4365/ad1686},
archivePrefix = {arXiv},
       eprint = {2312.02942},
 primaryClass = {astro-ph.SR},
       adsurl = {https://ui.adsabs.harvard.edu/abs/2024ApJS..270...28R}
}

@incollection{Salvadori+26,
    author = {{Salvadori}, S.},
    title = "{Data-driven constraints on the Pop III IMF}",
    booktitle = {Very Massive Stars in the Distant Universe},
    editor = {{Vink}, J. S.},
    publisher = {Springer},
    year = {in prep.},
}

@ARTICLE{Christlieb+02,
       author = {{Christlieb}, N. and {Bessell}, M.~S. and {Beers}, T.~C. and {Gustafsson}, B. and {Korn}, A. and {Barklem}, P.~S. and {Karlsson}, T. and {Mizuno-Wiedner}, M. and {Rossi}, S.},
        title = "{A stellar relic from the early Milky Way}",
      journal = {\nat},
         year = 2002,
        month = oct,
       volume = {419},
       number = {6910},
        pages = {904-906},
          doi = {10.1038/nature01142},
archivePrefix = {arXiv},
       eprint = {astro-ph/0211274},
 primaryClass = {astro-ph},
       adsurl = {https://ui.adsabs.harvard.edu/abs/2002Natur.419..904C}
}

@ARTICLE{Christlieb03,
       author = {{Christlieb}, Norbert},
        title = "{Finding the Most Metal-poor Stars of the Galactic Halo with the Hamburg/ESO Objective-prism Survey (With 6 Figures)}",
      journal = {Reviews in Modern Astronomy},
         year = 2003,
        month = jan,
       volume = {16},
        pages = {191},
          doi = {10.1002/9783527617647.ch8},
archivePrefix = {arXiv},
       eprint = {astro-ph/0308016},
 primaryClass = {astro-ph},
       adsurl = {https://ui.adsabs.harvard.edu/abs/2003RvMA...16..191C}
}

@ARTICLE{Frebel+05,
       author = {{Frebel}, Anna and {Aoki}, Wako and {Christlieb}, Norbert and {Ando}, Hiroyasu and {Asplund}, Martin and {Barklem}, Paul S. and {Beers}, Timothy C. and {Eriksson}, Kjell and {Fechner}, Cora and {Fujimoto}, Masayuki Y. and {Honda}, Satoshi and {Kajino}, Toshitaka and {Minezaki}, Takeo and {Nomoto}, Ken'ichi and {Norris}, John E. and {Ryan}, Sean G. and {Takada-Hidai}, Masahide and {Tsangarides}, Stelios and {Yoshii}, Yuzuru},
        title = "{Nucleosynthetic signatures of the first stars}",
      journal = {\nat},
         year = 2005,
        month = apr,
       volume = {434},
       number = {7035},
        pages = {871-873},
          doi = {10.1038/nature03455},
archivePrefix = {arXiv},
       eprint = {astro-ph/0503021},
 primaryClass = {astro-ph},
       adsurl = {https://ui.adsabs.harvard.edu/abs/2005Natur.434..871F}
}

@ARTICLE{Frebel+15,
       author = {{Frebel}, Anna and {Chiti}, Anirudh and {Ji}, Alexander P. and {Jacobson}, Heather R. and {Placco}, Vinicius M.},
        title = "{SD 1313-0019: Another Second-generation Star with [Fe/H] = -5.0, Observed with the Magellan Telescope}",
      journal = {\apjl},
         year = 2015,
        month = sep,
       volume = {810},
       number = {2},
          eid = {L27},
        pages = {L27},
          doi = {10.1088/2041-8205/810/2/L27},
archivePrefix = {arXiv},
       eprint = {1507.01973},
 primaryClass = {astro-ph.SR},
       adsurl = {https://ui.adsabs.harvard.edu/abs/2015ApJ...810L..27F}
}

@ARTICLE{Norris+07,
       author = {{Norris}, John E. and {Christlieb}, N. and {Korn}, A.~J. and {Eriksson}, K. and {Bessell}, M.~S. and {Beers}, Timothy C. and {Wisotzki}, L. and {Reimers}, D.},
        title = "{HE 0557-4840: Ultra-Metal-Poor and Carbon-Rich}",
      journal = {\apj},
         year = 2007,
        month = nov,
       volume = {670},
       number = {1},
        pages = {774-788},
          doi = {10.1086/521919},
archivePrefix = {arXiv},
       eprint = {0707.2657},
 primaryClass = {astro-ph},
       adsurl = {https://ui.adsabs.harvard.edu/abs/2007ApJ...670..774N}
}

@ARTICLE{Caffau+11,
       author = {{Caffau}, Elisabetta and {Bonifacio}, Piercarlo and {Fran{\c{c}}ois}, Patrick and {Sbordone}, Luca and {Monaco}, Lorenzo and {Spite}, Monique and {Spite}, Fran{\c{c}}ois and {Ludwig}, Hans-G. and {Cayrel}, Roger and {Zaggia}, Simone and {Hammer}, Fran{\c{c}}ois and {Randich}, Sofia and {Molaro}, Paolo and {Hill}, Vanessa},
        title = "{An extremely primitive star in the Galactic halo}",
      journal = {\nat},
         year = 2011,
        month = sep,
       volume = {477},
       number = {7362},
        pages = {67-69},
          doi = {10.1038/nature10377},
archivePrefix = {arXiv},
       eprint = {1203.2612},
 primaryClass = {astro-ph.GA},
       adsurl = {https://ui.adsabs.harvard.edu/abs/2011Natur.477...67C}
}

@ARTICLE{Caffau+13,
       author = {{Caffau}, E. and {Bonifacio}, P. and {Fran{\c{c}}ois}, P. and {Sbordone}, L. and {Spite}, M. and {Monaco}, L. and {Plez}, B. and {Spite}, F. and {Zaggia}, S. and {Ludwig}, H.-G. and {Cayrel}, R. and {Molaro}, P. and {Randich}, S. and {Hammer}, F. and {Hill}, V.},
        title = "{X-shooter GTO: evidence for a population of extremely metal-poor, alpha-poor stars}",
      journal = {\aap},
         year = 2013,
        month = dec,
       volume = {560},
          eid = {A15},
        pages = {A15},
          doi = {10.1051/0004-6361/201322213},
archivePrefix = {arXiv},
       eprint = {1309.4913},
 primaryClass = {astro-ph.GA},
       adsurl = {https://ui.adsabs.harvard.edu/abs/2013A&A...560A..15C}
}

@ARTICLE{Yong+13a,
       author = {{Yong}, David and {Norris}, John E. and {Bessell}, M.~S. and {Christlieb}, N. and {Asplund}, M. and {Beers}, Timothy C. and {Barklem}, P.~S. and {Frebel}, Anna and {Ryan}, S.~G.},
        title = "{The Most Metal-poor Stars. II. Chemical Abundances of 190 Metal-poor Stars Including 10 New Stars with [Fe/H] <= -3.5}",
      journal = {\apj},
         year = 2013,
        month = jan,
       volume = {762},
       number = {1},
          eid = {26},
        pages = {26},
          doi = {10.1088/0004-637X/762/1/26},
archivePrefix = {arXiv},
       eprint = {1208.3003},
 primaryClass = {astro-ph.GA},
       adsurl = {https://ui.adsabs.harvard.edu/abs/2013ApJ...762...26Y}
}

@ARTICLE{Yong+13b,
       author = {{Yong}, David and {Norris}, John E. and {Bessell}, M.~S. and {Christlieb}, N. and {Asplund}, M. and {Beers}, Timothy C. and {Barklem}, P.~S. and {Frebel}, Anna and {Ryan}, S.~G.},
        title = "{The Most Metal-poor Stars. III. The Metallicity Distribution Function and Carbon-enhanced Metal-poor Fraction}",
      journal = {\apj},
         year = 2013,
        month = jan,
       volume = {762},
       number = {1},
          eid = {27},
        pages = {27},
          doi = {10.1088/0004-637X/762/1/27},
archivePrefix = {arXiv},
       eprint = {1208.3016},
 primaryClass = {astro-ph.GA},
       adsurl = {https://ui.adsabs.harvard.edu/abs/2013ApJ...762...27Y}
}

@ARTICLE{Lee+13,
       author = {{Lee}, Young Sun and {Beers}, Timothy C. and {Masseron}, Thomas and {Plez}, Bertrand and {Rockosi}, Constance M. and {Sobeck}, Jennifer and {Yanny}, Brian and {Lucatello}, Sara and {Sivarani}, Thirupathi and {Placco}, Vinicius M. and {Carollo}, Daniela},
        title = "{Carbon-enhanced Metal-poor Stars in SDSS/SEGUE. I. Carbon Abundance Estimation and Frequency of CEMP Stars}",
      journal = {\aj},
         year = 2013,
        month = nov,
       volume = {146},
       number = {5},
          eid = {132},
        pages = {132},
          doi = {10.1088/0004-6256/146/5/132},
archivePrefix = {arXiv},
       eprint = {1310.3276},
 primaryClass = {astro-ph.SR},
       adsurl = {https://ui.adsabs.harvard.edu/abs/2013AJ....146..132L}
}

@ARTICLE{Placco+13,
       author = {{Placco}, Vinicius M. and {Frebel}, Anna and {Beers}, Timothy C. and {Karakas}, Amanda I. and {Kennedy}, Catherine R. and {Rossi}, Silvia and {Christlieb}, Norbert and {Stancliffe}, Richard J.},
        title = "{Metal-poor Stars Observed with the Magellan Telescope. I. Constraints on Progenitor Mass and Metallicity of AGB Stars Undergoing s-process Nucleosynthesis}",
      journal = {\apj},
         year = 2013,
        month = jun,
       volume = {770},
       number = {2},
          eid = {104},
        pages = {104},
          doi = {10.1088/0004-637X/770/2/104},
archivePrefix = {arXiv},
       eprint = {1304.7869},
 primaryClass = {astro-ph.GA},
       adsurl = {https://ui.adsabs.harvard.edu/abs/2013ApJ...770..104P}
}

@ARTICLE{Placco+14a,
       author = {{Placco}, Vinicius M. and {Frebel}, Anna and {Beers}, Timothy C. and {Christlieb}, Norbert and {Lee}, Young Sun and {Kennedy}, Catherine R. and {Rossi}, Silvia and {Santucci}, Rafael M.},
        title = "{Metal-poor Stars Observed with the Magellan Telescope. II. Discovery of Four Stars with [Fe/H] <= -3.5}",
      journal = {\apj},
         year = 2014,
        month = jan,
       volume = {781},
       number = {1},
          eid = {40},
        pages = {40},
          doi = {10.1088/0004-637X/781/1/40},
archivePrefix = {arXiv},
       eprint = {1311.5855},
 primaryClass = {astro-ph.SR},
       adsurl = {https://ui.adsabs.harvard.edu/abs/2014ApJ...781...40P}
}

@ARTICLE{Placco+14b,
       author = {{Placco}, Vinicius M. and {Frebel}, Anna and {Beers}, Timothy C. and {Stancliffe}, Richard J.},
        title = "{Carbon-enhanced Metal-poor Star Frequencies in the Galaxy: Corrections for the Effect of Evolutionary Status on Carbon Abundances}",
      journal = {\apj},
         year = 2014,
        month = dec,
       volume = {797},
       number = {1},
          eid = {21},
        pages = {21},
          doi = {10.1088/0004-637X/797/1/21},
archivePrefix = {arXiv},
       eprint = {1410.2223},
 primaryClass = {astro-ph.SR},
       adsurl = {https://ui.adsabs.harvard.edu/abs/2014ApJ...797...21P}
}

@ARTICLE{Keller+14,
       author = {{Keller}, S.~C. and {Bessell}, M.~S. and {Frebel}, A. and {Casey}, A.~R. and {Asplund}, M. and {Jacobson}, H.~R. and {Lind}, K. and {Norris}, J.~E. and {Yong}, D. and {Heger}, A. and {Magic}, Z. and {da Costa}, G.~S. and {Schmidt}, B.~P. and {Tisserand}, P.},
        title = "{A single low-energy, iron-poor supernova as the source of metals in the star SMSS J031300.36-670839.3}",
      journal = {\nat},
         year = 2014,
        month = feb,
       volume = {506},
       number = {7489},
        pages = {463-466},
          doi = {10.1038/nature12990},
archivePrefix = {arXiv},
       eprint = {1402.1517},
 primaryClass = {astro-ph.SR},
       adsurl = {https://ui.adsabs.harvard.edu/abs/2014Natur.506..463K}
}

@ARTICLE{Spite+14,
       author = {{Spite}, M. and {Spite}, F. and {Bonifacio}, P. and {Caffau}, E. and {Fran{\c{c}}ois}, P. and {Sbordone}, L.},
        title = "{The low Sr/Ba ratio on some extremely metal-poor stars}",
      journal = {\aap},
         year = 2014,
        month = nov,
       volume = {571},
          eid = {A40},
        pages = {A40},
          doi = {10.1051/0004-6361/201423658},
archivePrefix = {arXiv},
       eprint = {1410.0847},
 primaryClass = {astro-ph.SR},
       adsurl = {https://ui.adsabs.harvard.edu/abs/2014A&A...571A..40S}
}

@ARTICLE{Hansen+14,
       author = {{Hansen}, T. and {Hansen}, C.~J. and {Christlieb}, N. and {Yong}, D. and {Bessell}, M.~S. and {Garc{\'\i}a P{\'e}rez}, A.~E. and {Beers}, T.~C. and {Placco}, V.~M. and {Frebel}, A. and {Norris}, J.~E. and {Asplund}, M.},
        title = "{Exploring the Origin of Lithium, Carbon, Strontium, and Barium with Four New Ultra Metal-poor Stars}",
      journal = {\apj},
         year = 2014,
        month = jun,
       volume = {787},
       number = {2},
          eid = {162},
        pages = {162},
          doi = {10.1088/0004-637X/787/2/162},
archivePrefix = {arXiv},
       eprint = {1405.5846},
 primaryClass = {astro-ph.SR},
       adsurl = {https://ui.adsabs.harvard.edu/abs/2014ApJ...787..162H}
}

@ARTICLE{Hansen+15,
       author = {{Hansen}, T. and {Hansen}, C.~J. and {Christlieb}, N. and {Beers}, T.~C. and {Yong}, D. and {Bessell}, M.~S. and {Frebel}, A. and {Garc{\'\i}a P{\'e}rez}, A.~E. and {Placco}, V.~M. and {Norris}, J.~E. and {Asplund}, M.},
        title = "{An Elemental Assay of Very, Extremely, and Ultra-metal-poor Stars}",
      journal = {\apj},
         year = 2015,
        month = jul,
       volume = {807},
       number = {2},
          eid = {173},
        pages = {173},
          doi = {10.1088/0004-637X/807/2/173},
archivePrefix = {arXiv},
       eprint = {1506.00579},
 primaryClass = {astro-ph.SR},
       adsurl = {https://ui.adsabs.harvard.edu/abs/2015ApJ...807..173H}
}

@ARTICLE{Bonifacio+15,
       author = {{Bonifacio}, P. and {Caffau}, E. and {Spite}, M. and {Limongi}, M. and {Chieffi}, A. and {Klessen}, R.~S. and {Fran{\c{c}}ois}, P. and {Molaro}, P. and {Ludwig}, H.-G. and {Zaggia}, S. and {Spite}, F. and {Plez}, B. and {Cayrel}, R. and {Christlieb}, N. and {Clark}, P.~C. and {Glover}, S.~C.~O. and {Hammer}, F. and {Koch}, A. and {Monaco}, L. and {Sbordone}, L. and {Steffen}, M.},
        title = "{TOPoS . II. On the bimodality of carbon abundance in CEMP stars Implications on the early chemical evolution of galaxies}",
      journal = {\aap},
         year = 2015,
        month = jul,
       volume = {579},
          eid = {A28},
        pages = {A28},
          doi = {10.1051/0004-6361/201425266},
archivePrefix = {arXiv},
       eprint = {1504.05963},
 primaryClass = {astro-ph.GA},
       adsurl = {https://ui.adsabs.harvard.edu/abs/2015A&A...579A..28B}
}

@ARTICLE{Bonifacio+21,
       author = {{Bonifacio}, P. and {Monaco}, L. and {Salvadori}, S. and {Caffau}, E. and {Spite}, M. and {Sbordone}, L. and {Spite}, F. and {Ludwig}, H.-G. and {Di Matteo}, P. and {Haywood}, M. and {Fran{\c{c}}ois}, P. and {Koch-Hansen}, A.~J. and {Christlieb}, N. and {Zaggia}, S.},
        title = "{TOPoS. VI. The metal-weak tail of the metallicity distribution functions of the Milky Way and the Gaia-Sausage-Enceladus structure}",
      journal = {\aap},
         year = 2021,
        month = jul,
       volume = {651},
          eid = {A79},
        pages = {A79},
          doi = {10.1051/0004-6361/202140816},
archivePrefix = {arXiv},
       eprint = {2105.08360},
 primaryClass = {astro-ph.GA},
       adsurl = {https://ui.adsabs.harvard.edu/abs/2021A&A...651A..79B}
}

@ARTICLE{Starkenburg+18,
       author = {{Starkenburg}, Else and {Aguado}, David S. and {Bonifacio}, Piercarlo and {Caffau}, Elisabetta and {Jablonka}, Pascale and {Lardo}, Carmela and {Martin}, Nicolas and {S{\'a}nchez-Janssen}, Rub{\'e}n and {Sestito}, Federico and {Venn}, Kim A. and {Youakim}, Kris and {Allende Prieto}, Carlos and {Arentsen}, Anke and {Gentile}, Marc and {Gonz{\'a}lez Hern{\'a}ndez}, Jonay I. and {Kielty}, Collin and {Koppelman}, Helmer H. and {Longeard}, Nicolas and {Tolstoy}, Eline and {Carlberg}, Raymond G. and {C{\^o}t{\'e}}, Patrick and {Fouesneau}, Morgan and {Hill}, Vanessa and {McConnachie}, Alan W. and {Navarro}, Julio F.},
        title = "{The Pristine survey IV: approaching the Galactic metallicity floor with the discovery of an ultra-metal-poor star}",
      journal = {\mnras},
         year = 2018,
        month = dec,
       volume = {481},
       number = {3},
        pages = {3838-3852},
          doi = {10.1093/mnras/sty2276},
archivePrefix = {arXiv},
       eprint = {1807.04292},
 primaryClass = {astro-ph.SR},
       adsurl = {https://ui.adsabs.harvard.edu/abs/2018MNRAS.481.3838S}
}

@ARTICLE{Aguado+18a,
       author = {{Aguado}, David S. and {Gonz{\'a}lez Hern{\'a}ndez}, Jonay I. and {Allende Prieto}, Carlos and {Rebolo}, Rafael},
        title = "{J0815+4729: A Chemically Primitive Dwarf Star in the Galactic Halo Observed with Gran Telescopio Canarias}",
      journal = {\apjl},
         year = 2018,
        month = jan,
       volume = {852},
       number = {1},
          eid = {L20},
        pages = {L20},
          doi = {10.3847/2041-8213/aaa23a},
archivePrefix = {arXiv},
       eprint = {1712.06487},
 primaryClass = {astro-ph.SR},
       adsurl = {https://ui.adsabs.harvard.edu/abs/2018ApJ...852L..20A}
}

@ARTICLE{Aguado+18b,
       author = {{Aguado}, David S. and {Allende Prieto}, Carlos and {Gonz{\'a}lez Hern{\'a}ndez}, Jonay I. and {Rebolo}, Rafael},
        title = "{J0023+0307: A Mega Metal-poor Dwarf Star from SDSS/BOSS}",
      journal = {\apjl},
         year = 2018,
        month = feb,
       volume = {854},
       number = {2},
          eid = {L34},
        pages = {L34},
          doi = {10.3847/2041-8213/aaadb8},
archivePrefix = {arXiv},
       eprint = {1802.06240},
 primaryClass = {astro-ph.SR},
       adsurl = {https://ui.adsabs.harvard.edu/abs/2018ApJ...854L..34A}
}

@ARTICLE{Aguado+19,
       author = {{Aguado}, David S. and {Gonz{\'a}lez Hern{\'a}ndez}, Jonay I. and {Allende Prieto}, Carlos and {Rebolo}, Rafael},
        title = "{Back to the Lithium Plateau with the [Fe/H] < -6 Star J0023+0307}",
      journal = {\apjl},
         year = 2019,
        month = apr,
       volume = {874},
       number = {2},
          eid = {L21},
        pages = {L21},
          doi = {10.3847/2041-8213/ab1076},
archivePrefix = {arXiv},
       eprint = {1904.04892},
 primaryClass = {astro-ph.SR},
       adsurl = {https://ui.adsabs.harvard.edu/abs/2019ApJ...874L..21A}
}

@ARTICLE{Aguado+23a,
       author = {{Aguado}, D.~S. and {Caffau}, E. and {Molaro}, P. and {Allende Prieto}, C. and {Bonifacio}, P. and {Gonz{\'a}lez Hern{\'a}ndez}, J.~I. and {Rebolo}, R. and {Salvadori}, S. and {Zapatero Osorio}, M.~R. and {Cristiani}, S. and {Pepe}, F. and {Santos}, N.~C. and {Cupani}, G. and {Di Marcantonio}, P. and {D'Odorico}, V. and {Lovis}, C. and {Nunes}, N.~J. and {Martins}, C.~J.~A.~P. and {Milakovi}, D. and {Rodrigues}, J. and {Schmidt}, T.~M. and {Sozzetti}, A. and {Su{\'a}rez Mascare{\~n}o}, A.},
        title = "{The pristine nature of SMSS 1605‒1443 revealed by ESPRESSO}",
      journal = {\aap},
         year = 2023,
        month = jan,
       volume = {669},
          eid = {L4},
        pages = {L4},
          doi = {10.1051/0004-6361/202245392},
archivePrefix = {arXiv},
       eprint = {2301.02678},
 primaryClass = {astro-ph.SR},
       adsurl = {https://ui.adsabs.harvard.edu/abs/2023A&A...669L...4A}
}

@ARTICLE{Francois+18,
       author = {{Fran{\c{c}}ois}, P. and {Caffau}, E. and {Wanajo}, S. and {Aguado}, D. and {Spite}, M. and {Aoki}, M. and {Aoki}, W. and {Bonifacio}, P. and {Gallagher}, A.~J. and {Salvadori}, S. and {Spite}, F.},
        title = "{Chemical analysis of very metal-poor turn-off stars from SDSS-DR12}",
      journal = {\aap},
         year = 2018,
        month = nov,
       volume = {619},
          eid = {A10},
        pages = {A10},
          doi = {10.1051/0004-6361/201833824},
archivePrefix = {arXiv},
       eprint = {1808.09918},
 primaryClass = {astro-ph.GA},
       adsurl = {https://ui.adsabs.harvard.edu/abs/2018A&A...619A..10F}
}

@ARTICLE{Nordlander+19,
       author = {{Nordlander}, T. and {Bessell}, M.~S. and {Da Costa}, G.~S. and {Mackey}, A.~D. and {Asplund}, M. and {Casey}, A.~R. and {Chiti}, A. and {Ezzeddine}, R. and {Frebel}, A. and {Lind}, K. and {Marino}, A.~F. and {Murphy}, S.~J. and {Norris}, J.~E. and {Schmidt}, B.~P. and {Yong}, D.},
        title = "{The lowest detected stellar Fe abundance: the halo star SMSS J160540.18-144323.1}",
      journal = {\mnras},
         year = 2019,
        month = sep,
       volume = {488},
       number = {1},
        pages = {L109-L113},
          doi = {10.1093/mnrasl/slz109},
archivePrefix = {arXiv},
       eprint = {1904.07471},
 primaryClass = {astro-ph.SR},
       adsurl = {https://ui.adsabs.harvard.edu/abs/2019MNRAS.488L.109N}
}

@ARTICLE{Kirby+08,
       author = {{Kirby}, Evan N. and {Simon}, Joshua D. and {Geha}, Marla and {Guhathakurta}, Puragra and {Frebel}, Anna},
        title = "{Uncovering Extremely Metal-Poor Stars in the Milky Way's Ultrafaint Dwarf Spheroidal Satellite Galaxies}",
      journal = {\apjl},
         year = 2008,
        month = sep,
       volume = {685},
       number = {1},
        pages = {L43},
          doi = {10.1086/592432},
archivePrefix = {arXiv},
       eprint = {0807.1925},
 primaryClass = {astro-ph},
       adsurl = {https://ui.adsabs.harvard.edu/abs/2008ApJ...685L..43K}
}

@ARTICLE{KirbyCohen12,
       author = {{Kirby}, Evan N. and {Cohen}, Judith G.},
        title = "{Detailed Abundances of Two Very Metal-poor Stars in Dwarf Galaxies}",
      journal = {\aj},
         year = 2012,
        month = dec,
       volume = {144},
       number = {6},
          eid = {168},
        pages = {168},
          doi = {10.1088/0004-6256/144/6/168},
archivePrefix = {arXiv},
       eprint = {1209.3778},
 primaryClass = {astro-ph.GA},
       adsurl = {https://ui.adsabs.harvard.edu/abs/2012AJ....144..168K}
}

@ARTICLE{Norris+10a,
       author = {{Norris}, John E. and {Yong}, David and {Gilmore}, Gerard and {Wyse}, Rosemary F.~G.},
        title = "{Boo-1137{\textemdash}an Extremely Metal-Poor Star in the Ultra-Faint Dwarf Spheroidal Galaxy Bo{\"o}tes I}",
      journal = {\apj},
         year = 2010,
        month = mar,
       volume = {711},
       number = {1},
        pages = {350-360},
          doi = {10.1088/0004-637X/711/1/350},
archivePrefix = {arXiv},
       eprint = {0911.5350},
 primaryClass = {astro-ph.GA},
       adsurl = {https://ui.adsabs.harvard.edu/abs/2010ApJ...711..350N}
}

@ARTICLE{Norris+10b,
       author = {{Norris}, John E. and {Gilmore}, Gerard and {Wyse}, Rosemary F.~G. and {Yong}, David and {Frebel}, Anna},
        title = "{An Extremely Carbon-rich, Extremely Metal-poor Star in the Segue 1 System}",
      journal = {\apjl},
         year = 2010,
        month = oct,
       volume = {722},
       number = {1},
        pages = {L104-L109},
          doi = {10.1088/2041-8205/722/1/L104},
archivePrefix = {arXiv},
       eprint = {1008.0450},
 primaryClass = {astro-ph.GA},
       adsurl = {https://ui.adsabs.harvard.edu/abs/2010ApJ...722L.104N}
}

@ARTICLE{Norris+10c,
       author = {{Norris}, John E. and {Wyse}, Rosemary F.~G. and {Gilmore}, Gerard and {Yong}, David and {Frebel}, Anna and {Wilkinson}, Mark I. and {Belokurov}, V. and {Zucker}, Daniel B.},
        title = "{Chemical Enrichment in the Faintest Galaxies: The Carbon and Iron Abundance Spreads in the Bo{\"o}tes I Dwarf Spheroidal Galaxy and the Segue 1 System}",
      journal = {\apj},
         year = 2010,
        month = nov,
       volume = {723},
       number = {2},
        pages = {1632-1650},
          doi = {10.1088/0004-637X/723/2/1632},
archivePrefix = {arXiv},
       eprint = {1008.0137},
 primaryClass = {astro-ph.GA},
       adsurl = {https://ui.adsabs.harvard.edu/abs/2010ApJ...723.1632N}
}

@ARTICLE{Simon+11,
       author = {{Simon}, Joshua D. and {Geha}, Marla and {Minor}, Quinn E. and {Martinez}, Gregory D. and {Kirby}, Evan N. and {Bullock}, James S. and {Kaplinghat}, Manoj and {Strigari}, Louis E. and {Willman}, Beth and {Choi}, Philip I. and {Tollerud}, Erik J. and {Wolf}, Joe},
        title = "{A Complete Spectroscopic Survey of the Milky Way Satellite Segue 1: The Darkest Galaxy}",
      journal = {\apj},
         year = 2011,
        month = may,
       volume = {733},
       number = {1},
          eid = {46},
        pages = {46},
          doi = {10.1088/0004-637X/733/1/46},
archivePrefix = {arXiv},
       eprint = {1007.4198},
 primaryClass = {astro-ph.GA},
       adsurl = {https://ui.adsabs.harvard.edu/abs/2011ApJ...733...46S}
}

@ARTICLE{Yoon+19,
       author = {{Yoon}, Jinmi and {Beers}, Timothy C. and {Tian}, Di and {Whitten}, Devin D.},
        title = "{Origin of the CEMP-no Group Morphology in the Milky Way}",
      journal = {\apj},
         year = 2019,
        month = jun,
       volume = {878},
       number = {2},
          eid = {97},
        pages = {97},
          doi = {10.3847/1538-4357/ab1ead},
archivePrefix = {arXiv},
       eprint = {1904.02758},
 primaryClass = {astro-ph.SR},
       adsurl = {https://ui.adsabs.harvard.edu/abs/2019ApJ...878...97Y}
}

@ARTICLE{Lucatello+05,
       author = {{Lucatello}, Sara and {Tsangarides}, Stelios and {Beers}, Timothy C. and {Carretta}, Eugenio and {Gratton}, Raffaele G. and {Ryan}, Sean G.},
        title = "{The Binary Frequency Among Carbon-enhanced, s-Process-rich, Metal-poor Stars}",
      journal = {\apj},
         year = 2005,
        month = jun,
       volume = {625},
       number = {2},
        pages = {825-832},
          doi = {10.1086/428104},
archivePrefix = {arXiv},
       eprint = {astro-ph/0412422},
 primaryClass = {astro-ph},
       adsurl = {https://ui.adsabs.harvard.edu/abs/2005ApJ...625..825L}
}

@ARTICLE{Busso+99,
       author = {{Busso}, M. and {Gallino}, R. and {Wasserburg}, G.~J.},
        title = "{Nucleosynthesis in Asymptotic Giant Branch Stars: Relevance for Galactic Enrichment and Solar System Formation}",
      journal = {\araa},
         year = 1999,
        month = jan,
       volume = {37},
        pages = {239-309},
          doi = {10.1146/annurev.astro.37.1.239},
       adsurl = {https://ui.adsabs.harvard.edu/abs/1999ARA&A..37..239B}
}

@ARTICLE{Starkenburg+14,
       author = {{Starkenburg}, Else and {Shetrone}, Matthew D. and {McConnachie}, Alan W. and {Venn}, Kim A.},
        title = "{Binarity in carbon-enhanced metal-poor stars}",
      journal = {\mnras},
         year = 2014,
        month = jun,
       volume = {441},
       number = {2},
        pages = {1217-1229},
          doi = {10.1093/mnras/stu623},
archivePrefix = {arXiv},
       eprint = {1404.0385},
 primaryClass = {astro-ph.SR},
       adsurl = {https://ui.adsabs.harvard.edu/abs/2014MNRAS.441.1217S}
}

@ARTICLE{Tumlinson07a,
       author = {{Tumlinson}, Jason},
        title = "{Carbon-Enhanced Metal-poor Stars, the Cosmic Microwave Background, and the Stellar Initial Mass Function in the Early Universe}",
      journal = {\apjl},
         year = 2007,
        month = aug,
       volume = {664},
       number = {2},
        pages = {L63-L66},
          doi = {10.1086/520930},
archivePrefix = {arXiv},
       eprint = {0706.2903},
 primaryClass = {astro-ph},
       adsurl = {https://ui.adsabs.harvard.edu/abs/2007ApJ...664L..63T}
}

@ARTICLE{Tumlinson07b,
       author = {{Tumlinson}, Jason},
        title = "{Carbon-enhanced Hyper-Metal-poor Stars and the Stellar IMF at Low Metallicity}",
      journal = {\apj},
         year = 2007,
        month = aug,
       volume = {665},
       number = {2},
        pages = {1361-1370},
          doi = {10.1086/519917},
archivePrefix = {arXiv},
       eprint = {0707.0816},
 primaryClass = {astro-ph},
       adsurl = {https://ui.adsabs.harvard.edu/abs/2007ApJ...665.1361T}
}

@ARTICLE{Salvadori+07,
       author = {{Salvadori}, Stefania and {Schneider}, Raffaella and {Ferrara}, Andrea},
        title = "{Cosmic stellar relics in the Galactic halo}",
      journal = {\mnras},
         year = 2007,
        month = oct,
       volume = {381},
       number = {2},
        pages = {647-662},
          doi = {10.1111/j.1365-2966.2007.12133.x},
archivePrefix = {arXiv},
       eprint = {astro-ph/0611130},
 primaryClass = {astro-ph},
       adsurl = {https://ui.adsabs.harvard.edu/abs/2007MNRAS.381..647S}
}

@ARTICLE{Salvadori+15,
       author = {{Salvadori}, Stefania and {Sk{\'u}lad{\'o}ttir}, {\'A}sa and {Tolstoy}, Eline},
        title = "{Carbon-enhanced metal-poor stars in dwarf galaxies}",
      journal = {\mnras},
         year = 2015,
        month = dec,
       volume = {454},
       number = {2},
        pages = {1320-1331},
          doi = {10.1093/mnras/stv1969},
archivePrefix = {arXiv},
       eprint = {1506.03451},
 primaryClass = {astro-ph.GA},
       adsurl = {https://ui.adsabs.harvard.edu/abs/2015MNRAS.454.1320S}
}

@ARTICLE{deBennassuti+14,
       author = {{de Bennassuti}, Matteo and {Schneider}, Raffaella and {Valiante}, Rosa and {Salvadori}, Stefania},
        title = "{Decoding the stellar fossils of the dusty Milky Way progenitors}",
      journal = {\mnras},
         year = 2014,
        month = dec,
       volume = {445},
       number = {3},
        pages = {3039-3054},
          doi = {10.1093/mnras/stu1962},
archivePrefix = {arXiv},
       eprint = {1409.5798},
 primaryClass = {astro-ph.GA},
       adsurl = {https://ui.adsabs.harvard.edu/abs/2014MNRAS.445.3039D}
}

@ARTICLE{deBennassuti+17,
       author = {{de Bennassuti}, M. and {Salvadori}, S. and {Schneider}, R. and {Valiante}, R. and {Omukai}, K.},
        title = "{Limits on Population III star formation with the most iron-poor stars}",
      journal = {\mnras},
         year = 2017,
        month = feb,
       volume = {465},
       number = {1},
        pages = {926-940},
          doi = {10.1093/mnras/stw2687},
archivePrefix = {arXiv},
       eprint = {1610.05777},
 primaryClass = {astro-ph.GA},
       adsurl = {https://ui.adsabs.harvard.edu/abs/2017MNRAS.465..926D}
}

@ARTICLE{Graziani+15,
       author = {{Graziani}, L. and {Salvadori}, S. and {Schneider}, R. and {Kawata}, D. and {de Bennassuti}, M. and {Maselli}, A.},
        title = "{Galaxy formation with radiative and chemical feedback}",
      journal = {\mnras},
         year = 2015,
        month = may,
       volume = {449},
       number = {3},
        pages = {3137-3148},
          doi = {10.1093/mnras/stv494},
archivePrefix = {arXiv},
       eprint = {1502.07344},
 primaryClass = {astro-ph.GA},
       adsurl = {https://ui.adsabs.harvard.edu/abs/2015MNRAS.449.3137G}
}

@ARTICLE{Graziani+17,
       author = {{Graziani}, L. and {de Bennassuti}, M. and {Schneider}, R. and {Kawata}, D. and {Salvadori}, S.},
        title = "{The history of the dark and luminous side of Milky Way-like progenitors}",
      journal = {\mnras},
         year = 2017,
        month = jul,
       volume = {469},
       number = {1},
        pages = {1101-1116},
          doi = {10.1093/mnras/stx900},
archivePrefix = {arXiv},
       eprint = {1704.02983},
 primaryClass = {astro-ph.GA},
       adsurl = {https://ui.adsabs.harvard.edu/abs/2017MNRAS.469.1101G}
}

@ARTICLE{Hartwig+15,
       author = {{Hartwig}, Tilman and {Bromm}, Volker and {Klessen}, Ralf S. and {Glover}, Simon C.~O.},
        title = "{Constraining the primordial initial mass function with stellar archaeology}",
      journal = {\mnras},
         year = 2015,
        month = mar,
       volume = {447},
       number = {4},
        pages = {3892-3908},
          doi = {10.1093/mnras/stu2740},
archivePrefix = {arXiv},
       eprint = {1411.1238},
 primaryClass = {astro-ph.GA},
       adsurl = {https://ui.adsabs.harvard.edu/abs/2015MNRAS.447.3892H}
}

@ARTICLE{Hartwig+18a,
       author = {{Hartwig}, Tilman and {Yoshida}, Naoki and {Magg}, Mattis and {Frebel}, Anna and {Glover}, Simon C.~O. and {G{\'o}mez}, Facundo A. and {Griffen}, Brendan and {Ishigaki}, Miho N. and {Ji}, Alexander P. and {Klessen}, Ralf S. and {O'Shea}, Brian W. and {Tominaga}, Nozomu},
        title = "{Descendants of the first stars: the distinct chemical signature of second-generation stars}",
      journal = {\mnras},
         year = 2018,
        month = aug,
       volume = {478},
       number = {2},
        pages = {1795-1810},
          doi = {10.1093/mnras/sty1176},
archivePrefix = {arXiv},
       eprint = {1801.05044},
 primaryClass = {astro-ph.GA},
       adsurl = {https://ui.adsabs.harvard.edu/abs/2018MNRAS.478.1795H}
}

@ARTICLE{Hartwig+19,
       author = {{Hartwig}, Tilman and {Ishigaki}, Miho N. and {Klessen}, Ralf S. and {Yoshida}, Naoki},
        title = "{Fingerprint of the first stars: multi-enriched extremely metal-poor stars in the TOPoS survey}",
      journal = {\mnras},
         year = 2019,
        month = jan,
       volume = {482},
       number = {1},
        pages = {1204-1210},
          doi = {10.1093/mnras/sty2783},
archivePrefix = {arXiv},
       eprint = {1810.04713},
 primaryClass = {astro-ph.GA},
       adsurl = {https://ui.adsabs.harvard.edu/abs/2019MNRAS.482.1204H}
}

@ARTICLE{Hartwig+23,
       author = {{Hartwig}, Tilman and {Ishigaki}, Miho N. and {Kobayashi}, Chiaki and {Tominaga}, Nozomu and {Nomoto}, Ken'ichi},
        title = "{Machine Learning Detects Multiplicity of the First Stars in Stellar Archaeology Data}",
      journal = {\apj},
         year = 2023,
        month = mar,
       volume = {946},
       number = {1},
          eid = {20},
        pages = {20},
          doi = {10.3847/1538-4357/acbcc6},
archivePrefix = {arXiv},
       eprint = {2302.04366},
 primaryClass = {astro-ph.GA},
       adsurl = {https://ui.adsabs.harvard.edu/abs/2023ApJ...946...20H}
}

@ARTICLE{Ishigaki+18,
       author = {{Ishigaki}, Miho N. and {Tominaga}, Nozomu and {Kobayashi}, Chiaki and {Nomoto}, Ken'ichi},
        title = "{The Initial Mass Function of the First Stars Inferred from Extremely Metal-poor Stars}",
      journal = {\apj},
         year = 2018,
        month = apr,
       volume = {857},
       number = {1},
          eid = {46},
        pages = {46},
          doi = {10.3847/1538-4357/aab3de},
archivePrefix = {arXiv},
       eprint = {1801.07763},
 primaryClass = {astro-ph.SR},
       adsurl = {https://ui.adsabs.harvard.edu/abs/2018ApJ...857...46I}
}

@ARTICLE{Sarmento+19,
       author = {{Sarmento}, Richard and {Scannapieco}, Evan and {C{\^o}t{\'e}}, Benoit},
        title = "{Following the Cosmic Evolution of Pristine Gas. III. The Observational Consequences of the Unknown Properties of Population III Stars}",
      journal = {\apj},
         year = 2019,
        month = feb,
       volume = {871},
       number = {2},
          eid = {206},
        pages = {206},
          doi = {10.3847/1538-4357/aafa1a},
archivePrefix = {arXiv},
       eprint = {1901.03727},
 primaryClass = {astro-ph.GA},
       adsurl = {https://ui.adsabs.harvard.edu/abs/2019ApJ...871..206S}
}

@ARTICLE{Magg+18,
       author = {{Magg}, Mattis and {Hartwig}, Tilman and {Agarwal}, Bhaskar and {Frebel}, Anna and {Glover}, Simon C.~O. and {Griffen}, Brendan F. and {Klessen}, Ralf S.},
        title = "{Predicting the locations of possible long-lived low-mass first stars: importance of satellite dwarf galaxies}",
      journal = {\mnras},
         year = 2018,
        month = feb,
       volume = {473},
       number = {4},
        pages = {5308-5323},
          doi = {10.1093/mnras/stx2729},
archivePrefix = {arXiv},
       eprint = {1706.07054},
 primaryClass = {astro-ph.GA},
       adsurl = {https://ui.adsabs.harvard.edu/abs/2018MNRAS.473.5308M}
}

@ARTICLE{Magg+19,
       author = {{Magg}, Mattis and {Klessen}, Ralf S. and {Glover}, Simon C.~O. and {Li}, Haining},
        title = "{Observational constraints on the survival of pristine stars}",
      journal = {\mnras},
         year = 2019,
        month = jul,
       volume = {487},
       number = {1},
        pages = {486-490},
          doi = {10.1093/mnras/stz1210},
archivePrefix = {arXiv},
       eprint = {1903.08661},
 primaryClass = {astro-ph.GA},
       adsurl = {https://ui.adsabs.harvard.edu/abs/2019MNRAS.487..486M}
}

@ARTICLE{Rossi+21,
       author = {{Rossi}, Martina and {Salvadori}, Stefania and {Sk{\'u}lad{\'o}ttir}, {\'A}sa},
        title = "{Ultra-faint dwarf galaxies: unveiling the minimum mass of the first stars}",
      journal = {\mnras},
         year = 2021,
        month = jun,
       volume = {503},
       number = {4},
        pages = {6026-6044},
          doi = {10.1093/mnras/stab821},
archivePrefix = {arXiv},
       eprint = {2103.09834},
 primaryClass = {astro-ph.GA},
       adsurl = {https://ui.adsabs.harvard.edu/abs/2021MNRAS.503.6026R}
}

@ARTICLE{Rossi+23,
       author = {{Rossi}, Martina and {Salvadori}, Stefania and {Sk{\'u}lad{\'o}ttir}, {\'A}sa and {Vanni}, Irene},
        title = "{Understanding the origin of CEMP - no stars through ultra-faint dwarfs}",
      journal = {\mnras},
         year = 2023,
        month = jun,
       volume = {522},
       number = {1},
        pages = {L1-L5},
          doi = {10.1093/mnrasl/slad029},
archivePrefix = {arXiv},
       eprint = {2302.10210},
 primaryClass = {astro-ph.GA},
       adsurl = {https://ui.adsabs.harvard.edu/abs/2023MNRAS.522L...1R}
}

@ARTICLE{Rossi2025,
       author = {{Rossi}, Martina and {Salvadori}, Stefania and {Sk{\'u}lad{\'o}ttir}, {\'A}sa and {Vanni}, Irene and {Koutsouridou}, Ioanna},
        title = "{Hidden Population III Descendants in Ultrafaint Dwarf Galaxies}",
      journal = {\apj},
         year = 2025,
        month = jul,
       volume = {987},
       number = {2},
          eid = {121},
        pages = {121},
          doi = {10.3847/1538-4357/add5e9},
archivePrefix = {arXiv},
       eprint = {2406.12960},
 primaryClass = {astro-ph.GA},
       adsurl = {https://ui.adsabs.harvard.edu/abs/2025ApJ...987..121R}
}

@INPROCEEDINGS{Vanni+23a,
       author = {{Vanni}, I. and {Salvadori}, S. and {Sk{\'u}lad{\'o}ttir}, {\'A}.},
        title = "{Are all metal-poor stars of second-generation?}",
    booktitle = {Memorie della Societa Astronomica Italiana},
         year = 2023,
       volume = {94},
        month = sep,
        pages = {84},
          doi = {10.36116/MEMSAIT_94N2.2023.84},
archivePrefix = {arXiv},
       eprint = {2305.02358},
 primaryClass = {astro-ph.GA},
       adsurl = {https://ui.adsabs.harvard.edu/abs/2023MmSAI..94b..84V}
}

@ARTICLE{Vanni+23b,
       author = {{Vanni}, Irene and {Salvadori}, Stefania and {Sk{\'u}lad{\'o}ttir}, {\'A}sa and {Rossi}, Martina and {Koutsouridou}, Ioanna},
        title = "{Characterizing the true descendants of the first stars}",
      journal = {\mnras},
         year = 2023,
        month = dec,
       volume = {526},
       number = {2},
        pages = {2620-2644},
          doi = {10.1093/mnras/stad2910},
archivePrefix = {arXiv},
       eprint = {2309.07958},
 primaryClass = {astro-ph.GA},
       adsurl = {https://ui.adsabs.harvard.edu/abs/2023MNRAS.526.2620V}
}

@ARTICLE{Koutsouridou+23,
       author = {{Koutsouridou}, I. and {Salvadori}, S. and {Sk{\'u}lad{\'o}ttir}, {\'A}. and {Rossi}, M. and {Vanni}, I. and {Pagnini}, G.},
        title = "{The energy distribution of the first supernovae}",
      journal = {\mnras},
         year = 2023,
        month = oct,
       volume = {525},
       number = {1},
        pages = {190-210},
          doi = {10.1093/mnras/stad2304},
archivePrefix = {arXiv},
       eprint = {2309.00045},
 primaryClass = {astro-ph.GA},
       adsurl = {https://ui.adsabs.harvard.edu/abs/2023MNRAS.525..190K}
}

@ARTICLE{Koutsouridou+25,
       author = {{Koutsouridou}, I. and {Sk{\'u}lad{\'o}ttir}, {\'A}. and {Salvadori}, S.},
        title = "{Large databases of metal-poor stars corrected for three-dimensional and/or non-local thermodynamic equilibrium effects}",
      journal = {\aap},
         year = 2025,
        month = jul,
       volume = {699},
          eid = {A32},
        pages = {A32},
          doi = {10.1051/0004-6361/202554228},
archivePrefix = {arXiv},
       eprint = {2505.13607},
 primaryClass = {astro-ph.GA},
       adsurl = {https://ui.adsabs.harvard.edu/abs/2025A&A...699A..32K}
}

@ARTICLE{GalYam+09,
       author = {{Gal-Yam}, A. and {Mazzali}, P. and {Ofek}, E.~O. and {Nugent}, P.~E. and {Kulkarni}, S.~R. and {Kasliwal}, M.~M. and {Quimby}, R.~M. and {Filippenko}, A.~V. and {Cenko}, S.~B. and {Chornock}, R. and {Waldman}, R. and {Kasen}, D. and {Sullivan}, M. and {Beshore}, E.~C. and {Drake}, A.~J. and {Thomas}, R.~C. and {Bloom}, J.~S. and {Poznanski}, D. and {Miller}, A.~A. and {Foley}, R.~J. and {Silverman}, J.~M. and {Arcavi}, I. and {Ellis}, R.~S. and {Deng}, J.},
        title = "{Supernova 2007bi as a pair-instability explosion}",
      journal = {\nat},
         year = 2009,
        month = dec,
       volume = {462},
       number = {7273},
        pages = {624-627},
          doi = {10.1038/nature08579},
archivePrefix = {arXiv},
       eprint = {1001.1156},
 primaryClass = {astro-ph.CO},
       adsurl = {https://ui.adsabs.harvard.edu/abs/2009Natur.462..624G}
}

@ARTICLE{Moriya+10,
       author = {{Moriya}, Takashi and {Tominaga}, Nozomu and {Tanaka}, Masaomi and {Maeda}, Keiichi and {Nomoto}, Ken'ichi},
        title = "{A Core-collapse Supernova Model for the Extremely Luminous Type Ic Supernova 2007bi: An Alternative to the Pair-instability Supernova Model}",
      journal = {\apjl},
         year = 2010,
        month = jul,
       volume = {717},
       number = {2},
        pages = {L83-L86},
          doi = {10.1088/2041-8205/717/2/L83},
archivePrefix = {arXiv},
       eprint = {1004.2967},
 primaryClass = {astro-ph.HE},
       adsurl = {https://ui.adsabs.harvard.edu/abs/2010ApJ...717L..83M}
}

@ARTICLE{Quimby+11,
       author = {{Quimby}, R.~M. and {Kulkarni}, S.~R. and {Kasliwal}, M.~M. and {Gal-Yam}, A. and {Arcavi}, I. and {Sullivan}, M. and {Nugent}, P. and {Thomas}, R. and {Howell}, D.~A. and {Nakar}, E. and {Bildsten}, L. and {Theissen}, C. and {Law}, N.~M. and {Dekany}, R. and {Rahmer}, G. and {Hale}, D. and {Smith}, R. and {Ofek}, E.~O. and {Zolkower}, J. and {Velur}, V. and {Walters}, R. and {Henning}, J. and {Bui}, K. and {McKenna}, D. and {Poznanski}, D. and {Cenko}, S.~B. and {Levitan}, D.},
        title = "{Hydrogen-poor superluminous stellar explosions}",
      journal = {\nat},
         year = 2011,
        month = jun,
       volume = {474},
       number = {7352},
        pages = {487-489},
          doi = {10.1038/nature10095},
archivePrefix = {arXiv},
       eprint = {0910.0059},
 primaryClass = {astro-ph.CO},
       adsurl = {https://ui.adsabs.harvard.edu/abs/2011Natur.474..487Q}
}

@ARTICLE{Cooke+12,
       author = {{Cooke}, Jeff and {Sullivan}, Mark and {Gal-Yam}, Avishay and {Barton}, Elizabeth J. and {Carlberg}, Raymond G. and {Ryan-Weber}, Emma V. and {Horst}, Chuck and {Omori}, Yuuki and {D{\'\i}az}, C. Gonzalo},
        title = "{Superluminous supernovae at redshifts of 2.05 and 3.90}",
      journal = {\nat},
         year = 2012,
        month = nov,
       volume = {491},
       number = {7423},
        pages = {228-231},
          doi = {10.1038/nature11521},
archivePrefix = {arXiv},
       eprint = {1211.2003},
 primaryClass = {astro-ph.CO},
       adsurl = {https://ui.adsabs.harvard.edu/abs/2012Natur.491..228C}
}

@ARTICLE{Dessart+12,
       author = {{Dessart}, Luc and {Hillier}, D. John and {Waldman}, Roni and {Livne}, Eli and {Blondin}, St{\'e}phane},
        title = "{Superluminous supernovae: $^{56}$Ni power versus magnetar radiation}",
      journal = {\mnras},
         year = 2012,
        month = oct,
       volume = {426},
       number = {1},
        pages = {L76-L80},
          doi = {10.1111/j.1745-3933.2012.01329.x},
archivePrefix = {arXiv},
       eprint = {1208.1214},
 primaryClass = {astro-ph.SR},
       adsurl = {https://ui.adsabs.harvard.edu/abs/2012MNRAS.426L..76D}
}

@ARTICLE{Schulze+24,
       author = {{Schulze}, Steve and {Fransson}, Claes and {Kozyreva}, Alexandra and {Chen}, Ting-Wan and {Yaron}, Ofer and {Jerkstrand}, Anders and {Gal-Yam}, Avishay and {Sollerman}, Jesper and {Yan}, Lin and {Kangas}, Tuomas and {Leloudas}, Giorgos and {Omand}, Conor M.~B. and {Smartt}, Stephen J. and {Yang}, Yi and {Nicholl}, Matt and {Sarin}, Nikhil and {Yao}, Yuhan and {Brink}, Thomas G. and {Sharon}, Amir and {Rossi}, Andrea and {Chen}, Ping and {Chen}, Zhihao and {Cikota}, Aleksandar and {De}, Kishalay and {Drake}, Andrew J. and {Filippenko}, Alexei V. and {Fremling}, Christoffer and {Fr{\'e}our}, Laurane and {Fynbo}, Johan P.~U. and {Ho}, Anna Y.~Q. and {Inserra}, Cosimo and {Irani}, Ido and {Kuncarayakti}, Hanindyo and {Lunnan}, Ragnhild and {Mazzali}, Paolo and {Ofek}, Eran O. and {Palazzi}, Eliana and {Perley}, Daniel A. and {Pursiainen}, Miika and {Rothberg}, Barry and {Shingles}, Luke J. and {Smith}, Ken and {Taggart}, Kirsty and {Tartaglia}, Leonardo and {Zheng}, WeiKang and {Anderson}, Joseph P. and {Cassara}, Letizia and {Christensen}, Eric and {George Djorgovski}, S. and {Galbany}, Llu{\'\i}s and {Gkini}, Anamaria and {Graham}, Matthew J. and {Gromadzki}, Mariusz and {Groom}, Steven L. and {Hiramatsu}, Daichi and {Andrew Howell}, D. and {Kasliwal}, Mansi M. and {McCully}, Curtis and {M{\"u}ller-Bravo}, Tom{\'a}s E. and {Paiano}, Simona and {Paraskeva}, Emmanouela and {Pessi}, Priscila J. and {Polishook}, David and {Rau}, Arne and {Rigault}, Mickael and {Rusholme}, Ben},
        title = "{1100 days in the life of the supernova 2018ibb. The best pair-instability supernova candidate, to date}",
      journal = {\aap},
         year = 2024,
        month = mar,
       volume = {683},
          eid = {A223},
        pages = {A223},
          doi = {10.1051/0004-6361/202346855},
archivePrefix = {arXiv},
       eprint = {2305.05796},
 primaryClass = {astro-ph.HE},
       adsurl = {https://ui.adsabs.harvard.edu/abs/2024A&A...683A.223S}
}

@ARTICLE{WeinmannLilly+05,
       author = {{Weinmann}, Simone M. and {Lilly}, Simon J.},
        title = "{The Number and Observability of Population III Supernovae at High Redshifts}",
      journal = {\apj},
         year = 2005,
        month = may,
       volume = {624},
       number = {2},
        pages = {526-531},
          doi = {10.1086/428106},
archivePrefix = {arXiv},
       eprint = {astro-ph/0412248},
 primaryClass = {astro-ph},
       adsurl = {https://ui.adsabs.harvard.edu/abs/2005ApJ...624..526W}
}

@ARTICLE{Whalen+13b,
       author = {{Whalen}, Daniel J. and {Even}, Wesley and {Frey}, Lucille H. and {Smidt}, Joseph and {Johnson}, Jarrett L. and {Lovekin}, C.~C. and {Fryer}, Chris L. and {Stiavelli}, Massimo and {Holz}, Daniel E. and {Heger}, Alexander and {Woosley}, S.~E. and {Hungerford}, Aimee L.},
        title = "{Finding the First Cosmic Explosions. I. Pair-instability Supernovae}",
      journal = {\apj},
         year = 2013,
        month = nov,
       volume = {777},
       number = {2},
          eid = {110},
        pages = {110},
          doi = {10.1088/0004-637X/777/2/110},
archivePrefix = {arXiv},
       eprint = {1211.4979},
 primaryClass = {astro-ph.CO},
       adsurl = {https://ui.adsabs.harvard.edu/abs/2013ApJ...777..110W}
}

@ARTICLE{Wang+17,
       author = {{Wang}, Lifan and {Baade}, D. and {Baron}, E. and {Bernard}, S. and {Bromm}, V. and {Brown}, P. and {Clayton}, G. and {Cooke}, J. and {Croton}, D. and {Curtin}, C. and {Drout}, M. and {Doi}, M. and {Dominguez}, I. and {Finkelstein}, S. and {Gal-Yam}, A. and {Geil}, P. and {Heger}, A. and {Hoeflich}, P. and {Jian}, J. and {Krisciunas}, K. and {Koekemoer}, A. and {Lunnan}, R. and {Maeda}, K. and {Maund}, J. and {Modjaz}, M. and {Mould}, J. and {Nomoto}, K. and {Nugent}, P. and {Patat}, F. and {Pacucci}, F. and {Phillips}, M. and {Rest}, A. and {Regos}, E. and {Sand}, D. and {Sparks}, B. and {Spyromilio}, J. and {Staveley-Smith}, L. and {Suntzeff}, N. and {Uddin}, S. and {Villarroel}, B. and {Vinko}, J. and {Whalen}, D. and {Wheeler}, J. and {Wood-Vasey}, M. and {Yang}, Y. and {Yue}, Bin},
        title = "{A First Transients Survey with JWST: the FLARE project}",
      journal = {arXiv e-prints},
         year = 2017,
        month = oct,
          eid = {arXiv:1710.07005},
        pages = {arXiv:1710.07005},
          doi = {10.48550/arXiv.1710.07005},
archivePrefix = {arXiv},
       eprint = {1710.07005},
 primaryClass = {astro-ph.IM},
       adsurl = {https://ui.adsabs.harvard.edu/abs/2017arXiv171007005W}
}

@ARTICLE{Hartwig+18b,
       author = {{Hartwig}, Tilman and {Bromm}, Volker and {Loeb}, Abraham},
        title = "{Detection strategies for the first supernovae with JWST}",
      journal = {\mnras},
         year = 2018,
        month = sep,
       volume = {479},
       number = {2},
        pages = {2202-2213},
          doi = {10.1093/mnras/sty1576},
archivePrefix = {arXiv},
       eprint = {1711.05742},
 primaryClass = {astro-ph.GA},
       adsurl = {https://ui.adsabs.harvard.edu/abs/2018MNRAS.479.2202H}
}

@ARTICLE{Regos+20,
       author = {{Reg{\H{o}}s}, Enik{\H{o}} and {Vink{\'o}}, J{\'o}zsef and {Ziegler}, Bodo L.},
        title = "{Detecting Pair-instability Supernovae at z {\ensuremath{\lesssim}} 5 with the James Webb Space Telescope}",
      journal = {\apj},
         year = 2020,
        month = may,
       volume = {894},
       number = {2},
          eid = {94},
        pages = {94},
          doi = {10.3847/1538-4357/ab8636},
archivePrefix = {arXiv},
       eprint = {2002.07854},
 primaryClass = {astro-ph.HE},
       adsurl = {https://ui.adsabs.harvard.edu/abs/2020ApJ...894...94R}
}

@ARTICLE{LazarBromm22,
       author = {{Lazar}, Alexandres and {Bromm}, Volker},
        title = "{Probing the initial mass function of the first stars with transients}",
      journal = {\mnras},
         year = 2022,
        month = apr,
       volume = {511},
       number = {2},
        pages = {2505-2514},
          doi = {10.1093/mnras/stac176},
archivePrefix = {arXiv},
       eprint = {2110.11956},
 primaryClass = {astro-ph.HE},
       adsurl = {https://ui.adsabs.harvard.edu/abs/2022MNRAS.511.2505L}
}

@ARTICLE{Moriya+22a,
       author = {{Moriya}, Takashi J. and {Quimby}, Robert M. and {Robertson}, Brant E.},
        title = "{Discovering Supernovae at the Epoch of Reionization with the Nancy Grace Roman Space Telescope}",
      journal = {\apj},
         year = 2022,
        month = feb,
       volume = {925},
       number = {2},
          eid = {211},
        pages = {211},
          doi = {10.3847/1538-4357/ac415e},
archivePrefix = {arXiv},
       eprint = {2108.01801},
 primaryClass = {astro-ph.HE},
       adsurl = {https://ui.adsabs.harvard.edu/abs/2022ApJ...925..211M}
}

@ARTICLE{RomanObservationsTimeAllocationCommittee+25,
       author = {{Observations Time Allocation Committee}, Roman and {Community Survey Definition Committees}, Core},
        title = "{Roman Observations Time Allocation Committee: Final Report and Recommendations}",
      journal = {arXiv e-prints},
         year = 2025,
        month = may,
          eid = {arXiv:2505.10574},
        pages = {arXiv:2505.10574},
          doi = {10.48550/arXiv.2505.10574},
archivePrefix = {arXiv},
       eprint = {2505.10574},
 primaryClass = {astro-ph.IM},
       adsurl = {https://ui.adsabs.harvard.edu/abs/2025arXiv250510574O}
}

@ARTICLE{Moriya+22b,
       author = {{Moriya}, T.~J. and {Inserra}, C. and {Tanaka}, M. and {Cappellaro}, E. and {Della Valle}, M. and {Hook}, I. and {Kotak}, R. and {Longo}, G. and {Mannucci}, F. and {Mattila}, S. and {Tao}, C. and {Altieri}, B. and {Amara}, A. and {Auricchio}, N. and {Bonino}, D. and {Branchini}, E. and {Brescia}, M. and {Brinchmann}, J. and {Camera}, S. and {Capobianco}, V. and {Carbone}, C. and {Carretero}, J. and {Castellano}, M. and {Cavuoti}, S. and {Cimatti}, A. and {Cledassou}, R. and {Congedo}, G. and {Conselice}, C.~J. and {Conversi}, L. and {Copin}, Y. and {Corcione}, L. and {Courbin}, F. and {Cropper}, M. and {Da Silva}, A. and {Degaudenzi}, H. and {Douspis}, M. and {Dubath}, F. and {Duncan}, C.~A.~J. and {Dupac}, X. and {Dusini}, S. and {Ealet}, A. and {Farrens}, S. and {Ferriol}, S. and {Frailis}, M. and {Franceschi}, E. and {Fumana}, M. and {Garilli}, B. and {Gillard}, W. and {Gillis}, B. and {Giocoli}, C. and {Grazian}, A. and {Grupp}, F. and {Haugan}, S.~V.~H. and {Holmes}, W. and {Hormuth}, F. and {Hornstrup}, A. and {Jahnke}, K. and {Kermiche}, S. and {Kiessling}, A. and {Kilbinger}, M. and {Kitching}, T. and {Kurki-Suonio}, H. and {Ligori}, S. and {Lilje}, P.~B. and {Lloro}, I. and {Maiorano}, E. and {Mansutti}, O. and {Marggraf}, O. and {Markovic}, K. and {Marulli}, F. and {Massey}, R. and {McCracken}, H.~J. and {Melchior}, M. and {Meneghetti}, M. and {Meylan}, G. and {Moresco}, M. and {Moscardini}, L. and {Munari}, E. and {Niemi}, S.~M. and {Padilla}, C. and {Paltani}, S. and {Pasian}, F. and {Pedersen}, K. and {Pettorino}, V. and {Poncet}, M. and {Popa}, L. and {Raison}, F. and {Rhodes}, J. and {Riccio}, G. and {Rossetti}, E. and {Saglia}, R. and {Sartoris}, B. and {Schneider}, P. and {Secroun}, A. and {Seidel}, G. and {Sirignano}, C. and {Sirri}, G. and {Stanco}, L. and {Tallada-Cresp{\'\i}}, P. and {Taylor}, A.~N. and {Tereno}, I. and {Toledo-Moreo}, R. and {Torradeflot}, F. and {Wang}, Y. and {Zamorani}, G. and {Zoubian}, J. and {Andreon}, S. and {Scottez}, V. and {Morris}, P.~W.},
        title = "{Euclid: Searching for pair-instability supernovae with the Deep Survey}",
      journal = {\aap},
         year = 2022,
        month = oct,
       volume = {666},
          eid = {A157},
        pages = {A157},
          doi = {10.1051/0004-6361/202243810},
archivePrefix = {arXiv},
       eprint = {2204.08727},
 primaryClass = {astro-ph.HE},
       adsurl = {https://ui.adsabs.harvard.edu/abs/2022A&A...666A.157M}
}

@ARTICLE{Laureijs+11,
       author = {{Laureijs}, R. and {Amiaux}, J. and {Arduini}, S. and {Augu{\`e}res}, J.  -L. and {Brinchmann}, J. and {Cole}, R. and {Cropper}, M. and {Dabin}, C. and {Duvet}, L. and {Ealet}, A. and {Garilli}, B. and {Gondoin}, P. and {Guzzo}, L. and {Hoar}, J. and {Hoekstra}, H. and {Holmes}, R. and {Kitching}, T. and {Maciaszek}, T. and {Mellier}, Y. and {Pasian}, F. and {Percival}, W. and {Rhodes}, J. and {Saavedra Criado}, G. and {Sauvage}, M. and {Scaramella}, R. and {Valenziano}, L. and {Warren}, S. and {Bender}, R. and {Castander}, F. and {Cimatti}, A. and {Le F{\`e}vre}, O. and {Kurki-Suonio}, H. and {Levi}, M. and {Lilje}, P. and {Meylan}, G. and {Nichol}, R. and {Pedersen}, K. and {Popa}, V. and {Rebolo Lopez}, R. and {Rix}, H.  -W. and {Rottgering}, H. and {Zeilinger}, W. and {Grupp}, F. and {Hudelot}, P. and {Massey}, R. and {Meneghetti}, M. and {Miller}, L. and {Paltani}, S. and {Paulin-Henriksson}, S. and {Pires}, S. and {Saxton}, C. and {Schrabback}, T. and {Seidel}, G. and {Walsh}, J. and {Aghanim}, N. and {Amendola}, L. and {Bartlett}, J. and {Baccigalupi}, C. and {Beaulieu}, J.  -P. and {Benabed}, K. and {Cuby}, J.  -G. and {Elbaz}, D. and {Fosalba}, P. and {Gavazzi}, G. and {Helmi}, A. and {Hook}, I. and {Irwin}, M. and {Kneib}, J.  -P. and {Kunz}, M. and {Mannucci}, F. and {Moscardini}, L. and {Tao}, C. and {Teyssier}, R. and {Weller}, J. and {Zamorani}, G. and {Zapatero Osorio}, M.~R. and {Boulade}, O. and {Foumond}, J.~J. and {Di Giorgio}, A. and {Guttridge}, P. and {James}, A. and {Kemp}, M. and {Martignac}, J. and {Spencer}, A. and {Walton}, D. and {Bl{\"u}mchen}, T. and {Bonoli}, C. and {Bortoletto}, F. and {Cerna}, C. and {Corcione}, L. and {Fabron}, C. and {Jahnke}, K. and {Ligori}, S. and {Madrid}, F. and {Martin}, L. and {Morgante}, G. and {Pamplona}, T. and {Prieto}, E. and {Riva}, M. and {Toledo}, R. and {Trifoglio}, M. and {Zerbi}, F. and {Abdalla}, F. and {Douspis}, M. and {Grenet}, C. and {Borgani}, S. and {Bouwens}, R. and {Courbin}, F. and {Delouis}, J.  -M. and {Dubath}, P. and {Fontana}, A. and {Frailis}, M. and {Grazian}, A. and {Koppenh{\"o}fer}, J. and {Mansutti}, O. and {Melchior}, M. and {Mignoli}, M. and {Mohr}, J. and {Neissner}, C. and {Noddle}, K. and {Poncet}, M. and {Scodeggio}, M. and {Serrano}, S. and {Shane}, N. and {Starck}, J.  -L. and {Surace}, C. and {Taylor}, A. and {Verdoes-Kleijn}, G. and {Vuerli}, C. and {Williams}, O.~R. and {Zacchei}, A. and {Altieri}, B. and {Escudero Sanz}, I. and {Kohley}, R. and {Oosterbroek}, T. and {Astier}, P. and {Bacon}, D. and {Bardelli}, S. and {Baugh}, C. and {Bellagamba}, F. and {Benoist}, C. and {Bianchi}, D. and {Biviano}, A. and {Branchini}, E. and {Carbone}, C. and {Cardone}, V. and {Clements}, D. and {Colombi}, S. and {Conselice}, C. and {Cresci}, G. and {Deacon}, N. and {Dunlop}, J. and {Fedeli}, C. and {Fontanot}, F. and {Franzetti}, P. and {Giocoli}, C. and {Garcia-Bellido}, J. and {Gow}, J. and {Heavens}, A. and {Hewett}, P. and {Heymans}, C. and {Holland}, A. and {Huang}, Z. and {Ilbert}, O. and {Joachimi}, B. and {Jennins}, E. and {Kerins}, E. and {Kiessling}, A. and {Kirk}, D. and {Kotak}, R. and {Krause}, O. and {Lahav}, O. and {van Leeuwen}, F. and {Lesgourgues}, J. and {Lombardi}, M. and {Magliocchetti}, M. and {Maguire}, K. and {Majerotto}, E. and {Maoli}, R. and {Marulli}, F. and {Maurogordato}, S. and {McCracken}, H. and {McLure}, R. and {Melchiorri}, A. and {Merson}, A. and {Moresco}, M. and {Nonino}, M. and {Norberg}, P. and {Peacock}, J. and {Pello}, R. and {Penny}, M. and {Pettorino}, V. and {Di Porto}, C. and {Pozzetti}, L. and {Quercellini}, C. and {Radovich}, M. and {Rassat}, A. and {Roche}, N. and {Ronayette}, S. and {Rossetti}, E.},
        title = "{Euclid Definition Study Report}",
      journal = {arXiv e-prints},
         year = 2011,
        month = oct,
          eid = {arXiv:1110.3193},
        pages = {arXiv:1110.3193},
          doi = {10.48550/arXiv.1110.3193},
archivePrefix = {arXiv},
       eprint = {1110.3193},
 primaryClass = {astro-ph.CO},
       adsurl = {https://ui.adsabs.harvard.edu/abs/2011arXiv1110.3193L}
}

@ARTICLE{EuclidCollaboration+22,
       author = {{Euclid Collaboration} and {Schirmer}, M. and {Jahnke}, K. and {Seidel}, G. and {Aussel}, H. and {Bodendorf}, C. and {Grupp}, F. and {Hormuth}, F. and {Wachter}, S. and {Appleton}, P.~N. and {Barbier}, R. and {Brinchmann}, J. and {Carrasco}, J.~M. and {Castander}, F.~J. and {Coupon}, J. and {De Paolis}, F. and {Franco}, A. and {Ganga}, K. and {Hudelot}, P. and {Jullo}, E. and {Lan{\c{c}}on}, A. and {Nucita}, A.~A. and {Paltani}, S. and {Smadja}, G. and {Strafella}, F. and {Venancio}, L.~M.~G. and {Weiler}, M. and {Amara}, A. and {Auphan}, T. and {Auricchio}, N. and {Balestra}, A. and {Bender}, R. and {Bonino}, D. and {Branchini}, E. and {Brescia}, M. and {Capobianco}, V. and {Carbone}, C. and {Carretero}, J. and {Casas}, R. and {Castellano}, M. and {Cavuoti}, S. and {Cimatti}, A. and {Cledassou}, R. and {Congedo}, G. and {Conselice}, C.~J. and {Conversi}, L. and {Copin}, Y. and {Corcione}, L. and {Costille}, A. and {Courbin}, F. and {Da Silva}, A. and {Degaudenzi}, H. and {Douspis}, M. and {Dubath}, F. and {Dupac}, X. and {Dusini}, S. and {Ealet}, A. and {Farrens}, S. and {Ferriol}, S. and {Fosalba}, P. and {Frailis}, M. and {Franceschi}, E. and {Franzetti}, P. and {Fumana}, M. and {Garilli}, B. and {Gillard}, W. and {Gillis}, B. and {Giocoli}, C. and {Grazian}, A. and {Guzzo}, L. and {Haugan}, S.~V.~H. and {Hoekstra}, H. and {Holmes}, W. and {Hornstrup}, A. and {K{\"u}mmel}, M. and {Kermiche}, S. and {Kiessling}, A. and {Kilbinger}, M. and {Kitching}, T. and {Kohley}, R. and {Kunz}, M. and {Kurki-Suonio}, H. and {Laureijs}, R. and {Ligori}, S. and {Lilje}, P.~B. and {Lloro}, I. and {Maciaszek}, T. and {Maiorano}, E. and {Mansutti}, O. and {Marggraf}, O. and {Markovic}, K. and {Marulli}, F. and {Massey}, R. and {Maurogordato}, S. and {Mellier}, Y. and {Meneghetti}, M. and {Merlin}, E. and {Meylan}, G. and {Moresco}, M. and {Moscardini}, L. and {Munari}, E. and {Nakajima}, R. and {Nichol}, R.~C. and {Niemi}, S.~M. and {Padilla}, C. and {Pasian}, F. and {Pedersen}, K. and {Percival}, W.~J. and {Pettorino}, V. and {Pires}, S. and {Poncet}, M. and {Popa}, L. and {Pozzetti}, L. and {Prieto}, E. and {Raison}, F. and {Rhodes}, J. and {Rix}, H.-W. and {Roncarelli}, M. and {Rossetti}, E. and {Saglia}, R. and {Sartoris}, B. and {Scaramella}, R. and {Schneider}, P. and {Secroun}, A. and {Serrano}, S. and {Sirignano}, C. and {Sirri}, G. and {Stanco}, L. and {Tallada-Cresp{\'\i}}, P. and {Taylor}, A.~N. and {Teplitz}, H.~I. and {Tereno}, I. and {Toledo-Moreo}, R. and {Torradeflot}, F. and {Trifoglio}, M. and {Valentijn}, E.~A. and {Valenziano}, L. and {Wang}, Y. and {Weller}, J. and {Zamorani}, G. and {Zoubian}, J. and {Andreon}, S. and {Bardelli}, S. and {Boucaud}, A. and {Camera}, S. and {Farinelli}, R. and {Graci{\'a}-Carpio}, J. and {Maino}, D. and {Medinaceli}, E. and {Mei}, S. and {Morisset}, N. and {Polenta}, G. and {Renzi}, A. and {Romelli}, E. and {Tenti}, M. and {Vassallo}, T. and {Zacchei}, A. and {Zucca}, E. and {Baccigalupi}, C. and {Balaguera-Antol{\'\i}nez}, A. and {Biviano}, A. and {Blanchard}, A. and {Borgani}, S. and {Bozzo}, E. and {Burigana}, C. and {Cabanac}, R. and {Cappi}, A. and {Carvalho}, C.~S. and {Casas}, S. and {Castignani}, G. and {Colodro-Conde}, C. and {Cooray}, A.~R. and {Courtois}, H.~M. and {Crocce}, M. and {Cuby}, J.-G. and {Davini}, S. and {de la Torre}, S. and {Di Ferdinando}, D. and {Escartin}, J.~A. and {Farina}, M. and {Ferreira}, P.~G. and {Finelli}, F. and {Fotopoulou}, S. and {Galeotta}, S. and {Garcia-Bellido}, J. and {Gaztanaga}, E. and {George}, K. and {Gozaliasl}, G. and {Hook}, I.~M. and {Ili{\'c}}, S. and {Kansal}, V. and {Kashlinsky}, A. and {Keihanen}, E. and {Kirkpatrick}, C.~C. and {Lindholm}, V. and {Mainetti}, G. and {Maoli}, R. and {Martinelli}, M. and {Martinet}, N. and {Maturi}, M.},
        title = "{Euclid preparation. XVIII. The NISP photometric system}",
      journal = {\aap},
         year = 2022,
        month = jun,
       volume = {662},
          eid = {A92},
        pages = {A92},
          doi = {10.1051/0004-6361/202142897},
archivePrefix = {arXiv},
       eprint = {2203.01650},
 primaryClass = {astro-ph.IM},
       adsurl = {https://ui.adsabs.harvard.edu/abs/2022A&A...662A..92E}
}

@ARTICLE{Aoki+14,
       author = {{Aoki}, W. and {Tominaga}, N. and {Beers}, T.~C. and {Honda}, S. and {Lee}, Y.~S.},
        title = "{A chemical signature of first-generation very massive stars}",
      journal = {Science},
         year = 2014,
        month = aug,
       volume = {345},
       number = {6199},
        pages = {912-915},
          doi = {10.1126/science.1252633},
       adsurl = {https://ui.adsabs.harvard.edu/abs/2014Sci...345..912A}
}

@ARTICLE{Salvadori+19,
       author = {{Salvadori}, S. and {Bonifacio}, P. and {Caffau}, E. and {Korotin}, S. and {Andreevsky}, S. and {Spite}, M. and {Sk{\'u}lad{\'o}ttir}, {\'A}.},
        title = "{Probing the existence of very massive first stars}",
      journal = {\mnras},
         year = 2019,
        month = aug,
       volume = {487},
       number = {3},
        pages = {4261-4284},
          doi = {10.1093/mnras/stz1464},
archivePrefix = {arXiv},
       eprint = {1906.00994},
 primaryClass = {astro-ph.GA},
       adsurl = {https://ui.adsabs.harvard.edu/abs/2019MNRAS.487.4261S}
}

@ARTICLE{Aguado+23b,
       author = {{Aguado}, D.~S. and {Salvadori}, S. and {Sk{\'u}lad{\'o}ttir}, {\'A}. and {Caffau}, E. and {Bonifacio}, P. and {Vanni}, I. and {Gelli}, V. and {Koutsouridou}, I. and {Amarsi}, A.~M.},
        title = "{PISN-explorer: hunting the descendants of very massive first stars}",
      journal = {\mnras},
         year = 2023,
        month = mar,
       volume = {520},
       number = {1},
        pages = {866-878},
          doi = {10.1093/mnras/stad164},
archivePrefix = {arXiv},
       eprint = {2301.03604},
 primaryClass = {astro-ph.GA},
       adsurl = {https://ui.adsabs.harvard.edu/abs/2023MNRAS.520..866A}
}

@ARTICLE{Xing+23,
       author = {{Xing}, Qian-Fan and {Zhao}, Gang and {Liu}, Zheng-Wei and {Heger}, Alexander and {Han}, Zhan-Wen and {Aoki}, Wako and {Chen}, Yu-Qin and {Ishigaki}, Miho N. and {Li}, Hai-Ning and {Zhao}, Jing-Kun},
        title = "{A metal-poor star with abundances from a pair-instability supernova}",
      journal = {\nat},
         year = 2023,
        month = jun,
       volume = {618},
       number = {7966},
        pages = {712-715},
          doi = {10.1038/s41586-023-06028-1},
       adsurl = {https://ui.adsabs.harvard.edu/abs/2023Natur.618..712X}
}

@ARTICLE{JeenaBanerjee+24,
       author = {{Jeena}, S.~K. and {Banerjee}, Projjwal},
        title = "{Origin of LAMOST J1010+2358 Revisited}",
      journal = {The Open Journal of Astrophysics},
         year = 2024,
        month = oct,
       volume = {7},
          eid = {83},
        pages = {83},
          doi = {10.33232/001c.124113},
archivePrefix = {arXiv},
       eprint = {2408.02643},
 primaryClass = {astro-ph.SR},
       adsurl = {https://ui.adsabs.harvard.edu/abs/2024OJAp....7E..83J}
}

@ARTICLE{Skuladottir+24,
       author = {{Sk{\'u}lad{\'o}ttir}, {\'A}sa and {Koutsouridou}, Ioanna and {Vanni}, Irene and {Amarsi}, Anish M. and {Lucchesi}, Romain and {Salvadori}, Stefania and {Aguado}, David S.},
        title = "{On the Pair-instability Supernova Origin of J1010+2358}",
      journal = {\apjl},
         year = 2024,
        month = jun,
       volume = {968},
       number = {2},
          eid = {L23},
        pages = {L23},
          doi = {10.3847/2041-8213/ad4b1a},
archivePrefix = {arXiv},
       eprint = {2404.19086},
 primaryClass = {astro-ph.SR},
       adsurl = {https://ui.adsabs.harvard.edu/abs/2024ApJ...968L..23S}
}

@ARTICLE{Thibodeaux+24,
       author = {{Thibodeaux}, Pierre and {Ji}, Alexander P. and {Cerny}, William and {Kirby}, Evan N. and {Simon}, Joshua D.},
        title = "{LAMOST J1010+2358 is not a Pair-Instability Supernova Relic}",
      journal = {The Open Journal of Astrophysics},
         year = 2024,
        month = aug,
       volume = {7},
          eid = {66},
        pages = {66},
          doi = {10.33232/001c.122335},
archivePrefix = {arXiv},
       eprint = {2404.17078},
 primaryClass = {astro-ph.SR},
       adsurl = {https://ui.adsabs.harvard.edu/abs/2024OJAp....7E..66T}
}

@ARTICLE{Koutsouridou+24,
       author = {{Koutsouridou}, Ioanna and {Salvadori}, Stefania and {Sk{\'u}lad{\'o}ttir}, {\'A}sa},
        title = "{True Pair-instability Supernova Descendant: Implications for the First Stars' Mass Distribution}",
      journal = {\apjl},
         year = 2024,
        month = feb,
       volume = {962},
       number = {2},
          eid = {L26},
        pages = {L26},
          doi = {10.3847/2041-8213/ad2466},
archivePrefix = {arXiv},
       eprint = {2312.05309},
 primaryClass = {astro-ph.GA},
       adsurl = {https://ui.adsabs.harvard.edu/abs/2024ApJ...962L..26K}
}

@ARTICLE{Suda+08,
       author = {{Suda}, Takuma and {Katsuta}, Yutaka and {Yamada}, Shimako and {Suwa}, Tamon and {Ishizuka}, Chikako and {Komiya}, Yutaka and {Sorai}, Kazuo and {Aikawa}, Masayuki and {Fujimoto}, Masayuki Y.},
        title = "{Stellar Abundances for the Galactic Archeology (SAGA) Database --- Compilation of the Characteristics of Known Extremely Metal-Poor Stars}",
      journal = {\pasj},
         year = 2008,
        month = oct,
       volume = {60},
        pages = {1159},
          doi = {10.1093/pasj/60.5.1159},
archivePrefix = {arXiv},
       eprint = {0806.3697},
 primaryClass = {astro-ph},
       adsurl = {https://ui.adsabs.harvard.edu/abs/2008PASJ...60.1159S}
}

@ARTICLE{Suda+17,
       author = {{Suda}, Takuma and {Hidaka}, Jun and {Aoki}, Wako and {Katsuta}, Yutaka and {Yamada}, Shimako and {Fujimoto}, Masayuki Y. and {Ohtani}, Yukari and {Masuyama}, Miyu and {Noda}, Kazuhiro and {Wada}, Kentaro},
        title = "{Stellar Abundances for Galactic Archaeology Database. IV. Compilation of stars in dwarf galaxies}",
      journal = {\pasj},
         year = 2017,
        month = oct,
       volume = {69},
       number = {5},
          eid = {76},
        pages = {76},
          doi = {10.1093/pasj/psx059},
archivePrefix = {arXiv},
       eprint = {1703.10009},
 primaryClass = {astro-ph.GA},
       adsurl = {https://ui.adsabs.harvard.edu/abs/2017PASJ...69...76S}
}

@ARTICLE{Zhao+12,
       author = {{Zhao}, Gang and {Zhao}, Yong-Heng and {Chu}, Yao-Quan and {Jing}, Yi-Peng and {Deng}, Li-Cai},
        title = "{LAMOST spectral survey {\textemdash} An overview}",
      journal = {Research in Astronomy and Astrophysics},
         year = 2012,
        month = jul,
       volume = {12},
       number = {7},
        pages = {723-734},
          doi = {10.1088/1674-4527/12/7/002},
       adsurl = {https://ui.adsabs.harvard.edu/abs/2012RAA....12..723Z}
}

@ARTICLE{Karlsson+08,
       author = {{Karlsson}, Torgny and {Johnson}, Jarrett L. and {Bromm}, Volker},
        title = "{Uncovering the Chemical Signature of the First Stars in the Universe}",
      journal = {\apj},
         year = 2008,
        month = may,
       volume = {679},
       number = {1},
        pages = {6-16},
          doi = {10.1086/533520},
archivePrefix = {arXiv},
       eprint = {0709.4025},
 primaryClass = {astro-ph},
       adsurl = {https://ui.adsabs.harvard.edu/abs/2008ApJ...679....6K}
}

@ARTICLE{Whalen+08,
       author = {{Whalen}, Daniel and {van Veelen}, Bob and {O'Shea}, Brian W. and {Norman}, Michael L.},
        title = "{The Destruction of Cosmological Minihalos by Primordial Supernovae}",
      journal = {\apj},
         year = 2008,
        month = jul,
       volume = {682},
       number = {1},
        pages = {49-67},
          doi = {10.1086/589643},
archivePrefix = {arXiv},
       eprint = {0801.3698},
 primaryClass = {astro-ph},
       adsurl = {https://ui.adsabs.harvard.edu/abs/2008ApJ...682...49W}
}

@ARTICLE{Whalen+13a,
       author = {{Whalen}, Daniel J. and {Johnson}, Jarrett L. and {Smidt}, Joseph and {Meiksin}, Avery and {Heger}, Alexander and {Even}, Wesley and {Fryer}, Chris L.},
        title = "{The Supernova that Destroyed a Protogalaxy: Prompt Chemical Enrichment and Supermassive Black Hole Growth}",
      journal = {\apj},
         year = 2013,
        month = sep,
       volume = {774},
       number = {1},
          eid = {64},
        pages = {64},
          doi = {10.1088/0004-637X/774/1/64},
archivePrefix = {arXiv},
       eprint = {1305.6966},
 primaryClass = {astro-ph.CO},
       adsurl = {https://ui.adsabs.harvard.edu/abs/2013ApJ...774...64W}
}

@ARTICLE{Greif+10,
       author = {{Greif}, Thomas H. and {Glover}, Simon C.~O. and {Bromm}, Volker and {Klessen}, Ralf S.},
        title = "{The First Galaxies: Chemical Enrichment, Mixing, and Star Formation}",
      journal = {\apj},
         year = 2010,
        month = jun,
       volume = {716},
       number = {1},
        pages = {510-520},
          doi = {10.1088/0004-637X/716/1/510},
archivePrefix = {arXiv},
       eprint = {1003.0472},
 primaryClass = {astro-ph.CO},
       adsurl = {https://ui.adsabs.harvard.edu/abs/2010ApJ...716..510G}
}

@ARTICLE{Wise+12,
       author = {{Wise}, John H. and {Turk}, Matthew J. and {Norman}, Michael L. and {Abel}, Tom},
        title = "{The Birth of a Galaxy: Primordial Metal Enrichment and Stellar Populations}",
      journal = {\apj},
         year = 2012,
        month = jan,
       volume = {745},
       number = {1},
          eid = {50},
        pages = {50},
          doi = {10.1088/0004-637X/745/1/50},
archivePrefix = {arXiv},
       eprint = {1011.2632},
 primaryClass = {astro-ph.CO},
       adsurl = {https://ui.adsabs.harvard.edu/abs/2012ApJ...745...50W}
}

@ARTICLE{Wang+12,
       author = {{Wang}, F.~Y. and {Bromm}, Volker and {Greif}, Thomas H. and {Stacy}, Athena and {Dai}, Z.~G. and {Loeb}, Abraham and {Cheng}, K.~S.},
        title = "{Probing Pre-galactic Metal Enrichment with High-redshift Gamma-Ray Bursts}",
      journal = {\apj},
         year = 2012,
        month = nov,
       volume = {760},
       number = {1},
          eid = {27},
        pages = {27},
          doi = {10.1088/0004-637X/760/1/27},
archivePrefix = {arXiv},
       eprint = {1207.3879},
 primaryClass = {astro-ph.CO},
       adsurl = {https://ui.adsabs.harvard.edu/abs/2012ApJ...760...27W}
}

@ARTICLE{Ritter+12,
       author = {{Ritter}, Jeremy S. and {Safranek-Shrader}, Chalence and {Gnat}, Orly and {Milosavljevi{\'c}}, Milo{\v{s}} and {Bromm}, Volker},
        title = "{Confined Population III Enrichment and the Prospects for Prompt Second-generation Star Formation}",
      journal = {\apj},
         year = 2012,
        month = dec,
       volume = {761},
       number = {1},
          eid = {56},
        pages = {56},
          doi = {10.1088/0004-637X/761/1/56},
archivePrefix = {arXiv},
       eprint = {1203.2957},
 primaryClass = {astro-ph.CO},
       adsurl = {https://ui.adsabs.harvard.edu/abs/2012ApJ...761...56R}
}

@ARTICLE{Ritter+15,
       author = {{Ritter}, Jeremy S. and {Sluder}, Alan and {Safranek-Shrader}, Chalence and {Milosavljevi{\'c}}, Milo{\v{s}} and {Bromm}, Volker},
        title = "{Metal transport and chemical heterogeneity in early star forming systems}",
      journal = {\mnras},
         year = 2015,
        month = aug,
       volume = {451},
       number = {2},
        pages = {1190-1198},
          doi = {10.1093/mnras/stv982},
archivePrefix = {arXiv},
       eprint = {1408.0319},
 primaryClass = {astro-ph.GA},
       adsurl = {https://ui.adsabs.harvard.edu/abs/2015MNRAS.451.1190R}
}

@ARTICLE{Johnson+13_SN,
       author = {{Johnson}, Jarrett L. and {Whalen}, Daniel J. and {Even}, Wesley and {Fryer}, Chris L. and {Heger}, Alex and {Smidt}, Joseph and {Chen}, Ke-Jung},
        title = "{The Biggest Explosions in the Universe}",
      journal = {\apj},
         year = 2013,
        month = oct,
       volume = {775},
       number = {2},
          eid = {107},
        pages = {107},
          doi = {10.1088/0004-637X/775/2/107},
archivePrefix = {arXiv},
       eprint = {1304.4601},
 primaryClass = {astro-ph.CO},
       adsurl = {https://ui.adsabs.harvard.edu/abs/2013ApJ...775..107J}
}

@ARTICLE{Pan+13,
       author = {{Pan}, Liubin and {Scannapieco}, Evan and {Scalo}, Jon},
        title = "{Modeling the Pollution of Pristine Gas in the Early Universe}",
      journal = {\apj},
         year = 2013,
        month = oct,
       volume = {775},
       number = {2},
          eid = {111},
        pages = {111},
          doi = {10.1088/0004-637X/775/2/111},
archivePrefix = {arXiv},
       eprint = {1306.4663},
 primaryClass = {astro-ph.GA},
       adsurl = {https://ui.adsabs.harvard.edu/abs/2013ApJ...775..111P}
}

@ARTICLE{CookeMadau14,
       author = {{Cooke}, Ryan J. and {Madau}, Piero},
        title = "{Carbon-enhanced Metal-poor Stars: Relics from the Dark Ages}",
      journal = {\apj},
         year = 2014,
        month = aug,
       volume = {791},
       number = {2},
          eid = {116},
        pages = {116},
          doi = {10.1088/0004-637X/791/2/116},
archivePrefix = {arXiv},
       eprint = {1405.7369},
 primaryClass = {astro-ph.GA},
       adsurl = {https://ui.adsabs.harvard.edu/abs/2014ApJ...791..116C}
}

@ARTICLE{Jeon+15,
       author = {{Jeon}, Myoungwon and {Bromm}, Volker and {Pawlik}, Andreas H. and {Milosavljevi{\'c}}, Milo{\v{s}}},
        title = "{The first galaxies: simulating their feedback-regulated assembly}",
      journal = {\mnras},
         year = 2015,
        month = sep,
       volume = {452},
       number = {2},
        pages = {1152-1170},
          doi = {10.1093/mnras/stv1353},
archivePrefix = {arXiv},
       eprint = {1501.01002},
 primaryClass = {astro-ph.GA},
       adsurl = {https://ui.adsabs.harvard.edu/abs/2015MNRAS.452.1152J}
}

@ARTICLE{Ji+15,
       author = {{Ji}, Alexander P. and {Frebel}, Anna and {Bromm}, Volker},
        title = "{Preserving chemical signatures of primordial star formation in the first low-mass stars}",
      journal = {\mnras},
         year = 2015,
        month = nov,
       volume = {454},
       number = {1},
        pages = {659-674},
          doi = {10.1093/mnras/stv2052},
archivePrefix = {arXiv},
       eprint = {1508.06137},
 primaryClass = {astro-ph.GA},
       adsurl = {https://ui.adsabs.harvard.edu/abs/2015MNRAS.454..659J}
}

@ARTICLE{Sluder+16,
       author = {{Sluder}, Alan and {Ritter}, Jeremy S. and {Safranek-Shrader}, Chalence and {Milosavljevi{\'c}}, Milo{\v{s}} and {Bromm}, Volker},
        title = "{Abundance anomalies in metal-poor stars from Population III supernova ejecta hydrodynamics}",
      journal = {\mnras},
         year = 2016,
        month = feb,
       volume = {456},
       number = {2},
        pages = {1410-1423},
          doi = {10.1093/mnras/stv2587},
archivePrefix = {arXiv},
       eprint = {1505.07126},
 primaryClass = {astro-ph.GA},
       adsurl = {https://ui.adsabs.harvard.edu/abs/2016MNRAS.456.1410S}
}

@ARTICLE{Tarumi+20,
       author = {{Tarumi}, Yuta and {Hartwig}, Tilman and {Magg}, Mattis},
        title = "{Implications of Inhomogeneous Metal Mixing for Stellar Archaeology}",
      journal = {\apj},
         year = 2020,
        month = jul,
       volume = {897},
       number = {1},
          eid = {58},
        pages = {58},
          doi = {10.3847/1538-4357/ab960d},
archivePrefix = {arXiv},
       eprint = {2005.10401},
 primaryClass = {astro-ph.GA},
       adsurl = {https://ui.adsabs.harvard.edu/abs/2020ApJ...897...58T}
}

@ARTICLE{Magg+20,
       author = {{Magg}, Mattis and {Nordlander}, Thomas and {Glover}, Simon C.~O. and {Hansen}, Camilla J. and {Ishigaki}, Miho and {Heger}, Alexander and {Klessen}, Ralf S. and {Kobayashi}, Chiaki and {Nomoto}, Ken'ichi},
        title = "{A minimum dilution scenario for supernovae and consequences for extremely metal-poor stars}",
      journal = {\mnras},
         year = 2020,
        month = nov,
       volume = {498},
       number = {3},
        pages = {3703-3712},
          doi = {10.1093/mnras/staa2624},
archivePrefix = {arXiv},
       eprint = {2006.12517},
 primaryClass = {astro-ph.GA},
       adsurl = {https://ui.adsabs.harvard.edu/abs/2020MNRAS.498.3703M}
}

@ARTICLE{Magg+22,
       author = {{Magg}, Mattis and {Schauer}, Anna T.~P. and {Klessen}, Ralf S. and {Glover}, Simon C.~O. and {Tress}, Robin G. and {Jaura}, Ondrej},
        title = "{Metal Mixing in Minihalos: The Descendants of Pair-instability Supernovae}",
      journal = {\apj},
         year = 2022,
        month = apr,
       volume = {929},
       number = {2},
          eid = {119},
        pages = {119},
          doi = {10.3847/1538-4357/ac5aac},
archivePrefix = {arXiv},
       eprint = {2110.15372},
 primaryClass = {astro-ph.GA},
       adsurl = {https://ui.adsabs.harvard.edu/abs/2022ApJ...929..119M}
}

@ARTICLE{Kordt+26,
       author = {{Kordt}, Aron and {Glover}, Simon C.~O. and {Klessen}, Ralf S.},
        title = "{Metal Enrichment by the First Stars Exploding at the Lower Energy Limit of Pair-Instability Supernovae}",
      journal = {arXiv e-prints},
         year = 2026,
        month = may,
          eid = {arXiv:2605.11647},
        pages = {arXiv:2605.11647},
          doi = {10.48550/arXiv.2605.11647},
archivePrefix = {arXiv},
       eprint = {2605.11647},
 primaryClass = {astro-ph.GA},
       adsurl = {https://ui.adsabs.harvard.edu/abs/2026arXiv260511647K}
}

@ARTICLE{Gutcke+22,
       author = {{Gutcke}, Thales A. and {Pakmor}, R{\"u}diger and {Naab}, Thorsten and {Springel}, Volker},
        title = "{LYRA - II. Cosmological dwarf galaxy formation with inhomogeneous Population III enrichment}",
      journal = {\mnras},
         year = 2022,
        month = jun,
       volume = {513},
       number = {1},
        pages = {1372-1385},
          doi = {10.1093/mnras/stac867},
archivePrefix = {arXiv},
       eprint = {2110.06233},
 primaryClass = {astro-ph.GA},
       adsurl = {https://ui.adsabs.harvard.edu/abs/2022MNRAS.513.1372G}
}

@ARTICLE{Mead+25,
       author = {{Mead}, Jennifer and {Brauer}, Kaley and {Bryan}, Greg L. and {Mac Low}, Mordecai-Mark and {Ji}, Alexander P. and {Wise}, John H. and {Emerick}, Andrew and {Andersson}, Eric P. and {Frebel}, Anna and {C{\^o}t{\'e}}, Benoit},
        title = "{AEOS: Transport of Metals from Minihalos following Population III Stellar Feedback}",
      journal = {\apj},
         year = 2025,
        month = feb,
       volume = {980},
       number = {1},
          eid = {62},
        pages = {62},
          doi = {10.3847/1538-4357/ada3c1},
archivePrefix = {arXiv},
       eprint = {2411.14209},
 primaryClass = {astro-ph.GA},
       adsurl = {https://ui.adsabs.harvard.edu/abs/2025ApJ...980...62M}
}

@ARTICLE{Rey+25,
       author = {{Rey}, Martin P. and {Katz}, Harley and {Cadiou}, Corentin and {Sanati}, Mahsa and {Agertz}, Oscar and {Blaizot}, Jeremy and {Cameron}, Alex J. and {Choustikov}, Nicholas and {Devriendt}, Julien and {Hauk}, Uliana and {Ji}, Alexander P. and {Jones}, Gareth C. and {Kimm}, Taysun and {Laseter}, Isaac and {Martin-Alvarez}, Sergio and {Matsumoto}, Kosei and {Pearce}, Autumn and {Revaz}, Yves and {Rodriguez Montero}, Francisco and {Rosdahl}, Joki and {Saxena}, Aayush and {Slyz}, Adrianne and {Stiskalek}, Richard and {Storck}, Anatole and {Veenema}, Oscar and {Yee}, Wonjae},
        title = "{MEGATRON: how the first stars create an iron metallicity plateau in the smallest dwarf galaxies}",
      journal = {arXiv e-prints},
         year = 2025,
        month = oct,
          eid = {arXiv:2510.05232},
        pages = {arXiv:2510.05232},
          doi = {10.48550/arXiv.2510.05232},
archivePrefix = {arXiv},
       eprint = {2510.05232},
 primaryClass = {astro-ph.GA},
       adsurl = {https://ui.adsabs.harvard.edu/abs/2025arXiv251005232R}
}

@ARTICLE{Storck+26,
       author = {{Storck}, Anatole and {Katz}, Harley and {Devriendt}, Julien and {Slyz}, Adrianne and {Cadiou}, Corentin and {Choustikov}, Nicholas and {Rey}, Martin P. and {Saxena}, Aayush and {Agertz}, Oscar and {Kimm}, Taysun},
        title = "{MEGATRON: the environments of Population III stars at Cosmic Dawn and their connection to present-day galaxies}",
      journal = {\mnras},
         year = 2026,
        month = may,
       volume = {548},
       number = {1},
          eid = {stag529},
        pages = {stag529},
          doi = {10.1093/mnras/stag529},
archivePrefix = {arXiv},
       eprint = {2510.06853},
 primaryClass = {astro-ph.GA},
       adsurl = {https://ui.adsabs.harvard.edu/abs/2026MNRAS.548ag529S}
}

@ARTICLE{Samuel+26,
       author = {{Samuel}, Jenna and {Boylan-Kolchin}, Michael and {Feldmann}, Robert and {Hopkins}, Philip and {Sun}, Guochao and {Gandhi}, Pratik and {Venditti}, Alessandra and {Shen}, Xuejian and {Wetzel}, Andrew and {Moreno}, Jorge and {Munoz}, Julian and {Cochrane}, Rachel and {Faucher-Giguere}, Claude-Andre and {Bromm}, Volker and {Finkelstein}, Steven and {Straight}, Maria and {Painter}, Connor and {Stern}, Jonathan and {Bullock}, James},
        title = "{Resolving galaxy formation in the early Universe with BonFIRE and CampFIRE}",
      journal = {arXiv e-prints},
         year = 2026,
        month = may,
          eid = {arXiv:2605.24104},
        pages = {arXiv:2605.24104},
          doi = {10.48550/arXiv.2605.24104},
archivePrefix = {arXiv},
       eprint = {2605.24104},
 primaryClass = {astro-ph.GA},
       adsurl = {https://ui.adsabs.harvard.edu/abs/2026arXiv260524104S}
}

@ARTICLE{Simcoe+06,
       author = {{Simcoe}, Robert A.},
        title = "{High-Redshift Intergalactic C IV Abundance Measurements from the Near-Infrared Spectra of Two z \raisebox{-0.5ex}\textasciitilde 6 QSOs}",
      journal = {\apj},
         year = 2006,
        month = dec,
       volume = {653},
       number = {2},
        pages = {977-987},
          doi = {10.1086/508983},
archivePrefix = {arXiv},
       eprint = {astro-ph/0605710},
 primaryClass = {astro-ph},
       adsurl = {https://ui.adsabs.harvard.edu/abs/2006ApJ...653..977S}
}

@ARTICLE{Kobayashi+11,
       author = {{Kobayashi}, Chiaki and {Tominaga}, Nozomu and {Nomoto}, Ken'ichi},
        title = "{Chemical Enrichment in the Carbon-enhanced Damped Ly{\ensuremath{\alpha}} System by Population III Supernovae}",
      journal = {\apjl},
         year = 2011,
        month = apr,
       volume = {730},
       number = {2},
          eid = {L14},
        pages = {L14},
          doi = {10.1088/2041-8205/730/2/L14},
archivePrefix = {arXiv},
       eprint = {1101.1227},
 primaryClass = {astro-ph.CO},
       adsurl = {https://ui.adsabs.harvard.edu/abs/2011ApJ...730L..14K}
}

@ARTICLE{Cooke+17,
       author = {{Cooke}, Ryan J. and {Pettini}, Max and {Steidel}, Charles C.},
        title = "{Discovery of the most metal-poor damped Lyman-{\ensuremath{\alpha}} system}",
      journal = {\mnras},
         year = 2017,
        month = may,
       volume = {467},
       number = {1},
        pages = {802-811},
          doi = {10.1093/mnras/stx037},
archivePrefix = {arXiv},
       eprint = {1701.03103},
 primaryClass = {astro-ph.CO},
       adsurl = {https://ui.adsabs.harvard.edu/abs/2017MNRAS.467..802C}
}

@ARTICLE{Welsh+19,
       author = {{Welsh}, Louise and {Cooke}, Ryan and {Fumagalli}, Michele},
        title = "{Modelling the chemical enrichment of Population III supernovae: the origin of the metals in near-pristine gas clouds}",
      journal = {\mnras},
         year = 2019,
        month = aug,
       volume = {487},
       number = {3},
        pages = {3363-3376},
          doi = {10.1093/mnras/stz1526},
archivePrefix = {arXiv},
       eprint = {1906.00009},
 primaryClass = {astro-ph.SR},
       adsurl = {https://ui.adsabs.harvard.edu/abs/2019MNRAS.487.3363W}
}

@ARTICLE{Welsh+21,
       author = {{Welsh}, Louise and {Cooke}, Ryan and {Fumagalli}, Michele},
        title = "{The stochastic enrichment of Population II stars}",
      journal = {\mnras},
         year = 2021,
        month = jan,
       volume = {500},
       number = {4},
        pages = {5214-5228},
          doi = {10.1093/mnras/staa3342},
archivePrefix = {arXiv},
       eprint = {2010.10532},
 primaryClass = {astro-ph.GA},
       adsurl = {https://ui.adsabs.harvard.edu/abs/2021MNRAS.500.5214W}
}

@ARTICLE{Welsh+22,
       author = {{Welsh}, Louise and {Cooke}, Ryan and {Fumagalli}, Michele and {Pettini}, Max},
        title = "{Oxygen-enhanced Extremely Metal-poor Damped Ly{\ensuremath{\alpha}} Systems: A Signpost of the First Stars?}",
      journal = {\apj},
         year = 2022,
        month = apr,
       volume = {929},
       number = {2},
          eid = {158},
        pages = {158},
          doi = {10.3847/1538-4357/ac4503},
archivePrefix = {arXiv},
       eprint = {2201.08394},
 primaryClass = {astro-ph.GA},
       adsurl = {https://ui.adsabs.harvard.edu/abs/2022ApJ...929..158W}
}

@ARTICLE{Welsh+23,
       author = {{Welsh}, Louise and {Cooke}, Ryan and {Fumagalli}, Michele and {Pettini}, Max},
        title = "{Towards ultra metal-poor DLAs: linking the chemistry of the most metal-poor DLA to the first stars}",
      journal = {\mnras},
         year = 2023,
        month = oct,
       volume = {525},
       number = {1},
        pages = {527-541},
          doi = {10.1093/mnras/stad2181},
archivePrefix = {arXiv},
       eprint = {2307.03771},
 primaryClass = {astro-ph.GA},
       adsurl = {https://ui.adsabs.harvard.edu/abs/2023MNRAS.525..527W}
}

@ARTICLE{Welsh+24,
       author = {{Welsh}, Louise and {Cooke}, Ryan and {Fumagalli}, Michele and {Pettini}, Max and {Rudie}, Gwen C.},
        title = "{A survey of extremely metal-poor gas at cosmic noon: Evidence of elevated [O/Fe]}",
      journal = {\aap},
         year = 2024,
        month = nov,
       volume = {691},
          eid = {A285},
        pages = {A285},
          doi = {10.1051/0004-6361/202451147},
archivePrefix = {arXiv},
       eprint = {2409.07525},
 primaryClass = {astro-ph.GA},
       adsurl = {https://ui.adsabs.harvard.edu/abs/2024A&A...691A.285W}
}

@ARTICLE{Robert+22,
       author = {{Robert}, P. Fr{\'e}d{\'e}ric and {Murphy}, Michael T. and {O'Meara}, John M. and {Crighton}, Neil H.~M. and {Fumagalli}, Michele},
        title = "{Discovery of three new near-pristine absorption clouds at z = 2.6-4.4}",
      journal = {\mnras},
         year = 2022,
        month = aug,
       volume = {514},
       number = {3},
        pages = {3559-3578},
          doi = {10.1093/mnras/stac1550},
archivePrefix = {arXiv},
       eprint = {2206.02947},
 primaryClass = {astro-ph.CO},
       adsurl = {https://ui.adsabs.harvard.edu/abs/2022MNRAS.514.3559R}
}

@ARTICLE{Saccardi+23a,
       author = {{Saccardi}, Andrea and {Salvadori}, Stefania and {D'Odorico}, Valentina and {Cupani}, Guido and {Fumagalli}, Michele and {Berg}, Trystyn A.~M. and {Becker}, George D. and {Ellison}, Sara and {Lopez}, Sebastian},
        title = "{Evidence of First Stars-enriched Gas in High-redshift Absorbers}",
      journal = {\apj},
         year = 2023,
        month = may,
       volume = {948},
       number = {1},
          eid = {35},
        pages = {35},
          doi = {10.3847/1538-4357/acc39f},
archivePrefix = {arXiv},
       eprint = {2305.02346},
 primaryClass = {astro-ph.GA},
       adsurl = {https://ui.adsabs.harvard.edu/abs/2023ApJ...948...35S}
}

@ARTICLE{Saccardi+23b,
       author = {{Saccardi}, A. and {Vergani}, S.~D. and {De Cia}, A. and {D'Elia}, V. and {Heintz}, K.~E. and {Izzo}, L. and {Palmerio}, J.~T. and {Petitjean}, P. and {Rossi}, A. and {de Ugarte Postigo}, A. and {Christensen}, L. and {Konstantopoulou}, C. and {Levan}, A.~J. and {Malesani}, D.~B. and {M{\o}ller}, P. and {Ramburuth-Hurt}, T. and {Salvaterra}, R. and {Tanvir}, N.~R. and {Th{\"o}ne}, C.~C. and {Vejlgaard}, S. and {Fynbo}, J.~P.~U. and {Kann}, D.~A. and {Schady}, P. and {Watson}, D.~J. and {Wiersema}, K. and {Campana}, S. and {Covino}, S. and {De Pasquale}, M. and {Fausey}, H. and {Hartmann}, D.~H. and {van der Horst}, A.~J. and {Jakobsson}, P. and {Palazzi}, E. and {Pugliese}, G. and {Savaglio}, S. and {Starling}, R.~L.~C. and {Stratta}, G. and {Zafar}, T.},
        title = "{Dissecting the interstellar medium of a z = 6.3 galaxy. X-shooter spectroscopy and HST imaging of the afterglow and environment of the Swift GRB 210905A}",
      journal = {\aap},
         year = 2023,
        month = mar,
       volume = {671},
          eid = {A84},
        pages = {A84},
          doi = {10.1051/0004-6361/202244205},
archivePrefix = {arXiv},
       eprint = {2211.16524},
 primaryClass = {astro-ph.GA},
       adsurl = {https://ui.adsabs.harvard.edu/abs/2023A&A...671A..84S}
}

@ARTICLE{Davies+23,
       author = {{Davies}, Rebecca L. and {Ryan-Weber}, E. and {D'Odorico}, V. and {Bosman}, S.~E.~I. and {Meyer}, R.~A. and {Becker}, G.~D. and {Cupani}, G. and {Keating}, L.~C. and {Bischetti}, M. and {Davies}, F.~B. and {Eilers}, A.-C. and {Farina}, E.~P. and {Haehnelt}, M.~G. and {Pallottini}, A. and {Zhu}, Y.},
        title = "{Examining the decline in the C IV content of the Universe over 4.3 {\ensuremath{\lesssim}} z {\ensuremath{\lesssim}} 6.3 using the E-XQR-30 sample}",
      journal = {\mnras},
         year = 2023,
        month = may,
       volume = {521},
       number = {1},
        pages = {314-331},
          doi = {10.1093/mnras/stad294},
archivePrefix = {arXiv},
       eprint = {2303.02816},
 primaryClass = {astro-ph.GA},
       adsurl = {https://ui.adsabs.harvard.edu/abs/2023MNRAS.521..314D}
}

@ARTICLE{Christensen+23,
       author = {{Christensen}, L. and {Jakobsen}, P. and {Willott}, C. and {Arribas}, S. and {Bunker}, A. and {Charlot}, S. and {Maiolino}, R. and {Marshall}, M. and {Perna}, M. and {{\"U}bler}, H.},
        title = "{Metal enrichment and evolution in four z > 6.5 quasar sightlines observed with JWST/NIRSpec}",
      journal = {\aap},
         year = 2023,
        month = dec,
       volume = {680},
          eid = {A82},
        pages = {A82},
          doi = {10.1051/0004-6361/202347943},
archivePrefix = {arXiv},
       eprint = {2309.06470},
 primaryClass = {astro-ph.GA},
       adsurl = {https://ui.adsabs.harvard.edu/abs/2023A&A...680A..82C}
}

@ARTICLE{Zou+24,
       author = {{Zou}, Siwei and {Cai}, Zheng and {Wang}, Feige and {Fan}, Xiaohui and {Champagne}, Jaclyn B. and {Hennawi}, Joseph F. and {Schindler}, Jan-Torge and {Farina}, Emanuele Paolo and {Yang}, Jinyi and {Inayoshi}, Kohei and {Ba{\~n}ados}, Eduardo and {Bosman}, Sarah E.~I. and {Li}, Zihao and {Lin}, Xiaojing and {Wu}, Yunjing and {Sun}, Fengwu and {Guo}, Ziyi and {Kulkuarni}, Girish and {Habouzit}, M{\'e}lanie and {Charlot}, Stephane and {Chevallard}, Jacopo and {Connor}, Thomas and {Eilers}, Anna-Christina and {Jiang}, Linhua and {Jin}, Xiangyu and {Kakiichi}, Koki and {Li}, Mingyu and {Meyer}, Romain A. and {Walter}, Fabian and {Zhang}, Huanian},
        title = "{A SPectroscopic survey of biased halos In the Reionization Era (ASPIRE): Impact of Galaxies on the Circumgalactic Medium Metal Enrichment at z > 6 Using the JWST and VLT}",
      journal = {\apjl},
         year = 2024,
        month = mar,
       volume = {963},
       number = {1},
          eid = {L28},
        pages = {L28},
          doi = {10.3847/2041-8213/ad23e7},
archivePrefix = {arXiv},
       eprint = {2402.00113},
 primaryClass = {astro-ph.GA},
       adsurl = {https://ui.adsabs.harvard.edu/abs/2024ApJ...963L..28Z}
}

@ARTICLE{Sebastian+24,
       author = {{Sebastian}, Alma Maria and {Ryan-Weber}, Emma and {Davies}, Rebecca L. and {Becker}, George D. and {Keating}, Laura C. and {D'Odorico}, Valentina and {Meyer}, Romain A. and {Bosman}, Sarah E.~I. and {Cupani}, Guido and {Kulkarni}, Girish and {Haehnelt}, Martin G. and {Lai}, Samuel and {Eilers}, Anna-Christina and {Bischetti}, Manuela and {Gallerani}, Simona},
        title = "{E-XQR-30: The evolution of Mg II, C II, and O I across 2 < z < 6}",
      journal = {\mnras},
         year = 2024,
        month = may,
       volume = {530},
       number = {2},
        pages = {1829-1848},
          doi = {10.1093/mnras/stae789},
archivePrefix = {arXiv},
       eprint = {2403.10072},
 primaryClass = {astro-ph.GA},
       adsurl = {https://ui.adsabs.harvard.edu/abs/2024MNRAS.530.1829S}
}

@ARTICLE{Sodini+24,
       author = {{Sodini}, Alessio and {D'Odorico}, Valentina and {Salvadori}, Stefania and {Vanni}, Irene and {Bischetti}, Manuela and {Cupani}, Guido and {Davies}, Rebecca and {Becker}, George D. and {Ba{\~n}ados}, Eduardo and {Bosman}, Sarah and {Davies}, Frederick and {Paolo Farina}, Emanuele and {Ferrara}, Andrea and {Keating}, Laura and {Kulkarni}, Girish and {Lai}, Samuel and {Ryan-Weber}, Emma and {Sebastian}, Alma Maria and {Walter}, Fabian},
        title = "{Evidence of Pop III stars' chemical signature in neutral gas at z {\ensuremath{\sim}} 6. A study based on the E-XQR-30 spectroscopic sample}",
      journal = {\aap},
         year = 2024,
        month = jul,
       volume = {687},
          eid = {A314},
        pages = {A314},
          doi = {10.1051/0004-6361/202349062},
archivePrefix = {arXiv},
       eprint = {2404.10722},
 primaryClass = {astro-ph.GA},
       adsurl = {https://ui.adsabs.harvard.edu/abs/2024A&A...687A.314S}
}

@ARTICLE{Durocikova+25,
       author = {{{\v{D}}urov{\v{c}}{\'\i}kov{\'a}}, Dominika and {Eilers}, Anna-Christina and {Simcoe}, Robert A. and {Welsh}, Louise and {Meyer}, Romain A. and {Matthee}, Jorryt and {Ryan-Weber}, Emma V. and {Yue}, Minghao and {Katz}, Harley and {Satyavolu}, Sindhu and {Becker}, George and {Davies}, Frederick B. and {Farina}, Emanuele Paolo},
        title = "{An Extremely Metal-poor Ly{\ensuremath{\alpha}} Emitter Candidate at z = 6 Revealed through Absorption Spectroscopy}",
      journal = {\apjl},
         year = 2025,
        month = jul,
       volume = {987},
       number = {2},
          eid = {L33},
        pages = {L33},
          doi = {10.3847/2041-8213/ade71c},
archivePrefix = {arXiv},
       eprint = {2505.01499},
 primaryClass = {astro-ph.GA},
       adsurl = {https://ui.adsabs.harvard.edu/abs/2025ApJ...987L..33D}
}

@ARTICLE{Visbal+26,
       author = {{Visbal}, Eli and {Bryan}, Greg L. and {Haiman}, Zolt{\'a}n},
        title = "{Chemical signatures of Population III stars in damped Lyman-{\ensuremath{\alpha}} absorption systems at z ≍ 6}",
      journal = {\jcap},
         year = 2026,
        month = feb,
       volume = {2026},
       number = {2},
          eid = {077},
        pages = {077},
          doi = {10.1088/1475-7516/2026/02/077},
archivePrefix = {arXiv},
       eprint = {2506.14482},
 primaryClass = {astro-ph.GA},
       adsurl = {https://ui.adsabs.harvard.edu/abs/2026JCAP...02..077V}
}

@ARTICLE{Higginson+26,
       author = {{Higginson}, Jack and {Bordoloi}, Rongmon and {Simcoe}, Robert A. and {Matthee}, Jorryt and {Kashino}, Daichi and {Mackenzie}, Ruari and {Kramarenko}, Ivan and {Lilly}, Simon J. and {Eilers}, Anna-Christina and {Naidu}, Rohan P. and {Yue}, Minghao},
        title = "{EIGER. VIII. First Stars Signatures in the Connection between O I Absorption and Galaxies in the Epoch of Reionization}",
      journal = {\apj},
         year = 2026,
        month = mar,
       volume = {999},
       number = {1},
          eid = {49},
        pages = {49},
          doi = {10.3847/1538-4357/ae3e81},
archivePrefix = {arXiv},
       eprint = {2510.05220},
 primaryClass = {astro-ph.GA},
       adsurl = {https://ui.adsabs.harvard.edu/abs/2026ApJ...999...49H}
}

@ARTICLE{Schneider+04,
       author = {{Schneider}, R. and {Ferrara}, A. and {Salvaterra}, R.},
        title = "{Dust formation in very massive primordial supernovae}",
      journal = {\mnras},
         year = 2004,
        month = jul,
       volume = {351},
       number = {4},
        pages = {1379-1386},
          doi = {10.1111/j.1365-2966.2004.07876.x},
archivePrefix = {arXiv},
       eprint = {astro-ph/0307087},
 primaryClass = {astro-ph},
       adsurl = {https://ui.adsabs.harvard.edu/abs/2004MNRAS.351.1379S}
}

@ARTICLE{Chiaki+25,
       author = {{Chiaki}, Gen and {Nozawa}, Takaya and {Kobayashi}, Chiaki and {Tominaga}, Nozomu},
        title = "{Population III Supernovae as a dust factory I --- molecule formation and mixing/fallback in ejecta}",
      journal = {arXiv e-prints},
         year = 2025,
        month = apr,
          eid = {arXiv:2504.17506},
        pages = {arXiv:2504.17506},
          doi = {10.48550/arXiv.2504.17506},
archivePrefix = {arXiv},
       eprint = {2504.17506},
 primaryClass = {astro-ph.GA},
       adsurl = {https://ui.adsabs.harvard.edu/abs/2025arXiv250417506C}
}

@ARTICLE{Otaki+26,
       author = {{Otaki}, Koki and {Schneider}, Raffaella and {Graziani}, Luca and {Bonella}, Alessandro and {Marassi}, Stefania and {Limongi}, Marco and {Bianchi}, Simone},
        title = "{Effective supernova dust yields from rotating and nonrotating stellar progenitors}",
      journal = {\aap},
         year = 2026,
        month = apr,
       volume = {708},
          eid = {A136},
        pages = {A136},
          doi = {10.1051/0004-6361/202555906},
archivePrefix = {arXiv},
       eprint = {2506.11931},
 primaryClass = {astro-ph.GA},
       adsurl = {https://ui.adsabs.harvard.edu/abs/2026A&A...708A.136O}
}

@ARTICLE{Gebhardt+21,
       author = {{Gebhardt}, Karl and {Mentuch Cooper}, Erin and {Ciardullo}, Robin and {Acquaviva}, Viviana and {Bender}, Ralf and {Bowman}, William P. and {Castanheira}, Barbara G. and {Dalton}, Gavin and {Davis}, Dustin and {de Jong}, Roelof S. and {DePoy}, D.~L. and {Devarakonda}, Yaswant and {Dongsheng}, Sun and {Drory}, Niv and {Fabricius}, Maximilian and {Farrow}, Daniel J. and {Feldmeier}, John and {Finkelstein}, Steven L. and {Froning}, Cynthia S. and {Gawiser}, Eric and {Gronwall}, Caryl and {Herold}, Laura and {Hill}, Gary J. and {Hopp}, Ulrich and {House}, Lindsay R. and {Janowiecki}, Steven and {Jarvis}, Matthew and {Jeong}, Donghui and {Jogee}, Shardha and {Kakuma}, Ryota and {Kelz}, Andreas and {Kollatschny}, W. and {Komatsu}, Eiichiro and {Krumpe}, Mirko and {Landriau}, Martin and {Liu}, Chenxu and {Niemeyer}, Maja Lujan and {MacQueen}, Phillip and {Marshall}, Jennifer and {Mawatari}, Ken and {McLinden}, Emily M. and {Mukae}, Shiro and {Nagaraj}, Gautam and {Ono}, Yoshiaki and {Ouchi}, Masami and {Papovich}, Casey and {Sakai}, Nao and {Saito}, Shun and {Schneider}, Donald P. and {Schulze}, Andreas and {Shanmugasundararaj}, Khavvia and {Shetrone}, Matthew and {Sneden}, Chris and {Snigula}, Jan and {Steinmetz}, Matthias and {Thomas}, Benjamin P. and {Thomas}, Brianna and {Tuttle}, Sarah and {Urrutia}, Tanya and {Wisotzki}, Lutz and {Wold}, Isak and {Zeimann}, Gregory and {Zhang}, Yechi},
        title = "{The Hobby-Eberly Telescope Dark Energy Experiment (HETDEX) Survey Design, Reductions, and Detections}",
      journal = {\apj},
         year = 2021,
        month = dec,
       volume = {923},
       number = {2},
          eid = {217},
        pages = {217},
          doi = {10.3847/1538-4357/ac2e03},
archivePrefix = {arXiv},
       eprint = {2110.04298},
 primaryClass = {astro-ph.IM},
       adsurl = {https://ui.adsabs.harvard.edu/abs/2021ApJ...923..217G}
}

@ARTICLE{Fumagalli+24,
       author = {{Fumagalli}, Michele},
        title = "{The multiphase circumgalactic medium and its relation to galaxies: an observational perspective}",
      journal = {arXiv e-prints},
         year = 2024,
        month = aug,
          eid = {arXiv:2409.00174},
        pages = {arXiv:2409.00174},
          doi = {10.48550/arXiv.2409.00174},
archivePrefix = {arXiv},
       eprint = {2409.00174},
 primaryClass = {astro-ph.GA},
       adsurl = {https://ui.adsabs.harvard.edu/abs/2024arXiv240900174F}
}

@ARTICLE{MirzaKhanlari+25,
       author = {{Mirza Khanlari}, Mahan and {Gebhardt}, Karl and {Weiss}, Laurel H. and {Davis}, Dustin and {Mentuch Cooper}, Erin and {Qezlou}, Mahdi and {Lujan Niemeyer}, Maja and {Ciardullo}, Robin and {Schneider}, Donald P. and {Mukae}, Shiro and {Liu}, Chenxu and {Farrow}, Daniel and {Hill}, Gary J. and {Zeimann}, Gregory R. and {Kollatschny}, Wolfram},
        title = "{The HETDEX Survey: Probing Neutral Hydrogen in the Circumgalactic Medium of {\ensuremath{\sim}}88,000 Ly{\ensuremath{\alpha}} Emitters}",
      journal = {\apj},
         year = 2025,
        month = aug,
       volume = {989},
       number = {2},
          eid = {169},
        pages = {169},
          doi = {10.3847/1538-4357/adf10e},
archivePrefix = {arXiv},
       eprint = {2507.15942},
 primaryClass = {astro-ph.GA},
       adsurl = {https://ui.adsabs.harvard.edu/abs/2025ApJ...989..169K}
}

@ARTICLE{Yoshii+22,
       author = {{Yoshii}, Yuzuru and {Sameshima}, Hiroaki and {Tsujimoto}, Takuji and {Shigeyama}, Toshikazu and {Beers}, Timothy C. and {Peterson}, Bruce A.},
        title = "{Potential Signature of Population III Pair-instability Supernova Ejecta in the BLR Gas of the Most Distant Quasar at z = 7.54}",
      journal = {\apj},
         year = 2022,
        month = oct,
       volume = {937},
       number = {2},
          eid = {61},
        pages = {61},
          doi = {10.3847/1538-4357/ac8163},
archivePrefix = {arXiv},
       eprint = {2207.11909},
 primaryClass = {astro-ph.GA},
       adsurl = {https://ui.adsabs.harvard.edu/abs/2022ApJ...937...61Y}
}

@ARTICLE{Vanni+24,
       author = {{Vanni}, Irene and {Salvadori}, Stefania and {D'Odorico}, Valentina and {Becker}, George D. and {Cupani}, Guido},
        title = "{Chemical Diagnostics to Unveil Environments Enriched by First Stars}",
      journal = {\apjl},
         year = 2024,
        month = jun,
       volume = {967},
       number = {2},
          eid = {L22},
        pages = {L22},
          doi = {10.3847/2041-8213/ad46fa},
archivePrefix = {arXiv},
       eprint = {2402.18640},
 primaryClass = {astro-ph.GA},
       adsurl = {https://ui.adsabs.harvard.edu/abs/2024ApJ...967L..22V}
}

@ARTICLE{Nakane+24,
       author = {{Nakane}, Minami and {Ouchi}, Masami and {Nakajima}, Kimihiko and {Harikane}, Yuichi and {Tominaga}, Nozomu and {Takahashi}, Koh and {Kashino}, Daichi and {Yanagisawa}, Hiroto and {Watanabe}, Kuria and {Nomoto}, Ken'ichi and {Isobe}, Yuki and {Nishigaki}, Moka and {Ishigaki}, Miho N. and {Ono}, Yoshiaki and {Takeda}, Yui},
        title = "{Low [O/Fe] Ratio in a Luminous Galaxy at the Early Cosmic Epoch (z > 10): Signature of Short Delay Time or Bright Hypernovae/Pair-instability Supernovae?}",
      journal = {\apj},
         year = 2024,
        month = nov,
       volume = {976},
       number = {1},
          eid = {122},
        pages = {122},
          doi = {10.3847/1538-4357/ad84e8},
archivePrefix = {arXiv},
       eprint = {2407.14470},
 primaryClass = {astro-ph.GA},
       adsurl = {https://ui.adsabs.harvard.edu/abs/2024ApJ...976..122N}
}

@ARTICLE{Kobayashi+20,
       author = {{Kobayashi}, Chiaki and {Karakas}, Amanda I. and {Lugaro}, Maria},
        title = "{The Origin of Elements from Carbon to Uranium}",
      journal = {\apj},
         year = 2020,
        month = sep,
       volume = {900},
       number = {2},
          eid = {179},
        pages = {179},
          doi = {10.3847/1538-4357/abae65},
archivePrefix = {arXiv},
       eprint = {2008.04660},
 primaryClass = {astro-ph.GA},
       adsurl = {https://ui.adsabs.harvard.edu/abs/2020ApJ...900..179K}
}

\end{document}